\documentclass[reprint,prd,tightenlines,floatfix,
nofootinbib,eqsecnum,superscriptaddress,onecolumn]{revtex4-2}

\usepackage[T1]{fontenc}		% fnot encoding with polish letters !or OT4

\usepackage{amsmath,amsfonts,amssymb,amstext,mathrsfs}
\usepackage{mathpazo}

\usepackage[dvips]{graphicx}
\usepackage{epsf,float}
\usepackage{revsymb}

\usepackage{dcolumn}% Align table columns on decimal point
\usepackage{braket}
\usepackage{color,xcolor}
\usepackage{graphicx}
\usepackage{subfigure}
\usepackage{multirow}
\usepackage{tabularx}
\usepackage{pstricks}
\usepackage[section]{placeins}
\usepackage{booktabs}
\usepackage{array}

\usepackage{hyperref}
\usepackage[noabbrev]{cleveref}   

\usepackage{lineno}

\newcommand{\nn}{\nonumber}

\newcommand{\Pom}{\mathbb{P}}
\newcommand{\Ode}{\mathbb{O}}
\newcommand{\Reg}{\mathbb{R}}

\renewcommand\slash[1]{\not \! #1}

\usepackage[normalem]{ulem}  % \sout{old text} for strikeout

\begin{document}

%Title of paper
\title{Production of \boldmath{$\pi^{+} \pi^{-}$} pairs in diffractive photon-proton 
and in proton-proton collisions revisited, in particular concerning the Drell-S{\"o}ding contribution}

\author{Piotr Lebiedowicz}
%\orcid{0000-0003-1963-6263}
\email{Piotr.Lebiedowicz@ifj.edu.pl}
\affiliation{Institute of Nuclear Physics Polish Academy of Sciences, 
Radzikowskiego 152, PL-31342 Krak{\'o}w, Poland}

\author{Otto Nachtmann}
\email{O.Nachtmann@thphys.uni-heidelberg.de}
\affiliation{Institut f\"ur Theoretische Physik, Universit\"at Heidelberg,
Philosophenweg 16, D-69120 Heidelberg, Germany}

\author{Antoni Szczurek}
%\orcid{0000-0001-5247-8442}
\email{Antoni.Szczurek@ifj.edu.pl}
\affiliation{Institute of Nuclear Physics Polish Academy of Sciences, Radzikowskiego 152, PL-31342 Krak{\'o}w, Poland}
\affiliation{Institute of Physics, Faculty of Exact and Technical Sciences, University of Rzesz{\'o}w, 
Pigonia 1, PL-35310 Rzesz{\'o}w, Poland}

\begin{abstract}
We discuss the exclusive photoproduction of $\pi^{+}\pi^{-}$ pairs in photon-proton and in proton-proton collisions at high energies. The $\rho^{0}$, $\omega$, $f_{2}(1270)$, and non-resonant (Drell-S\"oding) contributions are considered. The calculation is based on the tensor-pomeron model that includes not only the dominant pomeron exchange but also reggeon and odderon exchanges. In the Drell-S\"oding contribution we have different subenergies for the $\pi^{+}p$ and $\pi^{-}p$ systems. In the method which we propose now we take this into account. Respecting the gauge-invariance constraints is then a nontrivial problem for which, however, we present a solution here. In the present paper we give in this way a substantial improvement of the calculations for real photoproduction of $\pi^{+}\pi^{-}$ from JHEP 01, 151 (2015), and we extend the calculations to low $Q^{2}$ electroproduction, $0 \leqslant Q^{2} \leqslant 0.5$ GeV$^{2}$. The photo- and electroproduction amplitudes are then the basis for the calculation of central exclusive production (CEP) of $\pi^{+}\pi^{-}$ pairs in proton-proton collisions, where at least one proton participates in the CEP via a virtual-photon emission. The revised model leads to enhanced cross sections and gives an increased skewing of the $\rho^{0}$ spectral shape. For the $pp \to pp \pi^{+}\pi^{-}$ reaction, we calculate differential cross sections as function of the two-pion invariant mass, pion transverse momentum and pion pseudorapidity. Predictions of proton-pion and proton-pion-pion invariant mass distributions and the distribution in the proton-proton four-momentum transfer squared are also presented. This research is relevant in the context of ALICE, ATLAS, CMS, and LHCb measurements in $pp$ collisions, even when the leading protons are not detected and instead only rapidity-gap conditions are checked experimentally. Our results can also serve as basis for the description of coherent $\pi^{+}\pi^{-}$ production in ultra-peripheral $p$A and AA collisions at the LHC. The formulas given in our paper can directly be used for the analysis of photoproduction and small-$Q^{2}$ electroproduction in $ep$ collisions at high energies. Such data exist from the HERA experiments and will be obtained in the future at the electron-ion colliders.
\end{abstract}

\maketitle

%--------------------------------------------------
\section{Introduction}
\label{sec:1}
%--------------------------------------------------

In this paper we study the production of $\pi^{+} \pi^{-}$ pairs 
in photon-proton collisions, and the central exclusive production (CEP)
of such pairs in proton-proton collisions,
\begin{eqnarray}
&&\gamma^{(*)} + p \to \pi^{+} + \pi^{-} + p\,,
\label{1.1}\\
&&p + p \to p + \pi^{+} + \pi^{-} + p\,.
\label{1.2}
\end{eqnarray}
Here $\gamma^{(*)}$ denotes a real or slightly virtual photon.
Out of all possible contributions to (\ref{1.2}) we consider only those involving
photon exchange at least from the side of one of the initial protons.

We are interested in the diffractive regime, that is, 
in high center-of-mass (c.m.) energies $(\sqrt{s})$
and small momentum transfers $(\sqrt{|t|})$. Our focus will be on $\pi^{+}\pi^{-}$ invariant
masses in the region of the $\rho^{0}(770)$ resonance and below.
An old problem there is to understand the shape of the $\rho^{0}$
resonance in the reactions (\ref{1.1}) and (\ref{1.2}).
Compared to the $\rho^{0}$ shape measured in $e^{+}e^{-}$ annihilation
there is a skewing of the $\rho^{0}$ shape observed in the reactions (\ref{1.1}) and (\ref{1.2}).
Already a long time ago this skewing was attributed to the interference
of the decay $\rho^{0} \to \pi^{+}\pi^{-}$
with the non-resonant production of $\pi^{+}\pi^{-}$,
the Drell-S\"oding term \cite{Drell:1960zz,Drell:1961zz,Soding:1965nh}.
In practice, the calculation of this term is a tricky problem,
not the least due to requirements of gauge invariance.
For general references concerning the reactions of the type (\ref{1.1}) and (\ref{1.2})
see \cite{Bauer:1977iq,Donnachie:2002,Close:2007,Fiore:2003iv,Szczurek:2004xe,
Forshaw:2012im,Berger:1972an,Ballam:1971wq,Park:1971ts,
Ballam:1971yd,Ballam:1972eq,Gladding:1973cac,
Struczinski:1975ik,Egloff:1979mg,Aston:1982hr,
ZEUS:1995bfs,H1:1996prv,ZEUS:1997rof,
Baur:2001jj,Bertulani:2005ru,Baltz:2007kq,Lebiedowicz:2009pj,Staszewski:2011bg,
Adler:2002sc,STAR:2007elq,Abelev:2008ew,Agakishiev:2011me,
Nystrand:2014vra}.

In our present paper we shall use the framework of the
tensor-pomeron model \cite{Ewerz:2013kda} which
has been constructed in order to describe soft high-energy hadronic reactions.
%This model was formulated in terms of effective propagators and vertices for the exchange objects: the pomeron, the odderon, and the reggeons.
The pomeron ($\Pom$) and the charge conjugation $C = +1$ reggeons
are considered as effective rank-two symmetric tensor exchanges
and the odderon ($\Ode$) and the $C = -1$ reggeons
as effective vector exchanges.
Due to the properties of these individual Regge exchanges 
the amplitudes obtained in this framework
naturally obey the rules of quantum field theory (QFT).
In particular, they exhibit the correct behavior under crossing 
and charge conjugation.
Moreover, in this approach, applications of
the vector-meson-dominance model to photon-induced reactions are
straightforward and do not lead to gauge-invariance problems; 
see section~4.3 of \cite{Ewerz:2013kda}.

Further details about the tensor-pomeron model, 
as well as many other results not discussed here,
can be found in \cite{Ewerz:2016onn,Britzger:2019lvc}.
In \cite{Ewerz:2016onn}, it was shown 
that the tensor-pomeron ansatz is the only one compatible 
with the STAR results from polarized elastic proton-proton scattering \cite{STAR:2012fiw}
when considering three different options for the soft pomeron: scalar, vector, and tensor.
In \cite{Britzger:2019lvc}, the authors provided further evidence against the hypothesis that the pomeron has vector character.
The tensor-pomeron model was applied also 
to various central exclusive production
(CEP) reactions in proton-proton collisions,
starting with \cite{Lebiedowicz:2013ika}; 
see \cite{Lebiedowicz:2016ioh,Lebiedowicz:2016zka,Lebiedowicz:2018sdt,Lebiedowicz:2018eui,Lebiedowicz:2019por,Lebiedowicz:2019boz,Lebiedowicz:2019jru,Lebiedowicz:2020yre,Lebiedowicz:2021pzd}.
In a recent paper \cite{Lebiedowicz:2025num} we discussed
CEP of $\eta$ and $\eta'$ mesons in the tensor-pomeron model.
There we also showed that with a scalar-pomeron ansatz
CEP of $\eta$, $\eta'$, and $f_{1}(1285)$ is not possible.
But these reactions are perfectly allowed 
in the tensor-pomeron model.

In \cite{Bolz:2014mya} photoproduction of a $\pi^{+} \pi^{-}$ pair (\ref{1.1})
was studied theoretically using the tensor-pomeron model \cite{Ewerz:2013kda}.
For the calculation of the Drell-S\"oding term 
a gauge invariant coupling Lagrangian was introduced
describing the $\gamma \pi \pi$, $\gamma \gamma \pi \pi$, 
$\Pom \pi \pi$, and $\Pom \gamma \pi \pi$ vertices;
see (B.66)--(B.71) of \cite{Bolz:2014mya}.
In the various diagrams arising in the calculation 
of the Drell-S\"oding term
in sections~2.5 and 2.6 of \cite{Bolz:2014mya} a common energy variable
was used in the respective Regge factors.
In this way a skewing of the $\rho^{0}$ shape was achieved
but compared to experiment it was not big enough;
see for instance \cite{H1:2020lzc,Bolz_Meson2021}.
The same model was then used by us in \cite{Lebiedowicz:2014bea}
for the reaction (\ref{1.2}).
In the present article,
we present an improved calculation of the Drell-S\"oding term
for the reactions (\ref{1.1}) and (\ref{1.2}).
Now we use for each diagram the appropriate energy variable
in the Regge factors and we present a new method how to incorporate
the gauge-invariance constraints in this framework.
We think that our new method is rather satisfactory
from the point of view of Quantum Field Theory (QFT).
In the present paper we shall not only discuss
the above improvement for the calculations
of the amplitudes of real photoproduction.
As main new results we shall present the amplitudes 
for (\ref{1.1}) which are needed
for electroproduction at small photon virtualities,
$0 \leqslant Q^{2} \leqslant 0.5$~GeV$^{2}$.

As mentioned above, diffractive photoproduction of $\pi^{+} \pi^{-}$ pairs
in the $pp \to pp \pi^{+} \pi^{-}$ reaction (\ref{1.2})
was studied in \cite{Lebiedowicz:2014bea}
within the tensor-pomeron approach.
There, the resonant (through the $\rho^{0} \equiv \rho(770)$ 
and $\rho(1450)$ mesons)
and non-resonant continuum (Drell-S\"oding) contributions 
to CEP of $\pi^{+} \pi^{-}$ pairs in $pp$ collisions were considered.
The model was then applied in \cite{Lebiedowicz:2016ryp} to estimate 
the size of the background from proton diffractive dissociation
to the $pp \to pp \rho^{0}$ reaction at LHC energies.
The ratio of integrated cross sections for
the $pp \to pN \rho^{0} \pi$ diffractive dissociation processes 
%where $pN \rho^{0} \pi$ stands for $pn \rho^{0} \pi^{+}$
%plus $pp \rho^{0} \pi^{0}$, 
to the $pp \to pp \rho^{0}$ reaction is of order (7-10)\% \cite{Lebiedowicz:2016ryp}.
Here $pN \rho^{0} \pi$ stands for $pn \rho^{0} \pi^{+}$ plus $pp \rho^{0} \pi^{0}$.
Thus, in experiments where the final state protons 
are not detected and only large rapidity gaps 
around the centrally produced $\pi^{+} \pi^{-}$ pairs are required,
diffractive dissociation reactions constitute an important and non-negligible contribution.
Experimental results for this type of measurement at the LHC
relating to the $pp \to pp \pi^{+} \pi^{-}$ reaction
have been published by the CMS Collaboration \cite{CMS:2020jbb}. 
The two-pion invariant mass distribution was also measured 
by the ALICE Collaboration in $pp$ collisions 
at $\sqrt{s} = 7$~TeV \cite{Schicker:2011ug}.
The purely hadronic diffractive resonant and non-resonant contributions 
for the $pp \to pp \pi^{+} \pi^{-}$ reaction
via the double-pomeron/reggeon exchanges
have been discussed in \cite{Lebiedowicz:2016ioh,Lebiedowicz:2019por}
and a comparison of theoretical results with 
the STAR, CMS, and ATLAS-ALFA experimental data  
can be found in \cite{Lebieodwicz_talk}.

More recently, in \cite{TOTEM:2024aso}, 
the CMS and TOTEM Collaborations 
published a study of CEP of $\pi^{+} \pi^{-}$ pairs 
in proton-proton collisions at $\sqrt{s} = 13$~TeV.
The CMS-TOTEM measurement was carried out
in the invariant mass regions,
$M_{\pi^{+} \pi^{-}} < 0.7$~GeV 
and $M_{\pi^{+} \pi^{-}} > 1.8$~GeV,
for pion rapidities $|{\rm y}_{\pi}|<2$,
and for proton transverse momenta, 
between 0.2 and 0.8~GeV.
The advantage of this measurement 
compared to that of \cite{CMS:2020jbb}
is that it is based on two intact tagged protons 
after interaction and thus free of assumptions on proton dissociation.
However, as we will demonstrate in our paper, 
requiring the above condition on the transverse momenta of the scattered protons 
leads to a significant reduction of the photoproduction contribution. 
As a result, the dominant production mechanism
becomes the double-pomeron exchange mechanism
which is not the topic of study in our present paper.

The proper model for the Drell-S{\"o}ding contribution
is also needed for the description of data 
from coherent photoproduction of $\rho^{0} \to \pi^{+} \pi^{-}$
measured in ultra-peripheral $p$A and AA collisions
%at a nucleon-nucleon center-of-mass energy 
%$\sqrt{s_{NN}} = 5.02$~TeV
by the STAR \cite{STAR:2017enh},
ALICE \cite{ALICE:2020ugp,ALICE:2021jnv},
ATLAS \cite{ATLAS:2025nac},
CMS \cite{CMS:2019awk},
and 
LHCb \cite{McNulty:2017ejl,LHCb:2025fzk} Collaborations.
%coherent photoproduction of $\rho^{0}$ meson
The interference of dipion continuum and
$\rho^{0}$ contributions
is a significant effect also there.
%However, the nonresonant contribution
%is usually described using a constant parameter;
%see, e.g., eq.~(4) and Table~2 of \cite{LHCb:2025fzk}
%and the $p$Pb collisions 
%at a nucleon-nucleon center-of-mass energy 
%$\sqrt{s_{NN}} = 8.16$~TeV \cite{McNulty:2017ejl}.

The numerical values of coupling constants occurring in effective vertices,
the functional form of vertex form factors with cutoff parameters,
and the parameters for the Regge trajectories
have been obtained in \cite{Ewerz:2013kda} 
from the comparison of the model 
to experimental data for soft high-energy reactions.
These parameter values should be considered as default values.
The list of these parameters and their default values for the model can be found 
in table~1 of appendix~B in \cite{Bolz:2014mya}.
It should be noted that in section~II of \cite{Lebiedowicz:2014bea}
we analyzed the $\gamma p \to \rho^{0} p$ reaction
comparing our theory with the data from HERA
and with a compilation of low-energy data
(see figure~4 of \cite{Lebiedowicz:2014bea} for references).
We found the parameter set~A for the $\Pom \rho \rho$ and
$f_{2 \Reg} \rho \rho$ coupling constants
that differs somewhat from the default values 
given in \cite{Bolz:2014mya};
see also the discussion in appendix~\ref{sec:appendixB} 
of the present paper.

%In this article we shell work with a technique
%similar to that presented in \cite{Lebiedowicz:2021byo}
%for the $\pi \pi \to \pi \pi \gamma$ processes
%(see the discussion in section~IV~A)
%that allows to improve the corresponding calculation of photoproduction
%of $\pi^{+} \pi^{-}$ pairs presented
%in \cite{Bolz:2014mya,Lebiedowicz:2014bea}.
%The results presented in \cite{Bolz_Meson2021} 
%suggest that the existing model
%does not describe the $d\sigma/dM_{\pi\pi}$ distribution 
%for the $\gamma p \to \pi^{+} \pi^{-} p$ reaction
%measured by the H1 Collaboration \cite{H1:2020lzc}.

Our paper is organized as follows.
In section~\ref{sec:2} we give analytic expressions
for the amplitudes for 
the $\gamma^{(*)} p \to \pi^{+} \pi^{-} p$ reaction
with pomeron and reggeon exchanges.
Section~\ref{sec:3} deals with the $pp \to pp \pi^{+} \pi^{-}$ reaction
for both resonant and non-resonant (Drell-S\"oding) contributions 
to CEP of $\pi^{+} \pi^{-}$ pairs.
The main purpose of our present paper is to provide these
theoretical expressions 
for the amplitudes of the reactions (\ref{1.1}) and (\ref{1.2}).
For (\ref{1.1}) there are data from the HERA experiments \cite{ZEUS:1995bfs,H1:1996prv,ZEUS:1997rof,H1:2020lzc,Bolz_Meson2021}.
We would like to invite experimentalists to reanalyze 
the HERA data in detail using our new formulas.
In this way one would also have a reference for the analysis
of the reaction~(\ref{1.1}) at the future electron-ion colliders,
EIC \cite{Accardi:2012qut,Aschenauer:2017jsk,AbdulKhalek:2021gbh,Anderle:2021wcy} 
and LHeC \cite{LHeCStudyGroup:2012zhm,Bordry:2018gri}.
Also there our formulas of section~\ref{sec:2} apply.
The formulas of section~\ref{sec:3} are for the reaction~(\ref{1.2})
which is currently being studied at the LHC.
We present in section~\ref{sec:4}
numerical results of our calculations 
for the $pp \to pp \pi^{+} \pi^{-}$ reaction.
These results are only meant to illustrate the type and size
of the effects which one gets by using our formulas for (\ref{1.2}).
Again we invite the experimentalists to make detailed analyses
of their data using our theoretical results 
presented in section~\ref{sec:3}.
In section~\ref{sec:4} we also refer to the H1 data 
\cite{H1:2020lzc}
for the $\gamma p \to \pi^{+} \pi^{-} p$ reaction
and preliminary results presented in \cite{Bolz_Meson2021}.
Section~\ref{sec:5} contains our conclusions.
In appendix~\ref{sec:appendixA} we discuss the kinematics and 
the propagators and vertices needed for our calculations.
We also present details of our calculations.
Many formulas which we use in our present paper are taken from \cite{Bolz:2014mya}
and we will refer for the explicit expressions to this paper.
In appendix~\ref{sec:appendixB}
we discuss uncertainties of our model by comparing it with the HERA data for the $\gamma p \to \rho^{0} p$ reaction
and with data for the $\pi^{-} p \to \pi^{-} p$ reaction.

%--------------------------------------------------
\section{\boldmath Theoretical formalism for the $\gamma^{(*)} p \to \pi^{+} \pi^{-} p$ reaction}
\label{sec:2}
%--------------------------------------------------

Here we consider the reaction
\begin{eqnarray}
\gamma^{(*)}(q, \mu) + p(p, \mathfrak{s}) \to \pi^{+}(k_{1}) + \pi^{-}(k_{2}) + p(p',\mathfrak{s'})\,,
\label{2.1}
\end{eqnarray}
where $\gamma^{(*)}$ stands for a real or slightly virtual photon.
We denote by $k_{1}$, $k_{2}$, $p$, $p'$, and $q$
the four-momenta of the involved particles,
by $\mu$ the vector index of $\gamma^{(*)}$,
and by $\mathfrak{s}$ and $\mathfrak{s'}$ the spin indices of the incoming and outgoing proton, respectively.
The matrix element for the reaction (\ref{2.1}) 
with a real photon of polarization vector $\epsilon$
is
\begin{eqnarray}
\braket{\pi^{+}(k_{1}),\pi^{-}(k_{2}),p(p',\mathfrak{s'})|{\cal T}|
\gamma(q, \epsilon),p(p,\mathfrak{s})} = \epsilon^{\mu}
{\cal M}_{\mu, \mathfrak{s'},\mathfrak{s}}(k_{1},k_{2},p',q,p)\,.
\label{2.2}
\end{eqnarray}
For a real photon $q^{2} = 0$ applies. 
But in the following we shall consider 
\begin{eqnarray}
{\cal M}_{\mu, \mathfrak{s'},\mathfrak{s}}(k_{1},k_{2},p',q,p)
\label{2.2a}
\end{eqnarray}
also for slightly virtual photons.
To be precise, we consider the region
\begin{eqnarray}
0 \leqslant Q^{2} = -q^{2} \leqslant 0.5~{\rm GeV}^{2} \,,
\label{2.2b}
\end{eqnarray}
which we motivate as follows. 
In \cite{Britzger:2019lvc} photoproduction and low-$x$ deep inelastic scattering (DIS)
were studied using the tensor-pomeron model with a soft and a hard tensor pomeron.
It was found that for high energies the total photoabsorption cross section,
where $Q^{2} = 0$,
and the DIS structure functions for 
$0 < Q^{2} \leqslant 0.5~{\rm GeV}^{2}$ are completely
dominated by the soft-pomeron exchange;
see figures~5, 8, and 10 of \cite{Britzger:2019lvc}.
We also note that in \cite{Lebiedowicz:2022xgi} the two-tensor-pomeron
model was applied to real and virtual Compton scattering on the proton
($\gamma^{(*)} p \to \gamma p$).
For real Compton scattering ($Q^{2} = 0$) the cross section
is dominated by the soft-pomeron exchange
with an additional contribution from reggeon exchange
at lower energies; see figures~3 and 4 of \cite{Lebiedowicz:2022xgi}.
In (\ref{2.2b}) we assume that these findings from 
\cite{Britzger:2019lvc,Lebiedowicz:2022xgi}
also hold for our reaction (\ref{2.1}) here.

\begin{figure}[tbp]
\centering
(a)\includegraphics[width=7.cm]{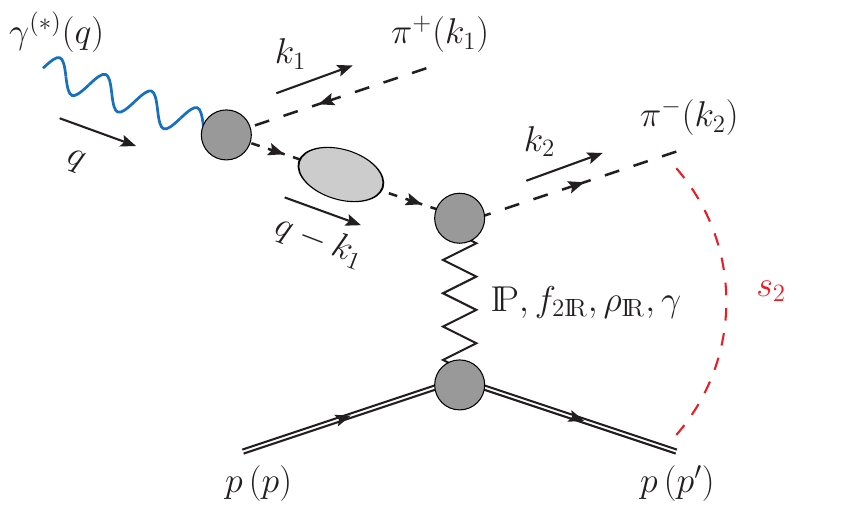}
(b)\includegraphics[width=7.cm]{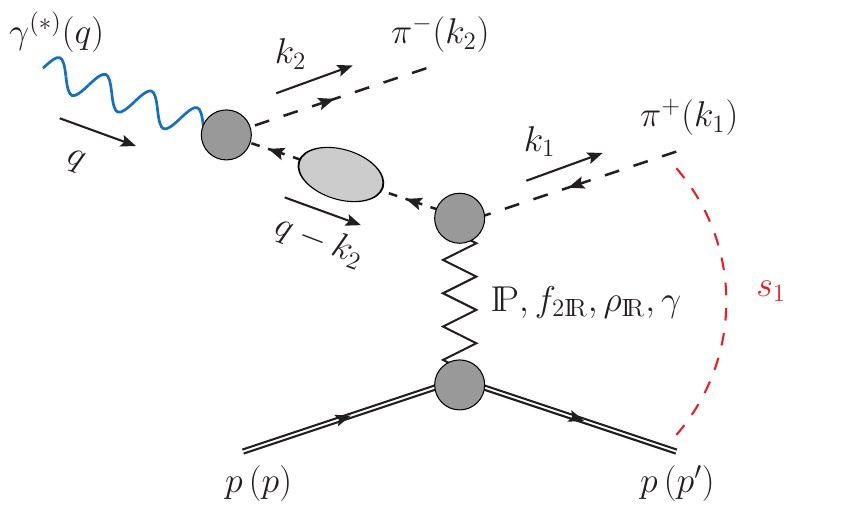}
(c)\includegraphics[width=4.6cm]{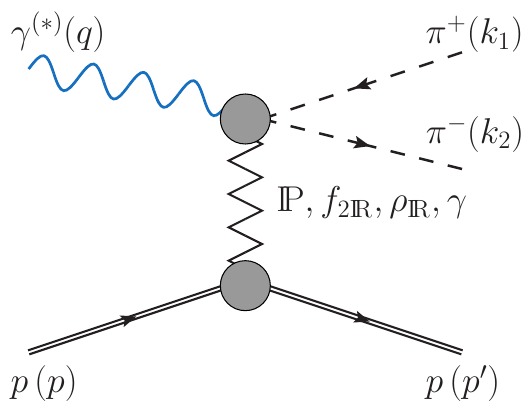}
\caption{\label{fig:100}
Diagrams for non-resonant production of a $\pi^{+} \pi^{-}$ pair in the
$\gamma^{(*)} p \to \pi^{+} \pi^{-} p$ reaction with
$\Pom$, $f_{2 \Reg}$, $\rho_{\Reg}$, and $\gamma$ exchange.
The variables $s_{1}$ and $s_{2}$ are defined 
in (\ref{A2}) of appendix~\ref{sec:appendixA}.}
\end{figure}

The list of the relevant diagrams for the reaction (\ref{2.1})
is given in figure~1 of \cite{Bolz:2014mya}.
We are interested in high energies and small momentum transfers,
that is, the regime of Regge exchanges.
In the present paper we shall first study the non-resonant
production of a $\pi^{+} \pi^{-}$ pair where the diagrams
are shown in figure~\ref{fig:100}.
There, the blobs denote the full vertices and the complete pion propagators.
We denote the contributions to the matrix element ${\cal M}_{\mu, \mathfrak{s'},\mathfrak{s}}$ (\ref{2.2a})
from the Drell-S{\"o}ding diagrams (a)-(c) of figure~\ref{fig:100}
with exchange of $\Pom, \ldots, \gamma$ by
\begin{eqnarray}
{\cal M}_{\mu, \mathfrak{s'},\mathfrak{s}}^{(a,b,c)} |_{\Pom}\,, \ldots, 
{\cal M}_{\mu, \mathfrak{s'},\mathfrak{s}}^{(a,b,c)} |_{\gamma}\,.
\label{2.2c}
\end{eqnarray}
%

%------------------------------------------------------------------
%-------------------------------------------------------------------

The diagrams with $\gamma$ exchange in figure~\ref{fig:100}
have been calculated in \cite{Bolz:2014mya} and we shall,
in essence, take over the result given there; 
see section~2.6 of \cite{Bolz:2014mya}.
The corresponding matrix element is discussed
in subsection~\ref{sec:2B6}.

In the following we set for the sum of the hadronic exchanges
$h = \Pom + f_{2 \Reg} + \rho_{\Reg}$,
\begin{eqnarray}
{\cal M}^{(\rm DS)}_{\mu, \mathfrak{s'},\mathfrak{s}}(k_{1},k_{2},p',q,p) |_{h} 
&= &
{\cal M}_{\mu, \mathfrak{s'},\mathfrak{s}}^{(a+b+c)}(k_{1},k_{2},p',q,p) |_{h}
 \nonumber \\
&= & {\cal M}_{\mu, \mathfrak{s'},\mathfrak{s}}^{(a)}(k_{1},k_{2},p',q,p) |_{h} + 
{\cal M}_{\mu, \mathfrak{s'},\mathfrak{s}}^{(b)}(k_{1},k_{2},p',q,p) |_{h} 
+
{\cal M}_{\mu, \mathfrak{s'},\mathfrak{s}}^{(c)}(k_{1},k_{2},p',q,p) |_{h}\,. \qquad
\label{2.3}
\end{eqnarray}
We have with the vertex functions, propagators, and amplitudes,
defined in appendix~\ref{sec:appendixA}
\begin{eqnarray}
{\cal M}_{\mu, \mathfrak{s'},\mathfrak{s}}^{(a)}(k_{1},k_{2},p',q,p) |_{h} &=&
- \Gamma_{\mu}^{(\gamma \pi^{+} \pi^{+})}(k_{1},k_{1}-q) 
\Delta_{F}[(k_{1}-q)^{2}]
{\cal M}_{\mathfrak{s'},\mathfrak{s}}^{(0, \,a)}|_{h}\nonumber \\
&=&
e \widehat{\Gamma}_{\mu}^{(\gamma \pi \pi)}(k_{1},k_{1}-q) 
\Delta_{F}[(k_{1}-q)^{2}]
{\cal M}_{\mathfrak{s'},\mathfrak{s}}^{(0, \,a)}|_{h}\,,\nonumber \\
{\cal M}_{\mathfrak{s'},\mathfrak{s}}^{(0, \,a)}|_{h} \equiv
{\cal M}_{\mathfrak{s'},\mathfrak{s}}^{(0, \,a)}(k_{2},p',q-k_{1},p)|_{h} &=&
{\cal M}_{\mathfrak{s'},\mathfrak{s}}^{(\pi^{-})}(k_{2},p',q-k_{1},p)|_{h}\,, \qquad
\label{2.4}
\end{eqnarray}
\begin{eqnarray}
{\cal M}_{\mu, \mathfrak{s'},\mathfrak{s}}^{(b)}(k_{1},k_{2},p',q,p) |_{h} &=&
- \Gamma_{\mu}^{(\gamma \pi^{-} \pi^{-})}(k_{2},k_{2}-q) 
\Delta_{F}[(k_{2}-q)^{2}]
{\cal M}_{\mathfrak{s'},\mathfrak{s}}^{(0, \,b)}|_{h}\nonumber \\
&=&
-e \widehat{\Gamma}_{\mu}^{(\gamma \pi \pi)}(k_{2},k_{2}-q) 
\Delta_{F}[(k_{2}-q)^{2}]
{\cal M}_{\mathfrak{s'},\mathfrak{s}}^{(0, \,b)}|_{h}\,,\nonumber \\
{\cal M}_{\mathfrak{s'},\mathfrak{s}}^{(0, \,b)}|_{h} \equiv
{\cal M}_{\mathfrak{s'},\mathfrak{s}}^{(0, \,b)}(k_{1},p',q-k_{2},p)|_{h} &=&
{\cal M}_{\mathfrak{s'},\mathfrak{s}}^{(\pi^{+})}(k_{1},p',q-k_{2},p)|_{h}\,. \qquad
\label{2.5}
\end{eqnarray}

Gauge invariance requires
\begin{eqnarray}
q^{\mu} {\cal M}^{(\rm DS)}_{\mu, \mathfrak{s'},\mathfrak{s}}(k_{1},k_{2},p',q,p)|_{h} &=&
q^{\mu} \Big{(} {\cal M}_{\mu, \mathfrak{s'},\mathfrak{s}}^{(a)}(k_{1},k_{2},p',q,p)|_{h} 
\nonumber \\
&&
+ {\cal M}_{\mu, \mathfrak{s'},\mathfrak{s}}^{(b)}(k_{1},k_{2},p',q,p)|_{h} 
\nonumber \\
&&
+ {\cal M}_{\mu, \mathfrak{s'},\mathfrak{s}}^{(c)}(k_{1},k_{2},p',q,p)|_{h} \Big{)}  = 0\,.
\label{2.6}
\end{eqnarray}
Using (\ref{A7a}) and the generalized Ward identity (\ref{A9})
we find
\begin{eqnarray}
q^{\mu} {\cal M}_{\mu, \mathfrak{s'},\mathfrak{s}}^{(a)}(k_{1},k_{2},p',q,p)|_{h} 
&=& -e {\cal M}_{\mathfrak{s'},\mathfrak{s}}^{(0, \,a)}|_{h}\,,  \nonumber \\
q^{\mu} {\cal M}_{\mu, \mathfrak{s'},\mathfrak{s}}^{(b)}(k_{1},k_{2},p',q,p)|_{h} 
&=& e {\cal M}_{\mathfrak{s'},\mathfrak{s}}^{(0, \,b)}|_{h}\,.
\label{2.7}
\end{eqnarray}
Therefore, we have
\begin{eqnarray}
q^{\mu} {\cal M}_{\mu, \mathfrak{s'},\mathfrak{s}}^{(c)}(k_{1},k_{2},p',q,p)|_{h} 
&=& e 
\Bigl[ 
{\cal M}_{\mathfrak{s'},\mathfrak{s}}^{(0, \,a)}(k_{2},p',q-k_{1},p)|_{h}
-{\cal M}_{\mathfrak{s'},\mathfrak{s}}^{(0, \,b)}(k_{1},p',q-k_{2},p)|_{h}
\Bigr] \,. \nn \\
\label{2.8}
\end{eqnarray}
The analogous relations hold for the individual exchanges
$\Pom$, $f_{2 \Reg}$, and $\rho_{\Reg}$ separately.

%--------------------------------------------------
\subsection{Calculation of the diagrams (a)--(c) of figure \ref{fig:100} for pomeron exchange}
\label{sec:2A}
%--------------------------------------------------

From (\ref{2.4}), (\ref{A21c}), and (\ref{A33}) we get
\begin{align}
{\cal M}_{\mu, \mathfrak{s'},\mathfrak{s}}^{(a)}(k_{1},k_{2},p',q,p)|_{\Pom}
=&\;
e \left[ \frac{(2 k_{1} - q)_{\mu}}{-2 k_{1} \cdot q + q^{2} + i \varepsilon}
F_{M}(q^{2}) - q_{\mu} \frac{1 - F_{M}(q^{2})}{q^{2}}
\right]
{\cal M}_{\mathfrak{s'},\mathfrak{s}}^{(0, \,a)}(k_{2},p',q-k_{1},p)|_{\Pom} 
\nonumber \\
=&\;
ie \left[
\frac{(2 k_{1} - q)_{\mu}}{-2 k_{1} \cdot q + q^{2} + i \varepsilon}
F_{M}(q^{2}) - q_{\mu} \frac{1 - F_{M}(q^{2})}{q^{2}}
\right]
{\cal F}_{\Pom \pi p}(2 \nu_{2}, t) \nn \\
& \times
\Big[
2(k_{2}-k_{1}+q)^{\nu}(k_{2}-k_{1}+q,p'+p)
\bar{u}_{\mathfrak{s'}}(p') \gamma_{\nu} u_{\mathfrak{s}}(p)
\nonumber \\
&
-(k_{2}-k_{1}+q)^{2} m_{p}
\bar{u}_{\mathfrak{s'}}(p') u_{\mathfrak{s}}(p)
\Big].
\label{2.9}
\end{align}
From (\ref{2.5}), (\ref{A21c}), and (\ref{A33}) we find
\begin{align}
{\cal M}_{\mu, \mathfrak{s'},\mathfrak{s}}^{(b)}(k_{1},k_{2},p',q,p)|_{\Pom}
=&
-e \left[
\frac{(2 k_{2} - q)_{\mu}}{-2 k_{2} \cdot q + q^{2} + i \varepsilon}
F_{M}(q^{2}) - q_{\mu} \frac{1 - F_{M}(q^{2})}{q^{2}}
\right] 
{\cal M}_{\mathfrak{s'},\mathfrak{s}}^{(0, \,b)}(k_{1},p',q-k_{2},p)|_{\Pom} 
\nonumber \\
=&
-ie \left[
\frac{(2 k_{2} - q)_{\mu}}{-2 k_{2} \cdot q + q^{2} + i \varepsilon}
F_{M}(q^{2}) - q_{\mu} \frac{1 - F_{M}(q^{2})}{q^{2}}
\right]
{\cal F}_{\Pom \pi p}(2 \nu_{1}, t)\nn \\
& \times
\Big[
2(k_{2}-k_{1}-q)^{\nu}(k_{2}-k_{1}-q,p'+p)
\bar{u}_{\mathfrak{s'}}(p') \gamma_{\nu} u_{\mathfrak{s}}(p) 
\nonumber \\
&
-(k_{2}-k_{1}-q)^{2} m_{p}
\bar{u}_{\mathfrak{s'}}(p') u_{\mathfrak{s}}(p)
\Big].
\label{2.10}
\end{align}
For $q^{\mu} {\cal M}_{\mu, \mathfrak{s'},\mathfrak{s}}^{(c)}|_{\Pom}$
we obtain from (\ref{2.8})--(\ref{2.10})
\begin{eqnarray}
q^{\mu} {\cal M}_{\mu, \mathfrak{s'},\mathfrak{s}}^{(c)}(k_{1},k_{2},p',q,p)|_{\Pom}
&=&
ie 
\bigg{\lbrace}
{\cal F}_{\Pom \pi p}(2 \nu_{2}, t)
\bigg{[}
2(k_{2}-k_{1}+q)^{\nu}(k_{2}-k_{1}+q,p'+p)
-\frac{1}{2} (k_{2}-k_{1}+q)^{2} (p'+p)^{\nu}
\bigg{]}
\nonumber \\
&& 
-{\cal F}_{\Pom \pi p}(2 \nu_{1}, t)
\bigg{[}
2(k_{2}-k_{1}-q)^{\nu}(k_{2}-k_{1}-q,p'+p)
-\frac{1}{2} (k_{2}-k_{1}-q)^{2} (p'+p)^{\nu}
\bigg{]}
\bigg{\rbrace} 
\nonumber \\
&& 
\times \bar{u}_{\mathfrak{s'}}(p') \gamma_{\nu} u_{\mathfrak{s}}(p) \,.
\label{2.11}
\end{eqnarray}

Now the challenge is to find an adequate solution for ${\cal M}_{\mu, \mathfrak{s'},\mathfrak{s}}^{(c)}|_{\Pom}$
from~(\ref{2.11}).
For this we shall write the r.h.s. of (\ref{2.11})
in a way that it is explicitly proportional to $q^{\mu}$.
We have from (\ref{A4})--(\ref{A5h}), and (\ref{A32})
\begin{eqnarray}
{\cal F}_{\Pom \pi p}(2 \nu_{2}, t) &=&
2 \beta_{\Pom \pi \pi}\, 3 \beta_{\Pom NN}\, F_{M}(t) F_{1}(t) 
\frac{1}{2}
\left(- \frac{i}{2}\alpha'_{\Pom}\right)^{\alpha_{\Pom}(t)-1} 
(4 \nu_{2})^{\alpha_{\Pom}(t)-2} \,, 
\label{2.12}\\
(4 \nu_{2})^{\alpha_{\Pom}(t)-2} &=&
(16 \nu_{2}^{2})^{\frac{\alpha_{\Pom}(t)-2}{2}} \nn\\
&=& [16 \bar{\nu}^{2} (1 - \varkappa)]^{\frac{\alpha_{\Pom}(t)-2}{2}}\nn\\
&=& (16 \bar{\nu}^{2})^{\frac{\alpha_{\Pom}(t)-2}{2}}
\left[
1 + \frac{2 - \alpha_{\Pom}(t)}{2} \varkappa \,
g\left( \frac{2 - \alpha_{\Pom}(t)}{2}, \varkappa \right)
\right].
\label{2.13}
\end{eqnarray}
Inserting (\ref{2.13}) in (\ref{2.12}) and using the explicit expression for 
$\varkappa$ from (\ref{A5}) we get
\begin{eqnarray}
{\cal F}_{\Pom \pi p}(2 \nu_{2}, t) =
{\cal F}_{\Pom \pi p}(2 \bar{\nu}, t)
\left[
1 + (2 - \alpha_{\Pom}(t))
\frac{(q,p+p')(p+p',k_{1}-k_{2})}{16 \bar{\nu}^{2}}
g\left( \frac{2 - \alpha_{\Pom}(t)}{2}, \varkappa \right)
\right]. \qquad \quad
\label{2.14}
\end{eqnarray}
In a completely analogous way we get
\begin{eqnarray}
{\cal F}_{\Pom \pi p}(2 \nu_{1}, t) =
{\cal F}_{\Pom \pi p}(2 \bar{\nu}, t) 
\left[
1 - (2 - \alpha_{\Pom}(t))
\frac{(q,p+p')(p+p',k_{1}-k_{2})}{16 \bar{\nu}^{2}}
g\left( \frac{2 - \alpha_{\Pom}(t)}{2}, -\varkappa \right)
\right]. \qquad \quad
\label{2.15}
\end{eqnarray}
Inserting (\ref{2.14}) and (\ref{2.15}) in (\ref{2.11}) we find
\begin{eqnarray}
&&q^{\mu} {\cal M}_{\mu, \mathfrak{s'},\mathfrak{s}}^{(c)}(k_{1},k_{2},p',q,p)|_{\Pom}
=
q^{\mu} ie {\cal F}_{\Pom \pi p}(2 \bar{\nu}, t)
\nonumber\\
&& \quad
\times
\bigg{\lbrace}
4 \delta_{\mu}^{\;\;\nu} (k_{2} - k_{1}, p' + p)
+ 4 (p'+p)_{\mu} (k_{2} - k_{1})^{\nu}
- 2 (k_{2} - k_{1})_{\mu} (p'+p)^{\nu}
+ (p'+p)_{\mu} (2 - \alpha_{\Pom}(t)) 
\frac{(p'+p,k_{1}-k_{2})}{16 \bar{\nu}^{2}}\nonumber\\
&& \quad
\times
\bigg{[}
g\left( \frac{2 - \alpha_{\Pom}(t)}{2}, \varkappa \right)
\bigg{(} 2 (k_{2} - k_{1} + q)^{\nu}(k_{2}-k_{1}+q,p'+p)
-\frac{1}{2} (k_{2}-k_{1}+q)^{2} (p'+p)^{\nu}
\bigg{)}
 \nonumber\\
&& \quad
+
g\left( \frac{2 - \alpha_{\Pom}(t)}{2}, -\varkappa \right)
\bigg{(} 2 (k_{2} - k_{1} - q)^{\nu}(k_{2}-k_{1}-q,p'+p)
-\frac{1}{2} (k_{2}-k_{1}-q)^{2} (p'+p)^{\nu}
\bigg{)}
\bigg{]}
\bigg{\rbrace}  
\bar{u}_{\mathfrak{s'}}(p') \gamma_{\nu} u_{\mathfrak{s}}(p) \,. \nonumber\\
\label{2.16}
\end{eqnarray}
With (\ref{2.16}) we have indeed written the r.h.s. of (\ref{2.11})
as an expression proportional to $q^{\mu}$.
Therefore, we get the simplest solution for ${\cal M}_{\mu, \mathfrak{s'},\mathfrak{s}}^{(c)}$
by dropping on both sides of (\ref{2.16}) the factor $q^{\mu}$.
Of course, this is only \textit{one} solution and there is
no claim that it is the general solution.
We could always add to this solution a term $\widetilde{{\cal M}}_{\mu, \mathfrak{s'},\mathfrak{s}}^{(c)}$
satisfying by itself
\begin{eqnarray}
q^{\mu} \widetilde{{\cal M}}_{\mu, \mathfrak{s'},\mathfrak{s}}^{(c)}= 0\,.
\label{2.17}
\end{eqnarray}
But we shall not do this and consider the simplest solution
for ${\cal M}_{\mu, \mathfrak{s'},\mathfrak{s}}^{(c)}$ as defined above as part of
our \textit{model assumptions}.

Collecting now everything together we get 
the amplitudes for the pomeron-exchange
contribution as follows.
For the diagram~(a) of figure~\ref{fig:100} we get
\begin{eqnarray}
{\cal M}_{\mu, \mathfrak{s'},\mathfrak{s}}^{(a)}(k_{1},k_{2},p',q,p)|_{\Pom}
&=&
e 
\left[
\frac{(2 k_{1} - q)_{\mu}}{- 2 k_{1} \cdot q + q^{2} + i \varepsilon}
F_{M}(q^{2}) - q_{\mu} \frac{1 - F_{M}(q^{2})}{q^{2}}
\right]
{\cal M}_{\mathfrak{s'},\mathfrak{s}}^{(0, \,a)}(k_{2},p',q-k_{1},p)|_{\Pom}\,, \nonumber \\ 
{\cal M}_{\mathfrak{s'},\mathfrak{s}}^{(0, \,a)}(k_{2},p',q-k_{1},p)|_{\Pom} &=&
i{\cal F}_{\Pom \pi p}(2 \bar{\nu}, t)
\left[
1 + (2 - \alpha_{\Pom}(t)) \frac{\varkappa}{2} \,
g\left( \frac{2 - \alpha_{\Pom}(t)}{2}, \varkappa \right)
\right] \nonumber\\
&& \times
\Big[
2(k_{2}-k_{1}+q)^{\nu}(k_{2}-k_{1}+q,p'+p)
\bar{u}_{\mathfrak{s'}}(p') \gamma_{\nu} u_{\mathfrak{s}}(p) 
\nonumber\\
&& 
-(k_{2}-k_{1}+q)^{2} m_{p}
\bar{u}_{\mathfrak{s'}}(p') u_{\mathfrak{s}}(p)
\Big] \,;
\label{2.18}
\end{eqnarray}
see (\ref{2.4}), (\ref{2.9}), and (\ref{2.14}).
Diagram~(b) of figure~\ref{fig:100} gives
\begin{eqnarray}
{\cal M}_{\mu, \mathfrak{s'},\mathfrak{s}}^{(b)}(k_{1},k_{2},p',q,p)|_{\Pom}
&=& -e \left[
\frac{ (2 k_{2} - q)_{\mu}}{-2 k_{2} \cdot q + q^{2} + i \varepsilon}
F_{M}(q^{2}) - q_{\mu} \frac{1 - F_{M}(q^{2})}{q^{2}}
\right] {\cal M}_{\mathfrak{s'},\mathfrak{s}}^{(0,\,b)}(k_{1},p',q-k_{2},p)|_{\Pom}\,, \nonumber \\
{\cal M}_{\mathfrak{s'},\mathfrak{s}}^{(0,\,b)}(k_{1},p',q-k_{2},p)|_{\Pom}&=&
i{\cal F}_{\Pom \pi p}(2 \bar{\nu}, t)
\left[
1 - (2 - \alpha_{\Pom}(t)) \frac{\varkappa}{2} \,
g\left( \frac{2 - \alpha_{\Pom}(t)}{2}, -\varkappa \right)
\right] \nonumber\\
&& \times
\Big[
2(k_{2}-k_{1}-q)^{\nu}(k_{2}-k_{1}-q,p'+p)
\bar{u}_{\mathfrak{s'}}(p') \gamma_{\nu} u_{\mathfrak{s}}(p)  
\nonumber \\
&&-
(k_{2}-k_{1}-q)^{2} m_{p}
\bar{u}_{\mathfrak{s'}}(p') u_{\mathfrak{s}}(p)
\Big] \,;
\label{2.19}
\end{eqnarray}
see (\ref{2.5}), (\ref{2.10}), and (\ref{2.15}).
For diagram~(c) of figure~\ref{fig:100} we get from (\ref{2.16})
as the simplest solution
\begin{eqnarray}
&&{\cal M}_{\mu, \mathfrak{s'},\mathfrak{s}}^{(c)}(k_{1},k_{2},p',q,p)|_{\Pom}
=
2ie 
{\cal F}_{\Pom \pi p}(2 \bar{\nu}, t)
\bigg{\lbrace}
2 \delta_{\mu}^{\;\;\nu} (k_{2} - k_{1}, p' + p)
+ 2 (p'+p)_{\mu} (k_{2} - k_{1})^{\nu}\nonumber\\
&& \quad 
+ (p'+p)_{\mu} (2 - \alpha_{\Pom}(t)) \frac{(p'+p,k_{1}-k_{2})}{16 \bar{\nu}^{2}}\nonumber\\
&& \quad  \times
\bigg{[}
g\left( \frac{2 - \alpha_{\Pom}(t)}{2}, \varkappa \right)
(k_{2} - k_{1} + q)^{\nu}(k_{2}-k_{1}+q,p'+p) \nonumber\\
&& \quad 
+
g\left( \frac{2 - \alpha_{\Pom}(t)}{2}, -\varkappa \right)
(k_{2} - k_{1} - q)^{\nu}(k_{2}-k_{1}-q,p'+p)
\bigg{]}
\bigg{\rbrace} 
\bar{u}_{\mathfrak{s'}}(p') \gamma_{\nu} u_{\mathfrak{s}}(p) \nonumber\\
&& \quad +
2ie {\cal F}_{\Pom \pi p}(2 \bar{\nu}, t)
\bigg{\lbrace}
- 2 (k_{2} - k_{1})_{\mu} 
- (p'+p)_{\mu} \frac{2 - \alpha_{\Pom}(t)}{2} \frac{(p'+p,k_{1}-k_{2})}{16 \bar{\nu}^{2}}\nonumber\\
&& \quad \times
\left[
g\left( \frac{2 - \alpha_{\Pom}(t)}{2}, \varkappa \right)
(k_{2}-k_{1}+q)^{2} +
g\left( \frac{2 - \alpha_{\Pom}(t)}{2},-\varkappa \right)
(k_{2}-k_{1}-q)^{2} 
\right] 
\bigg{\rbrace} \nonumber \\
&& \quad \times
m_{p} \bar{u}_{\mathfrak{s'}}(p') u_{\mathfrak{s}}(p)\,.
\label{2.20}
\end{eqnarray}
%

%--------------------------------------------------
\subsection{\boldmath The complete amplitude for $\gamma^{(*)} p \to \pi^{+} \pi^{-} p$}
\label{sec:2B}
%--------------------------------------------------

In this section we discuss the complete amplitude
${\cal M}_{\mu, \mathfrak{s'},\mathfrak{s}}(k_{1},k_{2},p',q,p)$
for the $\gamma^{(*)} p \to \pi^{+} \pi^{-} p$ reaction;
see (\ref{2.2a}).
The diagrams for this reaction are listed in figure~1 of \cite{Bolz:2014mya}.
%Note that all results for ${\cal M}_{\mu, \mathfrak{s'},\mathfrak{s}}$ from
%Section~2 of \cite{Bolz:2014mya} apply not only for photoproduction
%but also for production by a slightly virtual photon $\gamma^{*}$ (\ref{2.2b}).
%Thus, the results from the following subsections of \cite{Bolz:2014mya}
%are taken over or are improved by us here.
In section~2 of \cite{Bolz:2014mya} the results of the model were given
for the production reaction (\ref{1.1}) with real photons.
In the following we shall discuss the extension of these results
for the case of real and slightly virtual photons (\ref{2.2b}).

%--------------------------------------------------
\subsubsection{Vector-meson production}
\label{sec:2B1}
%--------------------------------------------------

%------------------------------------------------------------------
\begin{figure}[tbp]
\centering
\includegraphics[width=6cm]{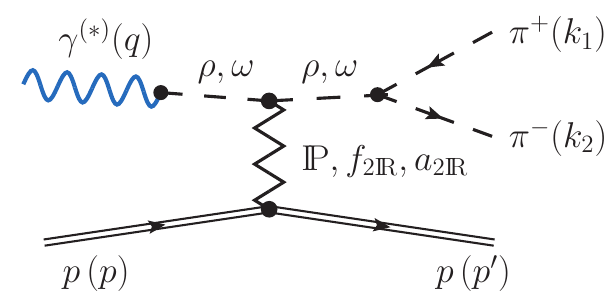}
\caption{\label{fig:400}
Diagram for resonant $\pi^{+} \pi^{-}$ production via $\rho$ and $\omega$ scattering on the proton. We include the $\rho$-$\omega$ interference effect only in the final state
via propagator mixing and the explicit $\omega \to \pi^{+} \pi^{-}$ decay.
In the initial state where the photon turns into
a $\rho$ or $\omega$ we neglect $\rho$-$\omega$ mixing.
We also assume absence of isospin violating effects in
the $\Pom$, $f_{2 \Reg}$, and $a_{2 \Reg}$ couplings
to the vector mesons.}
\end{figure}
%-------------------------------------------------------------------

Vector-meson production by real photons is treated extensively
in subsection~2.1 of \cite{Bolz:2014mya}.
The corresponding diagram for $\rho$ and $\omega$ production
by real and virtual photons
in the initial state is shown in figure~\ref{fig:400}.
This amplitude is denoted by
\begin{align}
{\cal M}^{(\rm res)}_{\mu, \mathfrak{s'},\mathfrak{s}}(k_{1},k_{2},p',q,p)\,.
\label{2.23.1}
\end{align}
The four-vector of the dipion system is defined by the four-vector
sum of the $\pi^{+}$ and $\pi^{-}$ as $k = k_{1} + k_{2}$.

For the calculation of this amplitude we use the vertices and propagators
of appendix~B of \cite{Bolz:2014mya}.
The result is as in (2.10)--(2.16) of \cite{Bolz:2014mya}
with a slight modification since we also consider virtual photons
(\ref{2.2b}) in the initial state.
We get
\begin{align}
{\cal M}^{(\rm res)}_{\mu, \mathfrak{s'},\mathfrak{s}}(k_{1},k_{2},p',q,p)|_{\Pom + f_{2 \Reg} + a_{2 \Reg}} =&
\sum_{\substack{V = \rho ,\, \omega,\\ V'= \rho,\, \omega}}
\left[
{\cal M}^{(V',V)}_{\mu, \mathfrak{s'},\mathfrak{s}}|_{\Pom}
+
{\cal M}^{(V',V)}_{\mu, \mathfrak{s'},\mathfrak{s}}|_{f_{2 \Reg}}
\right]
+
{\cal M}^{(\omega,\rho)}_{\mu, \mathfrak{s'},\mathfrak{s}}|_{a_{2 \Reg}}
+
{\cal M}^{(\rho,\omega)}_{\mu, \mathfrak{s'},\mathfrak{s}}|_{a_{2 \Reg}}\,,
\label{2.23.2}
\end{align}
where
\begin{align}
{\cal M}^{(V',V)}_{\mu, \mathfrak{s'},\mathfrak{s}}|_{\Pom} = &
\,
\frac{i}{4} \, e \,s \, F_1(t) \,F_M(t)\, 
\tilde{F}^{(V)}(k^2) \,g_{V' \pi \pi}
\left[
  \mathcal{K}_{\mu, \mathfrak{s'},\mathfrak{s}}^{(0, V', V)} V_{\Pom}^{(0,V)}
- \mathcal{K}_{\mu, \mathfrak{s'},\mathfrak{s}}^{(2, V', V)} V_{\Pom}^{(2,V)} \right]\nn \\
& \times
\tilde{F}^{(V)}(q^2) (- m_{V}^{2}) \Delta_{T}^{(V,V)}(q^{2})\,,
\label{2.23.3}\\
{\cal M}^{(V',V)}_{\mu, \mathfrak{s'},\mathfrak{s}}|_{f_{2 \Reg}}= &
\,
\frac{i}{4} \, e \,s \, F_1(t) \,F_M(t)\, 
\tilde{F}^{(V)}(k^2) \,g_{V' \pi \pi}
\left[
  \mathcal{K}_{\mu, \mathfrak{s'},\mathfrak{s}}^{(0, V', V)} V_{f_{2 \Reg}}^{(0,V)}
- \mathcal{K}_{\mu, \mathfrak{s'},\mathfrak{s}}^{(2, V', V)} V_{f_{2 \Reg}}^{(2,V)} \right]\nn \\
& \times
\tilde{F}^{(V)}(q^2) (- m_{V}^{2}) \Delta_{T}^{(V,V)}(q^{2})\,,
\label{2.23.4}
\end{align}
\begin{eqnarray}
{\cal M}^{(\omega,\rho)}_{\mu, \mathfrak{s'},\mathfrak{s}}|_{a_{2 \Reg}}
+ {\cal M}^{(\rho,\omega)}_{\mu, \mathfrak{s'},\mathfrak{s}}|_{a_{2 \Reg}} &=& 
\frac{i}{4} \, e \,s \, F_1(t) \,F_M(t)\,
\frac{g_{a_{2 \Reg} pp}}{M_{0}}(-is \alpha'_{a_{2 \Reg}})^{\alpha_{a_{2 \Reg}}(t)-1} \nn \\
&& 
\times \sum_{V' = \rho,\, \omega} g_{V' \pi \pi}
\bigg{\lbrace} 
\bigg{[}
  \mathcal{K}_{\mu, \mathfrak{s'},\mathfrak{s}}^{(0, V', \omega)} 
% V_{f_{2 \Reg}}^{(0,V)}
\tilde{F}^{(\omega)}(k^2) 
\tilde{F}^{(\rho)}(q^2) \frac{1}{\gamma_{\rho}}(-m_{\rho}^{2}) \Delta_{T}^{(\rho,\rho)}(q^{2}) \nn \\
&& 
+ \mathcal{K}_{\mu, \mathfrak{s'},\mathfrak{s}}^{(0, V', \rho)} 
% V_{f_{2 \Reg}}^{(2,V)} 
\tilde{F}^{(\rho)}(k^2) 
\tilde{F}^{(\omega)}(q^2) \frac{1}{\gamma_{\omega}}(-m_{\omega}^{2}) \Delta_{T}^{(\omega,\omega)}(q^{2}) 
\bigg{]}
 2 a_{a_{2 \Reg} \omega \rho} \nn \\
&&
- \bigg{[}
  \mathcal{K}_{\mu, \mathfrak{s'},\mathfrak{s}}^{(2, V', \omega)} 
% V_{f_{2 \Reg}}^{(0,V)}
\tilde{F}^{(\omega)}(k^2) 
\tilde{F}^{(\rho)}(q^2) \frac{1}{\gamma_{\rho}}(-m_{\rho}^{2}) \Delta_{T}^{(\rho,\rho)}(q^{2}) \nn \\
&&
+ \mathcal{K}_{\mu, \mathfrak{s'},\mathfrak{s}}^{(2, V', \rho)} 
% V_{f_{2 \Reg}}^{(2,V)} 
\tilde{F}^{(\rho)}(k^2) 
\tilde{F}^{(\omega)}(q^2) \frac{1}{\gamma_{\omega}}(-m_{\omega}^{2}) \Delta_{T}^{(\omega,\omega)}(q^{2}) 
\bigg{]} b_{a_{2 \Reg} \omega \rho}
\bigg{\rbrace}. \qquad \quad
\label{2.23.5}
\end{eqnarray}
The expressions for
$\mathcal{K}_{\mu, \mathfrak{s'},\mathfrak{s}}^{(i, V', V)}$
and 
$V_{\Pom, f_{2 \Reg}}^{(i,V)}$
$(i = 0, 2)$
are given in (2.13), and (2.14), (2.15) of \cite{Bolz:2014mya},
respectively.
The form factors
$\tilde{F}^{(\rho)}(k^2)$ and
$\tilde{F}^{(\omega)}(k^2)$ 
are given in (B.85), (B.86), and (B.88) of \cite{Bolz:2014mya},
respectively.
We have for $V = \rho, \omega$
%In \cite{Bolz:2014mya},
%the assumption
%$\tilde{F}^{(\omega)}(k^2) = \tilde{F}^{(\rho)}(k^2)$ 
%was used;
%see eqs.~(2.16) and (B.88) of \cite{Bolz:2014mya}.
%
\begin{equation}
\tilde{F}^{(V)}(k^2) 
= \left[ 1 + \frac{k^2(k^2-m^2_V)}{\Lambda^4_V} \right]^{-n_V}
\label{2.36.6}
\end{equation}
with $\Lambda_V$ a parameter in the range 2 to 5~GeV and $n_V > 0$.
We shall assume, as was done in \cite{Bolz:2014mya},
that $\Lambda_\rho = \Lambda_\omega$
and $n_\rho = n_\omega$.
In the following we shall set approximately
\begin{eqnarray}
-m_{\rho}^{2} \,\Delta_{T}^{(\rho,\rho)}(q^{2}) &\cong& 
-m_{\rho}^{2} \frac{1}{q^{2}-m_{\rho}^{2}} = 
\frac{m_{\rho}^{2}}{m_{\rho}^{2}-q^{2}} \,, \nn \\
-m_{\omega}^{2} \,\Delta_{T}^{(\omega,\omega)}(q^{2}) &\cong&
\frac{m_{\omega}^{2}}{m_{\omega}^{2}-q^{2}} \,.
\label{2.36.7}
\end{eqnarray}

For the propagator functions $\Delta_{T}^{(V',V)}(k^{2})$
needed for the calculation of
$\mathcal{K}_{\mu, \mathfrak{s'},\mathfrak{s}}^{(i, V', V)}$
$(V',V \in \{ \rho, \omega \})$
we insert the results given in \cite{Melikhov:2003hs}.
There, dispersion theory was used to calculate the $2 \times 2$
propagator matrix for the $\rho$-$\omega$ system.
The results are listed in (B.3)--(B.17) of \cite{Bolz:2014mya}.

It is clear that the production of higher-mass vector mesons
decaying to $\pi^{+} \pi^{-}$ could easily be added here,
as was done in \cite{Bolz:2014mya} for the $\rho(1450)$.

%There, $\rho^{0}$-$\omega$ interference is included. We also note that
%a dispersion-theoretic model for the $\rho^{0}$ propagator
%was used \cite{Melikhov:2003hs} which gave a very good fit to the $\rho^{0}$
%production in $e^{+} e^{-}$ annihilation.

%--------------------------------------------------
\subsubsection{\boldmath Production of $f_{2}$ by reggeon, photon, and odderon exchange}
\label{sec:2B2}
%--------------------------------------------------

\begin{figure}[tbp]
\centering
\includegraphics[width=6cm]{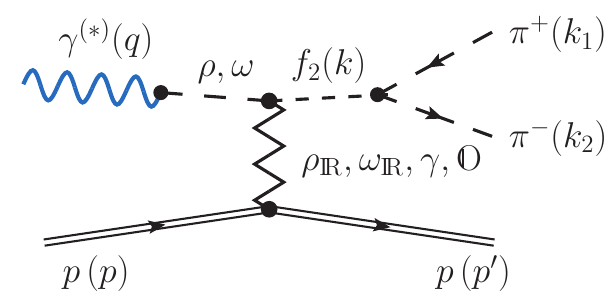}
\caption{\label{fig:401}
Production of $f_{2}$ by $\rho_{\Reg}$,
$\omega_{\Reg}$, $\gamma$, and $\Ode$ exchange.}
\end{figure}

Here we consider the production of $f_{2}$ by $\rho_{\Reg}$,
$\omega_{\Reg}$, photon, and odderon $(\Ode)$ exchange,
with the $f_{2}$ decaying to $\pi^{+} \pi^{-}$;
see figure~\ref{fig:401}.
%------------------------------------------------------------------
%-------------------------------------------------------------------

The corresponding matrix elements for the sum of $\rho_{\Reg}$
and $\omega_{\Reg}$ exchanges is obtained as follows:
\begin{align}
{\cal M}^{(f_{2})}_{\mu, \mathfrak{s'},\mathfrak{s}}(k_{1},k_{2},p',q,p)|_{\rho_{\Reg} + \omega_{\Reg}} = &
\,
\frac{1}{4} \, e \,s \,g_{f_{2} \pi \pi} \, F_1(t) \,F_M(t)\, 
%[F^{(f_{2} \pi \pi)}(k^2)]^{2} 
F^{(f_{2} \gamma\gamma)}(k^2) 
F^{(f_{2} \pi \pi)}(k^2) 
\nn \\
& \times 
\sum_{V = \rho, \,\omega}
\left[
  \mathcal{N}_{\mu, \mathfrak{s'},\mathfrak{s}}^{(0)} W_{V}^{(0)}
- \mathcal{N}_{\mu, \mathfrak{s'},\mathfrak{s}}^{(2)} W_{V}^{(2)} \right]
\tilde{F}^{(V)}(q^2) (- m_{V}^{2}) \Delta_{T}^{(V,V)}(q^{2}).
\label{2.23.8}
\end{align}
Here $\mathcal{N}_{\mu, \mathfrak{s'},\mathfrak{s}}^{(i)}$ and $W_{V}^{(i)}$ $(i = 0, 2)$ are given in
(2.19) and (2.20) of \cite{Bolz:2014mya},
respectively.
In the present paper, in the $\rho_{\Reg} \rho f_{2}$ and 
$\omega_{\Reg} \omega f_{2}$ vertices
given in (B.99) and (B.101) of \cite{Bolz:2014mya},
respectively,
we introduce $F^{(f_{2} \gamma \gamma)}(k^{2})$
instead of $F^{(f_{2} \pi \pi)}(k^{2})$.
We make the following assumptions
for the form factors:
\begin{eqnarray}
%F^{(f_{2} \pi \pi)}(s) = \exp\left(-\frac{(s-m_{f_{2}}^{2})^{2}}{\Lambda_{f_{2} \pi \pi}^{4}}\right)\,.
%\label{formfactor_f2_exp}
F^{(f_{2} \pi \pi)}(k^{2}) &=& \frac{\Lambda_{f_{2} \pi\pi}^{4}}{\Lambda_{f_{2} \pi\pi}^{4}+(k^{2}-m_{f_{2}}^{2})^{2}}\,,
\label{formfactor_f2pipi}\\
F^{(f_{2} \gamma \gamma)}(k^{2}) &=& \frac{\Lambda_{f_{2}\gamma\gamma}^{4}}{\Lambda_{f_{2}\gamma\gamma}^{4}+(k^{2}-m_{f_{2}}^{2})^{2}}\,,
\label{formfactor_f2gg}
%&&F^{(f_{2} \pi \pi)}(s) = \left(1+\frac{s-m_{f_{2}}^{2}}{\Lambda^{2}} \right) \exp\left(-\frac{s-%m_{f_{2}}^{2}}{\Lambda^{2}}\right)\,.
%\label{formfactor_f2_mix}
\end{eqnarray}
where $\Lambda_{f_{2} \pi\pi} = 2.0$~GeV and 
$\Lambda_{f_{2} \gamma\gamma} = 1.2$~GeV are estimated by comparing model results for the $\gamma \gamma \to \pi^{+} \pi^{-}$ reaction with the Belle data 
\cite{Mori:2007bu}.

For photon exchange we get
\begin{align}
{\cal M}^{(f_{2})}_{\mu, \mathfrak{s'},\mathfrak{s}}(k_{1},k_{2},p',q,p)|_{\gamma} = &
\,
\frac{1}{4} \, e \,s \,g_{f_{2} \pi \pi} \, 
F_M(t) \,F_M(q^{2})\, 
%[F^{(f_{2} \pi \pi)}(k^2)]^{2} 
F^{(f_{2} \gamma\gamma)}(k^2) 
F^{(f_{2} \pi \pi)}(k^2) 
\nn \\
& \times 
\Big{\lbrace} 
F_{1}(t)
\left[
  \mathcal{N}_{\mu, \mathfrak{s'},\mathfrak{s}}^{(0)} W_{\gamma}^{(0)}
- \mathcal{N}_{\mu, \mathfrak{s'},\mathfrak{s}}^{(2)} W_{\gamma}^{(2)} 
\right]
+
F_{2}(t)
\left[
  \mathcal{S}_{\mu, \mathfrak{s'},\mathfrak{s}}^{(0)} W_{\gamma}^{(0)}
- \mathcal{S}_{\mu, \mathfrak{s'},\mathfrak{s}}^{(2)} W_{\gamma}^{(2)} 
\right]
\Big{\rbrace} 
\,.
\label{2.23.9}
\end{align}
Here $\mathcal{N}_{\mu, \mathfrak{s'},\mathfrak{s}}^{(i)}$,
$\mathcal{S}_{\mu, \mathfrak{s'},\mathfrak{s}}^{(i)}$, 
and $W_{\gamma}^{(i)}$ $(i = 0, 2)$ are given in
(2.19), (2.22), and (2.23) of \cite{Bolz:2014mya},
respectively.

Very interesting is the production of $f_{2}$
by \textit{odderon} exchange.
We denote the corresponding amplitude by
\begin{align}
{\cal M}^{(f_{2})}_{\mu, \mathfrak{s'},\mathfrak{s}}(k_{1},k_{2},p',q,p)|_{\Ode} \,.
\label{2.23.10}
\end{align}
The odderon has been introduced on theoretical grounds
in \cite{Lukaszuk:1973nt,Joynson:1975az}. Experimental evidence for odderon effects
in $pp \to pp$  and $\bar{p} p \to \bar{p} p$ scattering
have been presented in \cite{TOTEM:2016lxj,TOTEM:2017sdy,TOTEM:2018hki,D0:2020tig}.
But the status of the odderon remains controversial;
see for instance \cite{Donnachie:2019ciz,Donnachie:2022aiq}.
Exclusive photoproduction of mesons 
with charge conjugation $C = +1$ offers
the possibility to observe odderon effects,
as was first discussed in \cite{Schafer:1992pq,Barakhovsky:1991ra}.
Here we consider production of the $C = +1$ tensor meson
$f_{2}(1270)$ which was discussed in view of odderon effects in \cite{Berger:2000wt}.
In the original tensor-pomeron model for
real photons producing an $f_{2}$ by odderon exchange,
as used in \cite{Bolz:2014mya},
the odderon was assumed to couple to hadrons
like a vector and to be describable 
by a single Regge pole.
In our study of $pp$ and $\bar{p}p$ elastic scattering
in \cite{Lebiedowicz:2022nnn} we found, however, 
that a double-pole ansatz for the odderon 
gave a better description of the data.
The effective odderon propagator and vertices
which we use are listed in (\ref{A46})--(\ref{A51}) 
of appendix~A of our present paper.
From these we get for the odderon-exchange amplitude
of figure~\ref{fig:401}, setting $k = k_{1} + k_{2}$,
\begin{align}
{\cal M}^{(f_{2})}_{\mu, \mathfrak{s'},\mathfrak{s}}(k_{1},k_{2},p',q,p)|_{\Ode} = & \,
\Gamma^{(f_{2} \pi \pi)\, \kappa'\lambda'}(k_{1},k_{2})\,
\Delta^{(f_{2})\,\kappa\lambda}_{\kappa'\lambda'}(k)\,
\Gamma^{(\Ode \gamma f_{2})}(k,q)_{\mu \nu \kappa \lambda}\,
\tilde{\Delta}^{(\Ode)\, \nu \nu'}(s,t)\,
\bar{u}_{\mathfrak{s'}}(p') \Gamma^{(\Ode pp)}_{\nu'}(p',p) u_{\mathfrak{s}}(p)\,.
\label{2.23.11}
\end{align}
With the odderon propagator and vertices from (\ref{A46})--(\ref{A51}),
the $f_{2}$ propagator and $f_{2} \pi \pi$ vertex
from (B.23) and (B.58)--(B.60) of \cite{Bolz:2014mya},
respectively, we get
\begin{align}
{\cal M}^{(f_{2})}_{\mu, \mathfrak{s'},\mathfrak{s}}(k_{1},k_{2},p',q,p)|_{\Ode} = &
\frac{1}{4} \, e \,s \,g_{f_{2} \pi \pi} \, 
F_{1}(t) \, F_M(t) \,F_M(q^{2})\, 
%[F^{(f_{2} \pi \pi)}(k^2)]^{2} 
F^{(f_{2} \gamma\gamma)}(k^2) 
F^{(f_{2} \pi \pi)}(k^2) 
\nn \\
& \times 
\left[
  \mathcal{N}_{\mu, \mathfrak{s'},\mathfrak{s}}^{(0)} W_{\Ode}^{(0)}
- \mathcal{N}_{\mu, \mathfrak{s'},\mathfrak{s}}^{(2)} W_{\Ode}^{(2)} 
\right]
\left[
C_{1} + C_{2} \ln(-is \alpha'_{\Ode})
\right]
\,.
\label{2.23.12}
\end{align}
Here $\mathcal{N}_{\mu, \mathfrak{s'},\mathfrak{s}}^{(0,\,2)}$
and $W_{\Ode}^{(0,\,2)}$ are defined in
(2.19) and (2.25) of \cite{Bolz:2014mya},
respectively.

%--------------------------------------------------
\subsubsection{\boldmath Non-resonant production of $\pi^{+} \pi^{-}$ by pomeron and $f_{2 \Reg}$ exchange}
\label{sec:2B5}
%--------------------------------------------------
Here we have the reactions for which we give
an improved result in the present paper.
The pomeron-exchange contribution is given in (\ref{2.18})--(\ref{2.20}):
\begin{eqnarray}
{\cal M}^{(\rm DS)}_{\mu, \mathfrak{s'},\mathfrak{s}}(k_{1},k_{2},p',q,p)|_{\Pom} = 
{\cal M}_{\mu, \mathfrak{s'},\mathfrak{s}}^{(a+b+c)}(k_{1},k_{2},p',q,p)|_{\Pom}\,.
\label{2.21}
\end{eqnarray}
Also for the other exchanges the complete result is the sum
of the contributions of the diagrams $(a+b+c)$
of figure~\ref{fig:100}.

For the amplitudes with the $f_{2 \Reg}$ exchange
we have the same structure as in (\ref{2.18})--(\ref{2.20})
and (\ref{2.21})
but with the replacements
\begin{eqnarray}
{\cal F}_{\Pom \pi p}(2 \bar{\nu}, t) 
&\to& {\cal F}_{f_{2 \Reg} \pi p}(2 \bar{\nu}, t)\,, \nonumber \\
\alpha_{\Pom}(t) 
&\to& \alpha_{f_{2 \Reg}}(t) \,;
\label{2.22}
\end{eqnarray}
see (\ref{A17}), (\ref{A18}), (\ref{A21}), (\ref{A22}),
(\ref{A24})--(\ref{A27}), and (\ref{A32}).
In this way we get
\begin{eqnarray}
{\cal M}^{(\rm DS)}_{\mu, \mathfrak{s'},\mathfrak{s}}(k_{1},k_{2},p',q,p)|_{f_{2 \Reg}} = 
{\cal M}_{\mu, \mathfrak{s'},\mathfrak{s}}^{(a+b+c)}(k_{1},k_{2},p',q,p)|_{f_{2 \Reg}}\,.
\label{2.25a}
\end{eqnarray}

Note that with $\Pom$ and $f_{2 \Reg}$ exchange we have in figure~\ref{fig:100}
the fusion of a photon with charge conjugation $C = -1$
and a $C = +1$ object, $\Pom$ or $f_{2 \Reg}$,
giving a $\pi^{+} \pi^{-}$ pair.
Therefore, the amplitudes must satisfy
\begin{eqnarray}
{\cal M}^{(\rm DS)}_{\mu, \mathfrak{s'},\mathfrak{s}}(k_{1},k_{2},p',q,p)|_{\Pom, f_{2 \Reg}} = 
- {\cal M}^{(\rm DS)}_{\mu, \mathfrak{s'},\mathfrak{s}}(k_{2},k_{1},p',q,p)|_{\Pom, f_{2 \Reg}}\,.
\label{2.22b}
\end{eqnarray}
This is easily checked from (\ref{2.18})--(\ref{2.20}).

%--------------------------------------------------
\subsubsection{\boldmath Non-resonant production of $\pi^{+} \pi^{-}$ by $\rho_{\Reg}$ and photon exchange}
\label{sec:2B6}
%--------------------------------------------------

The photon-exchange contribution 
to ${\cal M}_{\mu, \mathfrak{s'},\mathfrak{s}}$ (\ref{2.2a}),
${\cal M}^{(\rm DS)}_{\mu, \mathfrak{s'},\mathfrak{s}}(k_{1},k_{2},p',q,p)|_{\gamma}$,
can be taken over directly from (2.27) of \cite{Bolz:2014mya}.
In detail:
the result given in (2.27) of \cite{Bolz:2014mya} corresponds
to the production of pointlike pions.
In reality pions are extended objects and a simple way to take
this into account is to introduce form factors.
Here we follow this phenomenological way,
leaving a more detailed investigation of the reaction
$\gamma \gamma \to \pi^{+} \pi^{-}$ to a separate study.
We get then
\begin{eqnarray}
{\cal M}^{(\rm DS)}_{\mu, \mathfrak{s'},\mathfrak{s}}(k_{1},k_{2},p',q,p) |_{\gamma} 
&= &
e^{3} F_{M}(q^{2})
\bigg{[}
(q - 2 k_{1})_{\mu} (q - k_{1} + k_{2})_{\nu} 
\frac{1}{\hat{t} - m_{\pi}^{2}} 
\nonumber \\
&& + 
(q - 2 k_{2})_{\mu} (q + k_{1} - k_{2})_{\nu}
\frac{1}{\hat{u} - m_{\pi}^{2}} 
- 2 g_{\mu \nu} \bigg{]} F(\hat{t},\hat{u},\hat{s}) \,
F_{M}(t) \,\frac{1}{t}
\nonumber \\
&& \times
\bar{u}_{\mathfrak{s'}}(p') 
\bigg{[}\gamma^{\nu} F_{1}(t) + \frac{i}{2 m_{p}} \sigma^{\nu \lambda} (p'-p)_{\lambda} F_{2}(t)
\bigg{]}
u_{\mathfrak{s}}(p)\,,
\label{2.5.1}
\end{eqnarray}
where
\begin{eqnarray}
\hat{t} &=& (q - k_{1})^{2}\,, \nonumber \\
\hat{u} &=& (q - k_{2})^{2}\,, \nonumber \\
\hat{s} &=& (k_{1} + k_{2})^{2} = M_{\pi\pi}^{2}\,.
\label{2.5.2}
\end{eqnarray}
$F_{M}(t)$ is defined in (\ref{A21b}),
and $F(\hat{t},\hat{u},\hat{s})$ as
\begin{eqnarray}
F(\hat{t},\hat{u},\hat{s}) 
= \frac{[F(\hat{t})]^{2} + [F(\hat{u})]^{2}}{1+[\tilde{F}(\hat{s})]^{2}}
\label{2.5.3}
\end{eqnarray}
with the exponential parametrization
\begin{eqnarray}
&&F(\hat{t}) = \exp\left( \frac{\hat{t}-m_{\pi}^{2}}{\Lambda_{\pi}^{2}} \right)\,, \nn \\
&&F(\hat{u}) = \exp\left( \frac{\hat{u}-m_{\pi}^{2}}{\Lambda_{\pi}^{2}} \right)\,,\nn \\
&&\tilde{F}(\hat{s}) = \exp\left( \frac{-(\hat{s}-4m_{\pi}^{2})}{\Lambda_{\pi}^{2}} \right)\,.
\label{2.5.4}
\end{eqnarray}
Here, we take $\Lambda_{\pi} = 1.7$~GeV,
estimated from the comparison of the model results
for the $\gamma \gamma \to \pi^{+} \pi^{-}$ reaction with the experimental data
\cite{Boyer:1990vu,Behrend:1992hy,Heister:2003ae,Nakazawa:2004gu,Mori:2007bu}.

For the $\rho_{\Reg}$-exchange contribution
we find now with the methods of our present paper
the following results:
\begin{align}
{\cal M}^{(\rm DS)}_{\mu, \mathfrak{s'},\mathfrak{s}}(k_{1},k_{2},p',q,p)|_{\rho_{\Reg}} =
{\cal M}_{\mu, \mathfrak{s'},\mathfrak{s}}^{(a+b+c)}(k_{1},k_{2},p',q,p)|_{\rho_{\Reg}}
\,,
\label{2.26b}
\end{align}
where
\begin{align}
{\cal M}_{\mu, \mathfrak{s'},\mathfrak{s}}^{(a)}(k_{1},k_{2},p',q,p)|_{\rho_{\Reg}}
=& \; e 
\left[ 
\frac{ (2 k_{1} - q)_{\mu}}{-2 k_{1} \cdot q + q^{2} + i \varepsilon}
F_{M}(q^{2}) - q_{\mu} \frac{1 - F_{M}(q^{2})}{q^{2}}
\right] 
{\cal M}_{\mathfrak{s'},\mathfrak{s}}^{(0,\,a)}(k_{2},p',q-k_{1},p)|_{\rho_{\Reg}}\,, 
\nonumber \\
{\cal M}_{\mathfrak{s'},\mathfrak{s}}^{(0,\,a)}(k_{2},p',q-k_{1},p)|_{\rho_{\Reg}}=& \;
{\cal F}_{\rho_{\Reg} \pi p}(2 \bar{\nu}, t)
\left[
1 + (1 - \alpha_{\rho_{\Reg}}(t)) \frac{\varkappa}{2} \,
g\left( \frac{1 - \alpha_{\rho_{\Reg}}(t)}{2}, \varkappa \right)
\right] 
(k_{2}-k_{1}+q)^{\nu}
\bar{u}_{\mathfrak{s'}}(p') \gamma_{\nu} u_{\mathfrak{s}}(p)
\,;
\label{2.23}\\
{\cal M}_{\mu, \mathfrak{s'},\mathfrak{s}}^{(b)}(k_{1},k_{2},p',q,p)|_{\rho_{\Reg}}
=& -e 
\left[
\frac{ (2 k_{2} - q)_{\mu}}{-2 k_{2} \cdot q + q^{2} + i \varepsilon}
F_{M}(q^{2}) - q_{\mu} \frac{1 - F_{M}(q^{2})}{q^{2}}
\right] 
{\cal M}_{\mathfrak{s'},\mathfrak{s}}^{(0,\,b)}(k_{1},p',q-k_{2},p)|_{\rho_{\Reg}}\,,
 \nonumber\\
{\cal M}_{\mathfrak{s'},\mathfrak{s}}^{(0,\,b)}(k_{1},p',q-k_{2},p)|_{\rho_{\Reg}}=& \;
{\cal F}_{\rho_{\Reg} \pi p}(2 \bar{\nu}, t)
\left[
1 - (1 - \alpha_{\rho_{\Reg}}(t)) \frac{\varkappa}{2} \,
g\left( \frac{1 - \alpha_{\rho_{\Reg}}(t)}{2}, -\varkappa \right)
\right]
(k_{2}-k_{1}-q)^{\nu}
\bar{u}_{\mathfrak{s'}}(p') \gamma_{\nu} u_{\mathfrak{s}}(p)
\,;
\label{2.24}\\
{\cal M}_{\mu, \mathfrak{s'},\mathfrak{s}}^{(c)}(k_{1},k_{2},p',q,p)|_{\rho_{\Reg}}
=& \;
e {\cal F}_{\rho_{\Reg} \pi p}(2 \bar{\nu}, t)
\bigg{\lbrace}
2 \delta_{\mu}^{\;\;\nu}
+ (p'+p)_{\mu} (1 - \alpha_{\rho_{\Reg}}(t)) 
\frac{(p'+p,k_{1}-k_{2})}{16 \bar{\nu}^{2}}\nonumber\\
&\times
\bigg[
g\left( \frac{1 - \alpha_{\rho_{\Reg}}(t)}{2}, \varkappa \right)
(k_{2} - k_{1} + q)^{\nu}
+
g\left( \frac{1 - \alpha_{\rho_{\Reg}}(t)}{2}, -\varkappa \right)
(k_{2} - k_{1} - q)^{\nu}
\bigg]
\bigg{\rbrace} \nn \\
& \times
\bar{u}_{\mathfrak{s'}}(p') \gamma_{\nu} u_{\mathfrak{s}}(p) \,.
\label{2.25}
\end{align}
Note that for the $C = -1$ exchange of $\rho_{\Reg}$
we have 
\begin{eqnarray}
{\cal M}^{(\rm DS)}_{\mu, \mathfrak{s'},\mathfrak{s}}(k_{1},k_{2},p',q,p)|_{\rho_{\Reg}} = 
{\cal M}^{(\rm DS)}_{\mu, \mathfrak{s'},\mathfrak{s}}(k_{2},k_{1},p',q,p)|_{\rho_{\Reg}}\,.
\label{2.26}
\end{eqnarray}

This concludes our discussion of all contributions 
to ${\cal M}_{\mu, \mathfrak{s'},\mathfrak{s}}$
from the diagrams of figure~1 of \cite{Bolz:2014mya}.

%--------------------------------------------------
\section{\boldmath Amplitudes for the $pp \to pp \pi^{+} \pi^{-}$ reaction}
\label{sec:3}
%--------------------------------------------------

In this section we shall discuss production 
of $\pi^{+} \pi^{-}$ in proton-proton collisions
\begin{eqnarray}
p(p_{a},\lambda_{a}) + p(p_{b},\lambda_{b}) \to
p(p_{1},\lambda_{1}) + \pi^{+}(p_{3}) + \pi^{-}(p_{4}) + p(p_{2},\lambda_{2}) \,,
\label{3.1}
\end{eqnarray}
where $p_{a,b}$, $p_{1,2}$ and 
$\lambda_{a,b}$, $\lambda_{1,2} \in \{1/2, -1/2\}$ 
denote the four-momenta and helicities of the protons, 
and $p_{3,4}$ denote the four-momenta of the charged pions, respectively.
Here we use the notation as in figure~2 and
section~III of \cite{Lebiedowicz:2014bea}.

The kinematic variables for (\ref{3.1}) are as follows.
We have energy-momentum conservation
\begin{eqnarray}
p_{a} + p_{b} = p_{1} + p_{2} + p_{3} + p_{4}\,,
\label{3.2}
\end{eqnarray}
and we set
\begin{eqnarray}
p_{34} &=& p_{3} + p_{4}\,, \quad M_{\pi \pi}^{2} = p_{34}^{2}\,, \nn \\
q_{1} &=& p_{a} - p_{1}\,, \quad t_{1} = q_{1}^{2}\,, \nn \\
q_{2} &=& p_{b} - p_{2}\,, \quad t_{2} = q_{2}^{2}\,, \nn \\
s &=& (p_{a} + p_{b})^{2}\,,\nn \\
s_{1} &=& (p_{a} + q_{2})^{2}\,,\nn \\
s_{2} &=& (p_{b} + q_{1})^{2}\,.
\label{3.3}
\end{eqnarray}

\begin{figure}[tbp]
\centering
(a)\includegraphics[width=6cm]{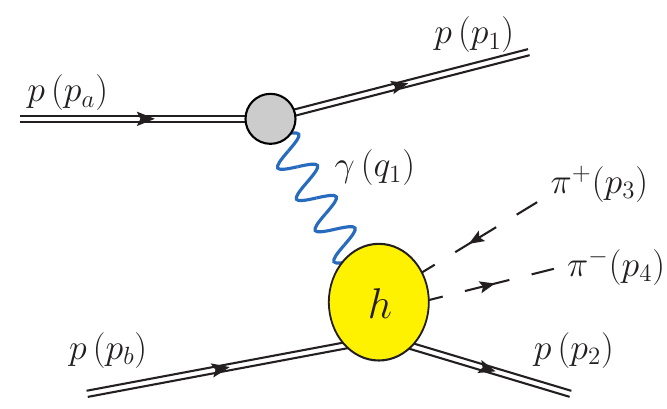}
(b)\includegraphics[width=5.6cm]{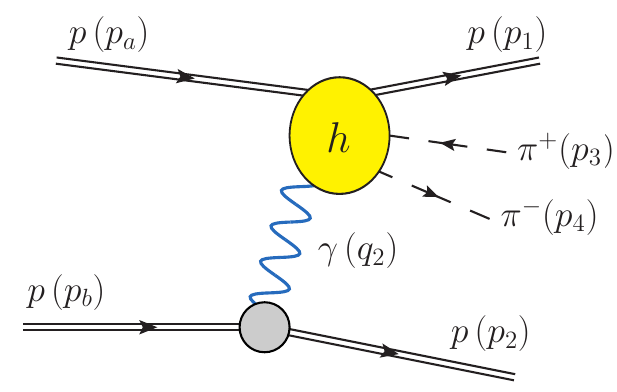}
(c)\includegraphics[width=6cm]{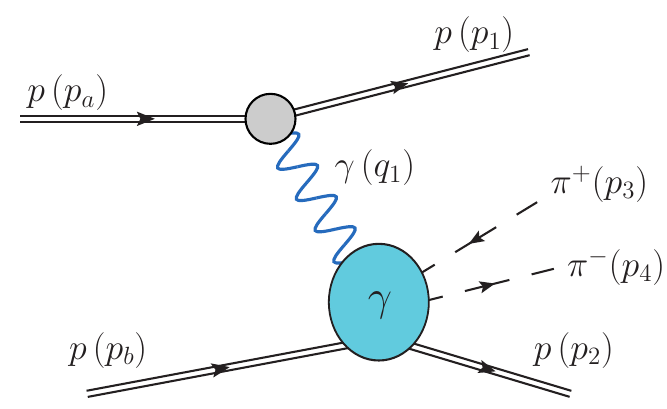}
\caption{\label{fig:500}
Diagrams for the reaction (\ref{3.1}) in the diffractive regime
where we require photon exchange at least for one of the incoming protons.
The diagrams (a) and (b) describe the reactions where
one proton emits a photon and the other proton reacts
hadronically indicated by the blob with the $h$.
Diagram (c) stands for the $\gamma \gamma$ fusion contribution
where both incoming protons emit a photon.
Note that we have two diagrams $\gamma$-$h$ and $h$-$\gamma$,
but only one diagram $\gamma$-$\gamma$.
In principle there are also the diagrams with $p_{1}$
and $p_{2}$ interchanged.
But in the kinematic regime considered by us these
give very small contributions and we neglect them here.}
\end{figure}

We consider diffractive reactions with CEP (central
exclusive production) of a $\pi^{+}\pi^{-}$ pair.
That is, we require large~$s$ and small
$|t_{1}|, |t_{2}| \lesssim {\cal O}(1 \, {\rm GeV}^{2})$.
%But in the present paper we require more.
%In order to have photon exchange at least on one side
%of the reaction we ask to have at least one of 
%$|t_{1}|$ and $|t_{2}| \lesssim {\cal O}(0.05 \, {\rm GeV}^{2})$.
%Then the contribution of other, purely hadronic diffractive, 
%processes is strongly suppressed.
In this kinematic regime we shall study $\pi^{+}\pi^{-}$ production
where at least one of the protons emits a virtual photon;
see figure~\ref{fig:500}.
But there also purely hadronic diagrams will contribute
where only pomeron, odderon, and reggeon exchanges are involved in the production of the pion pair.
These latter reactions have been discussed in
the tensor-pomeron approach in
\cite{Lebiedowicz:2016ioh,Lebiedowicz:2019por}.
In nature, of course, the amplitudes for these two types of processes
will interfere.
Thus, in experiments one will have to deal with all contributions
simultanously.
But these contributions have different $t$-dependences,
with the photonic ones being peaked at very small $|t|$.
This is well known and we shall discuss this below in detail.
Thus, with suitable cuts one will be able to enhance either the photonic
or the hadronic contributions to 
central exclusive $\pi^{+}\pi^{-}$ production.
But further studies of this point go beyond the scope of the present article,
since this will depend on the detailed possibilities of a concrete experiment.

The general diagrams for the reaction (\ref{3.1}),
where at least one of the protons emits a virtual photon,
are as shown in figure~\ref{fig:500}.
%------------------------------------------------------------------
%-------------------------------------------------------------------

We denote the ${\cal T}$-matrix element for the reaction (\ref{3.1}) by
\begin{eqnarray}
{\cal M}_{pp \to pp \pi^{+} \pi^{-}} =
\bra{\pi^{+}(p_{3}),\pi^{-}(p_{4}),p(p_{1}),p(p_{2})}{\cal T}\ket{p(p_{a}),p(p_{b})} \,.
\label{3.4} 
\end{eqnarray}
From diagram~(a) of figure~\ref{fig:500} we get
\begin{eqnarray}
{\cal M}^{(\gamma h)}_{pp \to pp \pi^{+} \pi^{-}} &=&
\bar{u}(p_{1}, \lambda_{1}) 
\Gamma^{(\gamma pp)\,\mu}(p_{1},p_{a}) 
u(p_{a}, \lambda_{a}) \frac{1}{t_{1}} \nn \\
&&\times
\Big{[} 
{\cal M}^{(\rm res)}_{\mu, \lambda_{2}, \lambda_{b}}(p_{3},p_{4},p_{2},q_{1},p_{b})|_{\Pom + f_{2 \Reg} + a_{2 \Reg}} \nn \\
&&+
{\cal M}^{(f_{2})}_{\mu, \lambda_{2}, \lambda_{b}}(p_{3},p_{4},p_{2},q_{1},p_{b})|_{\rho_{\Reg} + \omega_{\Reg} + \Ode} \nn \\
&&+
{\cal M}^{(\rm DS)}_{\mu, \lambda_{2}, \lambda_{b}}(p_{3},p_{4},p_{2},q_{1},p_{b})|_{\Pom + f_{2 \Reg} + \rho_{\Reg}}
\Big{]} \,.
\label{3.5} 
\end{eqnarray}
Here $\Gamma^{(\gamma pp)\,\mu}$ is given in (\ref{A12a}),
${\cal M}^{(\rm res)}_{\mu, \lambda_{2}, \lambda_{b}}|_{\Pom + f_{2 \Reg} + a_{2 \Reg}}$
in (\ref{2.23.2}),
${\cal M}^{(f_{2})}_{\mu, \lambda_{2}, \lambda_{b}}|_{\rho_{\Reg} + \omega_{\Reg}}$
in (\ref{2.23.8}),
\newline
${\cal M}^{(f_{2})}_{\mu, \lambda_{2}, \lambda_{b}}|_{\Ode}$
in (\ref{2.23.12}), 
${\cal M}^{(\rm DS)}_{\mu, \lambda_{2}, \lambda_{b}}|_{\Pom + f_{2 \Reg}}$
in (\ref{2.21}), (\ref{2.25a}), 
and
${\cal M}^{(\rm DS)}_{\mu, \lambda_{2}, \lambda_{b}}|_{\rho_{ \Reg}}$
in (\ref{2.26b}).

In a completely analogous way we get from figure~\ref{fig:500}~(b)
\begin{eqnarray}
{\cal M}^{(h \gamma)}_{pp \to pp \pi^{+} \pi^{-}} &=&
\bar{u}(p_{2}, \lambda_{2}) 
\Gamma^{(\gamma pp)\,\mu}(p_{2},p_{b}) 
u(p_{b}, \lambda_{b}) \frac{1}{t_{2}} \nn \\
&&\times
\Big{[} 
{\cal M}^{(\rm res)}_{\mu, \lambda_{1}, \lambda_{a}}(p_{3},p_{4},p_{1},q_{2},p_{a})|_{\Pom + f_{2 \Reg} + a_{2 \Reg}} \nn \\
&&+
{\cal M}^{(f_{2})}_{\mu, \lambda_{1}, \lambda_{a}}(p_{3},p_{4},p_{1},q_{2},p_{a})|_{\rho_{\Reg} + \omega_{\Reg} + \Ode} \nn \\
&&+
{\cal M}^{(\rm DS)}_{\mu, \lambda_{1}, \lambda_{a}}(p_{3},p_{4},p_{1},q_{2},p_{a})|_{\Pom + f_{2 \Reg} + \rho_{\Reg}}
\Big{]} \,.
\label{3.6} 
\end{eqnarray}
Finally we have from figure~\ref{fig:500}~(c) the $\gamma \gamma$-fusion contribution
\begin{eqnarray}
{\cal M}^{(\gamma \gamma)}_{pp \to pp \pi^{+} \pi^{-}} &=&
\bar{u}(p_{1}, \lambda_{1}) 
\Gamma^{(\gamma pp)\,\mu}(p_{1},p_{a}) 
u(p_{a}, \lambda_{a}) \frac{1}{t_{1}}  \nn \\
&&\times
\Big{[} 
{\cal M}^{(f_{2})}_{\mu, \lambda_{2}, \lambda_{b}}(p_{3},p_{4},p_{2},q_{1},p_{b})|_{\gamma} 
+
{\cal M}^{(\rm DS)}_{\mu, \lambda_{2}, \lambda_{b}}(p_{3},p_{4},p_{2},q_{1},p_{b})|_{\gamma}
\Big{]} \,.
\label{3.7} 
\end{eqnarray}

The complete amplitude for (\ref{3.1}) is then
\begin{eqnarray}
{\cal M}_{pp \to pp \pi^{+} \pi^{-}} =
{\cal M}^{(\gamma h)}_{pp \to pp \pi^{+} \pi^{-}} +
{\cal M}^{(h \gamma)}_{pp \to pp \pi^{+} \pi^{-}} +
{\cal M}^{(\gamma \gamma)}_{pp \to pp \pi^{+} \pi^{-}} \,.
\label{3.8} 
\end{eqnarray}
With (\ref{3.8}) we have given the complete amplitude for CEP
of $\pi^{+} \pi^{-}$ in the diffractive regime as specified above.

Eventually we should also
include absorption corrections 
due to the proton-proton interactions to the Born amplitudes
discussed above.
How to treat approximately these effects was discussed 
in section~III~C of \cite{Lebiedowicz:2014bea}.
The absorption reduces the cross section for photoproduction processes by about 10\% at LHC energies.
In this work, the focus is on the non-resonant Drell-S\"oding term,
rather than a precise study of absorptive effects.
These effects depend on the kinematic conditions 
in a particular experiment.
Therefore, in comparing our model results to experimental data, 
these effects should be taken into account 
at the amplitude level.

%--------------------------------------------------
\section{\boldmath Results for the $pp \to pp \pi^{+} \pi^{-}$ reaction and discussion of $\gamma p \to \pi^{+}\pi^{-}p$}
\label{sec:4}
%--------------------------------------------------

In this section we briefly discuss
a preliminary comparison of the tensor pomeron model 
with data for the $\gamma p \to \pi^{+}\pi^{-} p$ reaction 
from \cite{H1:2020lzc,Bolz_Meson2021}
and we present selected results
for the reaction $pp \to pp \pi^{+} \pi^{-}$.

%--------------------------------------------------
\subsection{\boldmath Discussion of $\gamma p \to \pi^{+}\pi^{-}p$}
\label{sec:4.1}
%--------------------------------------------------
In this section we discuss our theoretical approach
for the reaction $\gamma p \to \pi^{+}\pi^{-} p$
with real photons.
It is clearly beyond the scope of the article to give a detailed comparison 
of our theory with the available experimental data.
In our opinion this can only be done by experimentalists,
or together with them.
Thus, we shall here only discuss a preliminary comparison
of our model with some selected results
from the experiment \cite{H1:2020lzc}.
For this we consider \cite{Bolz_Meson2021} from the H1 Collaboration.
There, H1 data from \cite{H1:2020lzc} were compared to 
the tensor-pomeron model results 
for the $\gamma p \to \pi^{+} \pi^{-} p$ reaction
based on the model of \cite{Bolz:2014mya}.
This comparison shows that a better description 
of the H1 data can be achieved by modifying 
the Drell-S\"oding terms
(see sections 2.5 and 2.6 of \cite{Bolz:2014mya})
in such a way that the energy dependence
in the Regge factors (propagators) is changed from
$s \to s_{\rm eff} = s/2$.
This procedure was suggested by the authors of \cite{Bolz:2014mya}.
After this substitution, the energy dependence of the three diagrams
corresponding to the production of $\pi^{+}\pi^{-}$ by pomeron/reggeon exchange remains the same.
As expected, 
the gauge invariance requirements are satisfied as well.
However, the Drell-S\"oding model based on this somewhat arbitrary 
procedure should be considered as an effective model.
In the present work, 
we developed an improved version of the Drell-S\"oding mechanism, 
accounting for the correct kinematics and Regge variables, 
as discussed in detail in section~\ref{sec:2A}.

%-------------------------------------------------------------
\begin{figure}[tbp]
\centering
\includegraphics[width=0.46\textwidth]{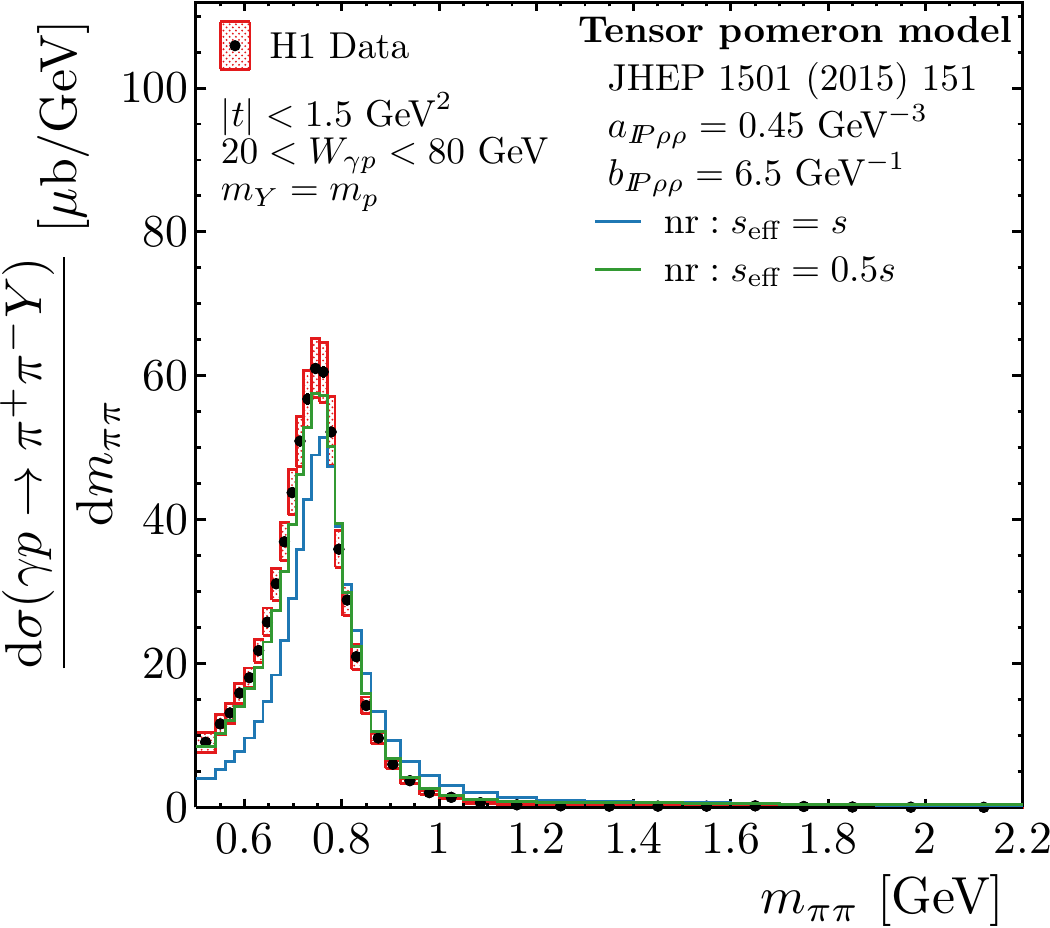}
  \caption{\label{fig:H1}
The invariant mass $M_{\pi\pi}$ distribution measured by the H1 Collaboration \cite{H1:2020lzc}
compared to the results of the tensor-pomeron model \cite{Bolz:2014mya}
in its original form (blue histogram) and with the modification
of replacing in the Regge propagators of the Drell-S\"oding term
$s$ by $s_{\rm eff} = s/2$ (green histogram).
In the figure ``nr'' stands for ``non-resonant''.
In the calculations both resonant and non-resonant contributions
for the $\gamma p \to \pi^{+} \pi^{-} p$ reaction
were included.
This figure is taken from \cite{Bolz_Meson2021}.}
\end{figure}
%--------------------------------------------------------------
Here we show in figure~\ref{fig:H1}, as an example,
the distribution of the $\pi^{+}\pi^{-}$ invariant mass $M_{\pi\pi}$
presented in \cite{Bolz_Meson2021}.
It is clear from figure~\ref{fig:H1} that the prescription to
use $s_{\rm eff} = s/2$ 
in the Drell-S\"oding term gives a much better representation
of the observed $M_{\pi\pi}$ distribution than the original
model \cite{Bolz:2014mya}.
In the section~\ref{sec:4.2} we shall show
that the $M_{\pi\pi}$ distributions for the non-resonant term
with the above $s_{\rm eff}$ prescription
and with our QFT model from section~\ref{sec:2A}
are very similar.
See figure~\ref{fig:1} below.
Thus, we can take the preliminary comparison of
the tensor-pomeron model 
with the \textit{ad hoc} $s_{\rm eff}$ prescription
for the Drell-S\"oding term as an indication that our
present model with the correct QFT calculation for this term
will be quite close to the data.
Of course, as already mentioned, we invite the experimentalists to
make a detailed comparison of the data from \cite{H1:2020lzc}
to our present tensor-pomeron model.

%--------------------------------------------------
\subsection{\boldmath Results for $pp \to pp \pi^{+} \pi^{-}$}
\label{sec:4.2}
%--------------------------------------------------
Here we present selected results 
for the $pp \to pp \pi^{+} \pi^{-}$ reaction
calculated at $\sqrt{s}=13$~TeV.
In the calculations, the resonant
and two-pion continuum (Drell-S\"oding) 
contributions are taken into account.
As already specified after eq.~(\ref{3.3}) we consider
the case where at least one photon exchange is involved.
%that is, at least one proton gets only a very small deviation.
%In this kinematic regime
%the purely hadronic diffractive 
%$\Pom \Pom$, $\Pom \Reg$, and $\Reg \Reg$ contributions, 
%that were discussed e.g. in \cite{Lebiedowicz:2016ioh},
%are much smaller than the ones with photon exchange.
That is, we concentrate here on the calculation of the
$\gamma \Pom$, $\gamma \Reg$, and $\gamma \gamma$ contributions.
%discussed here extensively.
As we mentioned already in the Introduction our purpose here
is to discuss the main characteristics and the magnitudes of the
effects which are produced by our improved version
of the Drell-S\"oding term.
We are inviting the experimentalists to apply our formulas
for the analysis of their data.
Of course, only such a comparison of our
theoretical results with real data
will tell us if our theory is viable.

In our analysis we use the approximation
\begin{eqnarray}
\bar{u}_{\mathfrak{s'}}(p') \gamma^{\mu} u_{\mathfrak{s}}(p) =
(p' + p)^{\mu} \, \delta_{\mathfrak{s'},\mathfrak{s}}\,,
\label{4.1}
\end{eqnarray}
and only the Dirac coupling in the photon-proton vertex function (\ref{A12a}) is taken into account.
In a first approximation we neglect absorption effects.
The Drell-S\"oding term is calculated 
using the propagators and vertices presented in
Appendix~\ref{sec:appendixA}.
In the calculations presented here, 
we use the $F_{M}$ form factor (\ref{A21b})
with $m_{0}^{2} = 0.5~{\rm GeV}^{2}$,
i.e. assuming the same functional form in the photon-pion-pion and the pomeron/reggeon-pion-pion vertices.
The resonant ($\rho^{0}(770) \to \pi^{+} \pi^{-}$) term 
is treated here exactly as in section~III~A of \cite{Lebiedowicz:2014bea}.
For the $\Pom \rho \rho$ and $f_{2 \Reg} \rho \rho$ coupling constants 
we have taken the parameter 
set~A given in (2.15) of \cite{Lebiedowicz:2014bea},
which corresponds to set~B (\ref{B2})
in the present paper, 
and we have used the following parameters 
$\Lambda_{V} = 2$~GeV and $n_{V} = 0.5$
occurring in the $\tilde{F}^{(V)}$ form factor (\ref{2.36.6}),
where $V = \rho, \omega$.

\begin{figure}[tbp]
\centering
\includegraphics[width=0.46\textwidth]{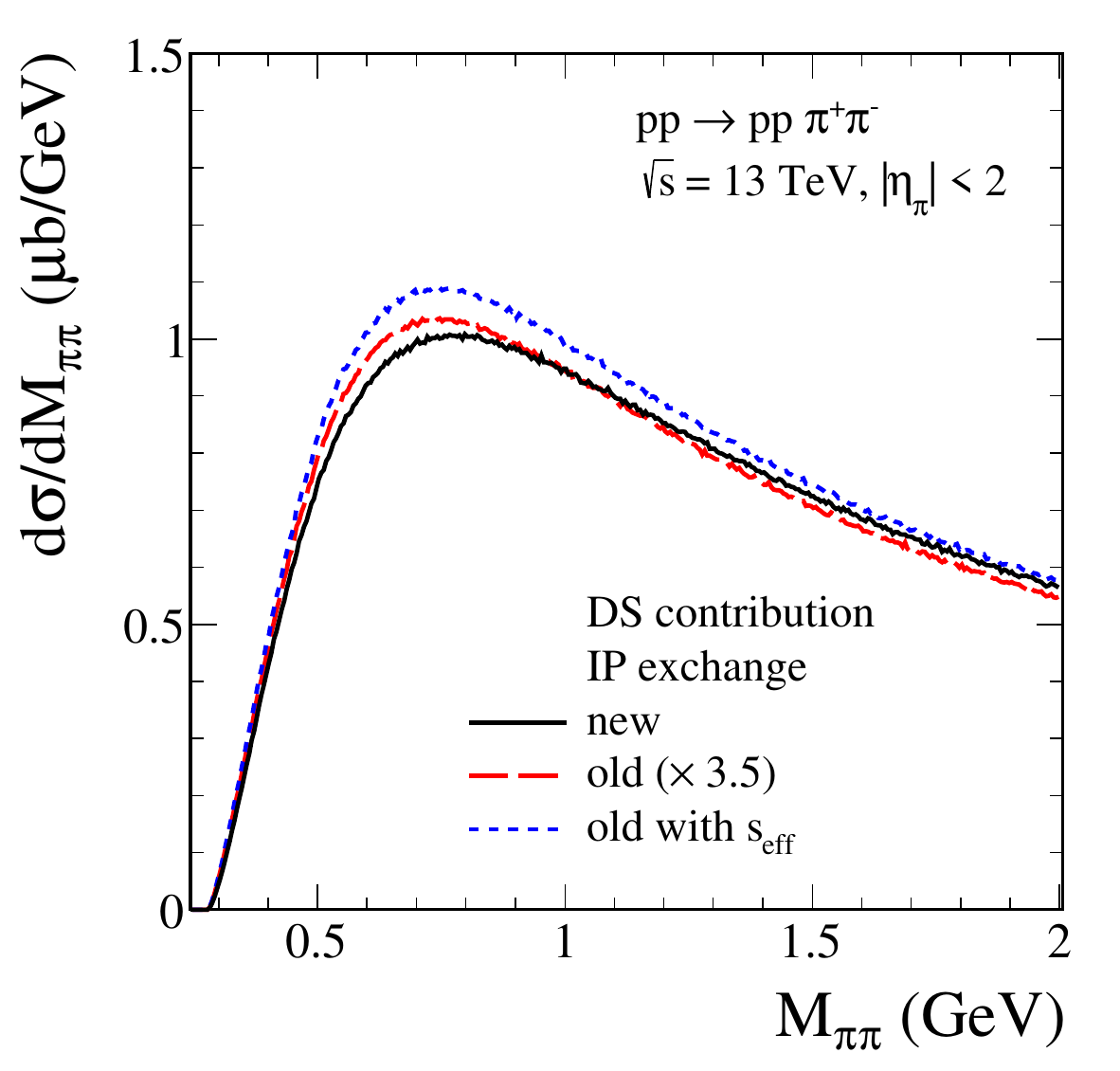}
\includegraphics[width=0.46\textwidth]{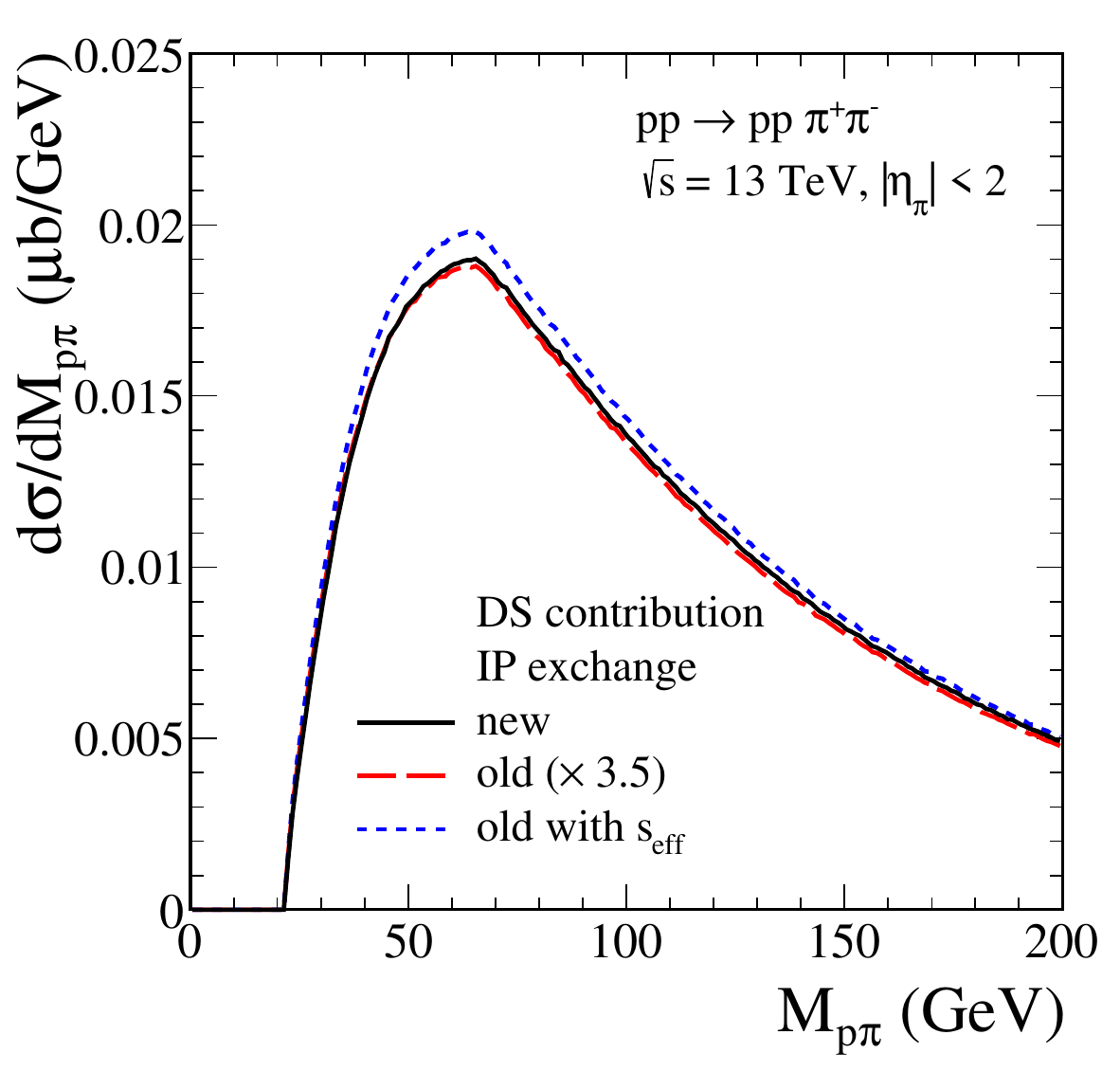}
\includegraphics[width=0.46\textwidth]{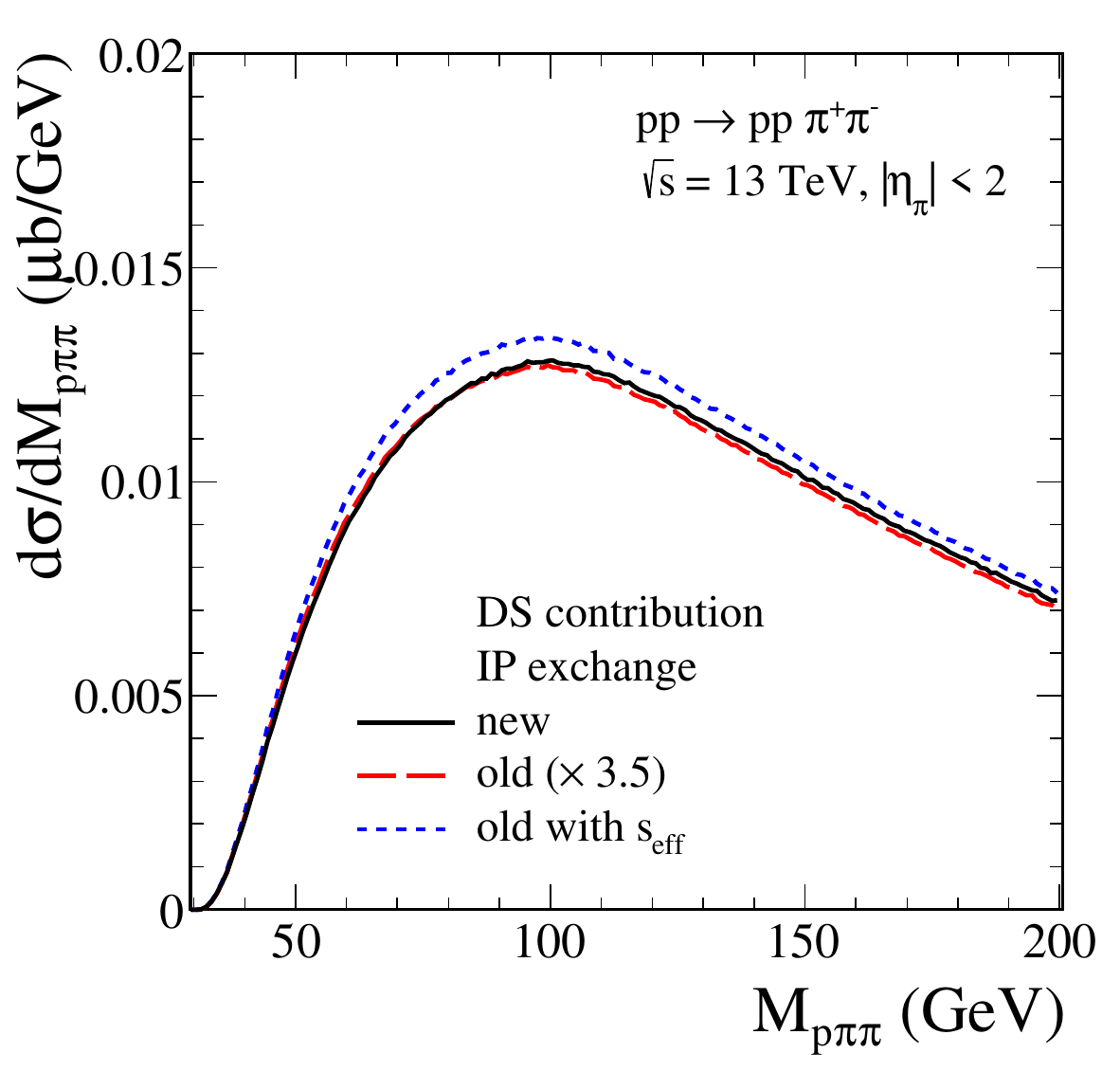}\includegraphics[width=0.46\textwidth]{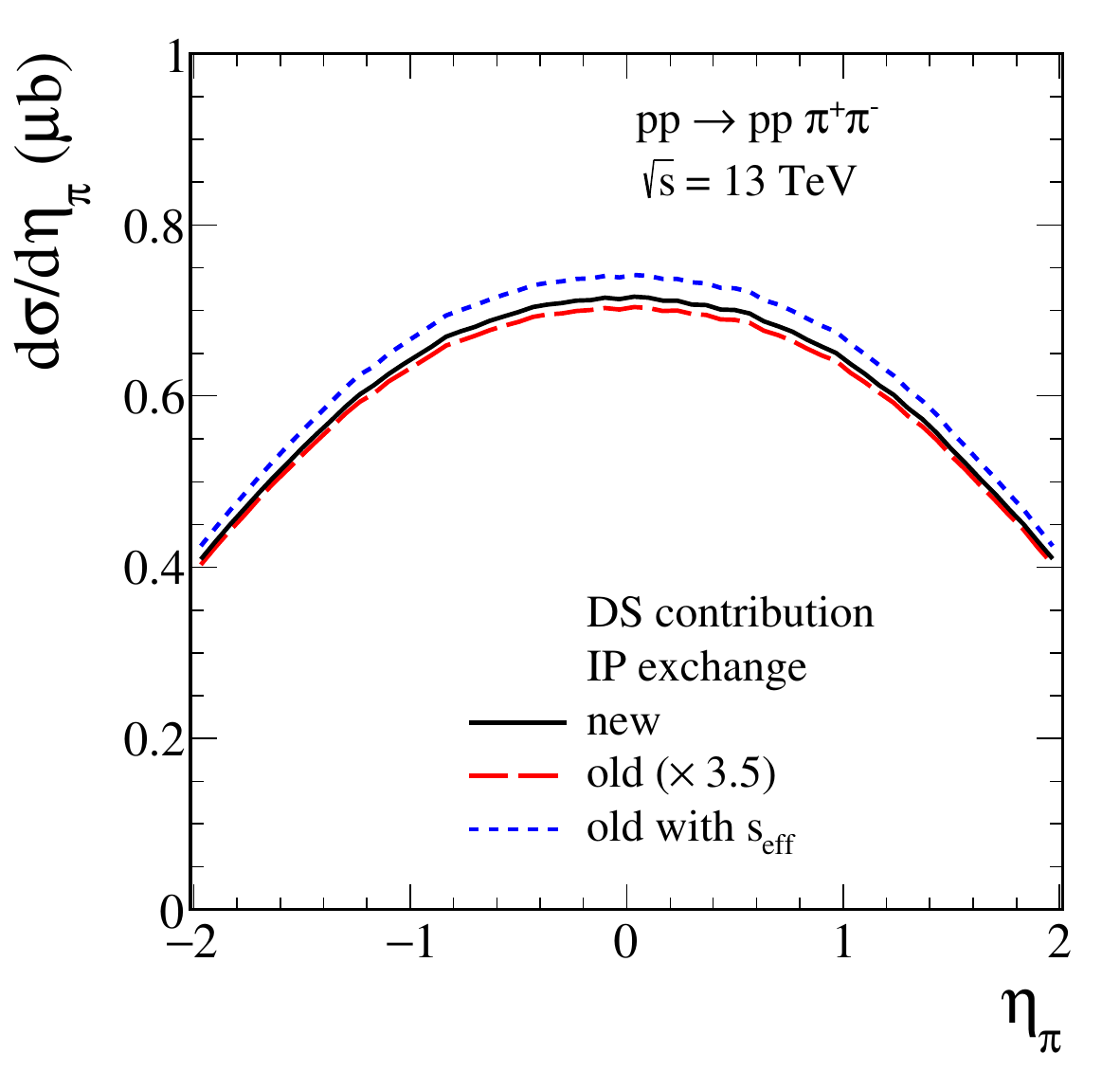}
  \caption{\label{fig:1}
The differential distributions for the $pp \to pp \pi^{+} \pi^{-}$ reaction
at $\sqrt{s}=13$~TeV and for $|\eta_{\pi}| < 2$.
%The calculations were done 
Shown are results for the Drell-S\"oding contribution,
taking only pomeron exchange into account,
%for the revised/improved model 
for three models explained in the text.
The results depicted by the red long-dashed lines,
marked ``old ($\times 3.5$)'', 
correspond to the original model \cite{Bolz:2014mya,Lebiedowicz:2014bea} but
multiplied by a factor of 3.5.
The blue dotted lines correspond to the model \cite{Bolz:2014mya,Lebiedowicz:2014bea}
but with $s$ replaced by $s_{\rm eff} = s/2$
in the pomeron propagators.
The black solid lines are our present results.
%and for the model from \cite{Lebiedowicz:2014bea} (see the red line).
No absorption effects are included here.}
\end{figure}

In figure~\ref{fig:1} we compare our new results 
for the Drell-S\"oding (DS) contribution 
(see the black solid lines)
with the ``old'' DS results based on 
the model discussed in \cite{Lebiedowicz:2014bea}
(see the red long-dashed lines).
For simplicity, we do not include in the calculations 
subleading reggeon exchanges.
%The $\rho(770)$ term is the same in both cases.
From this comparison, 
we can see that the cross section for our new DS contribution 
is larger by a factor of about 3.5 compared to the old one.
In order to improve the old model, one can use
the energy variable $s_{\rm eff} = s/2$ 
instead of $s$ in the Regge propagators.
We see from figure~\ref{fig:H1} which is reproduced from
\cite{Bolz_Meson2021} 
that this procedure leads to a reasonable description of the data
for the $\gamma p \to \pi^{+} \pi^{-} p$ reaction
measured by the H1 Collaboration \cite{H1:2020lzc}.
In figure~\ref{fig:1},
the improved old DS results are shown
by the blue dotted lines and denoted as ``old with $s_{\rm eff}$''.
One can see that the new and the improved old approaches
yield similar characteristics
for the $pp \to pp \pi^{+} \pi^{-}$ reaction.
We emphasize that the factor about 3.5 between the old and the new calculations
of the DS term is essentially due to a correct treatment of the Regge kinematics
as explained in section~\ref{sec:2A}.
The parameters like the $\Pom \pi \pi$ and $f_{2 \Reg} \pi \pi$ coupling constants, $\beta_{\Pom \pi \pi}$ (\ref{A17}) 
and $g_{f_{2 \Reg} \pi \pi}$ (\ref{A18}), 
respectively, are not changed.
%-------------------------------------------------------------
%--------------------------------------------------------------

\begin{figure}[tbp]
\centering
(a)\includegraphics[width=0.46\textwidth]{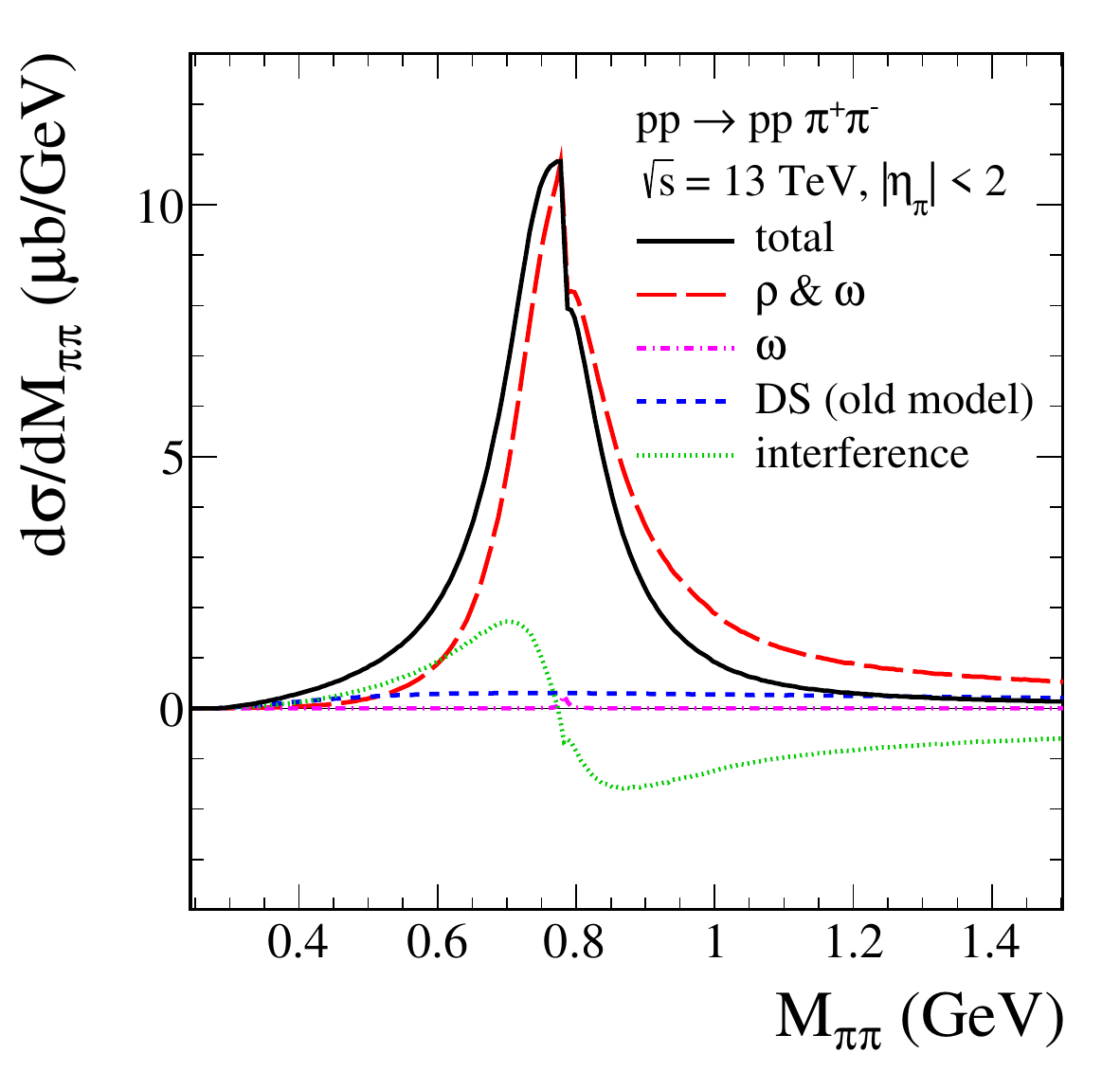}
(b)\includegraphics[width=0.46\textwidth]{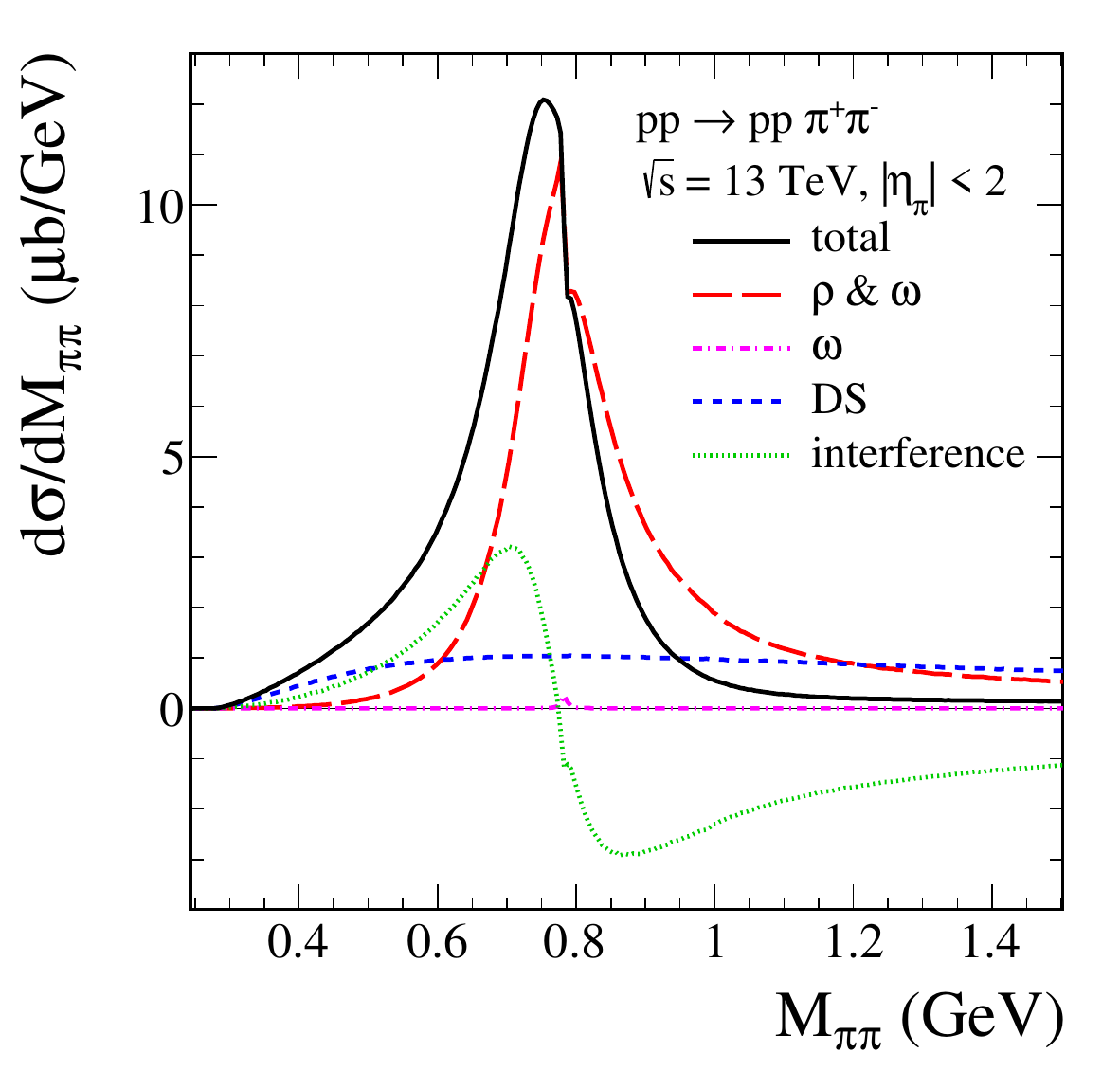}
\caption{\label{fig:2a}
The differential distributions
for the $pp \to pp \pi^{+} \pi^{-}$ reaction
at $\sqrt{s}=13$~TeV and for $|\eta_{\pi}| < 2$.
The full model (total) and individual contributions
from vector-meson and Drell-S\"oding (DS) productions
are shown.
In the panel~(a), 
the DS result corresponds to the model from \cite{Lebiedowicz:2014bea},
while in the panel~(b) to our new model.
%in the right panel we show the results with improved non-resonant model.
The resonant contributions are the same in both cases.
The interference term between ($\rho\,\& \,\omega$) and DS contributions 
is shown by the green dotted line.
No absorption effects are included here.}
\end{figure}

In figure~\ref{fig:2a} we show the distributions
of the two-pion invariant mass for the reaction (\ref{1.2}),
taking into account resonance production of
$\rho^{0}$ and $\omega$ and the Drell-S\"oding (DS) contribution.
In the panel~(a) we show the results 
for the DS contribution from \cite{Lebiedowicz:2014bea}
including the $\Pom$, $f_{2 \Reg}$, and $\rho_{\Reg}$ exchanges.
In the panel~(b) we show the results with our improved DS model.
The skewing of the $\rho(770)$ line-shape caused 
by the interference of 
the $\rho^{0}$ and $\pi^{+} \pi^{-}$ continuum 
contributions is much more significant
within the revised model.
The $\rho$-$\omega$ interference effect is also clearly exposed.
%------------------------------------------------------------------------
%---------------------------------

\begin{figure}[tbp]
\centering
(a)\includegraphics[width=0.46\textwidth]{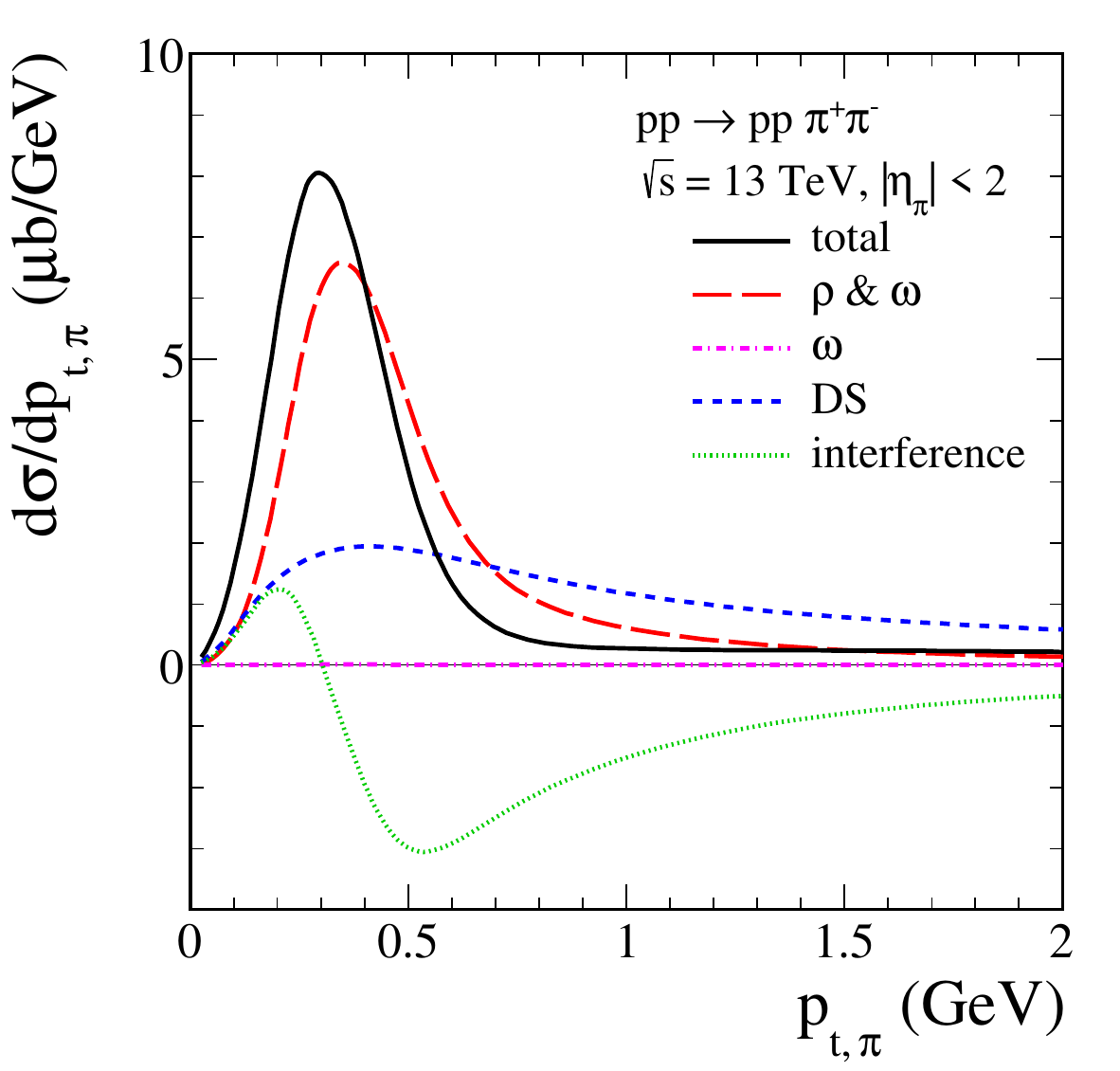}
(b)\includegraphics[width=0.46\textwidth]{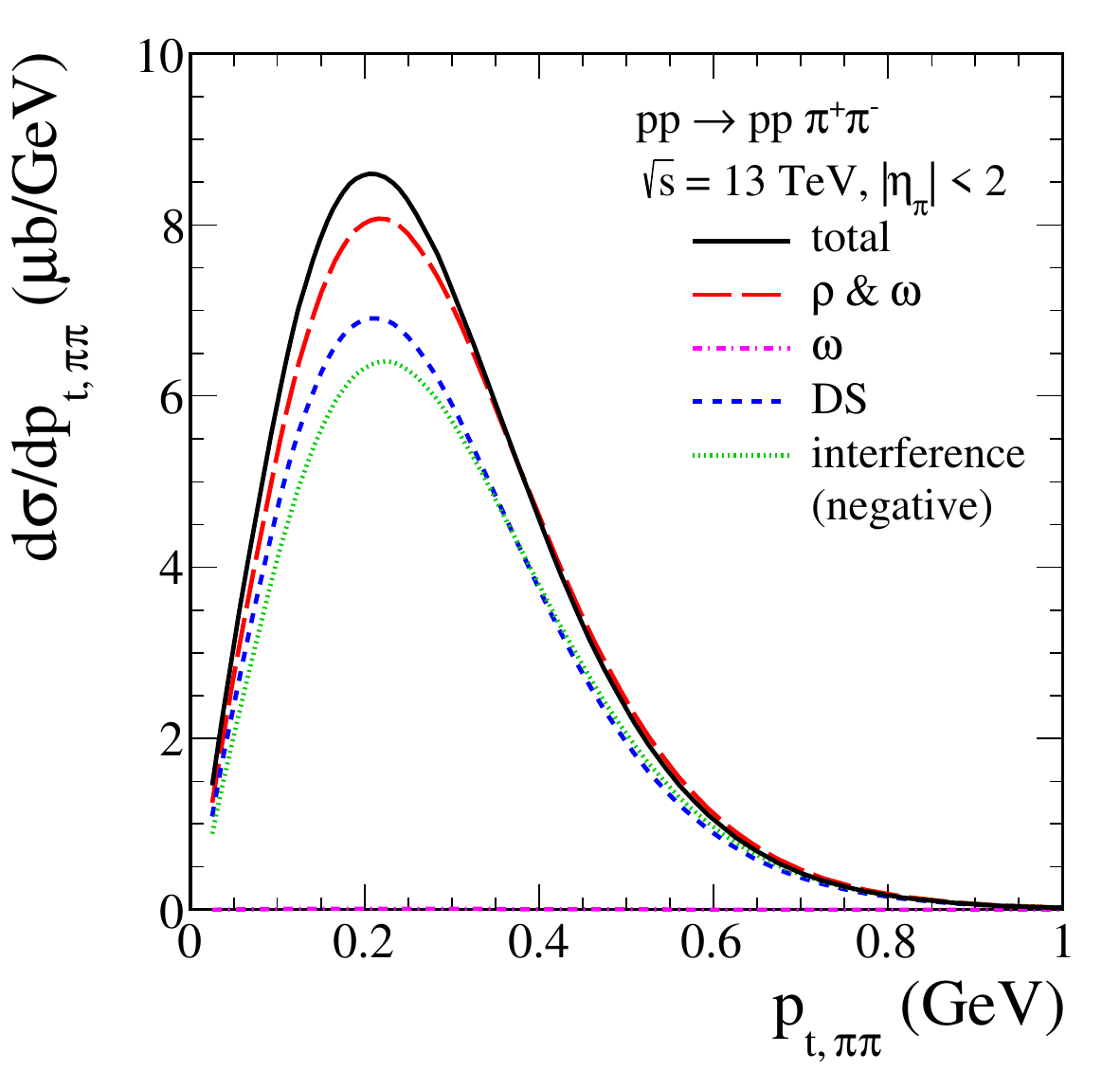}
(c)\includegraphics[width=0.46\textwidth]{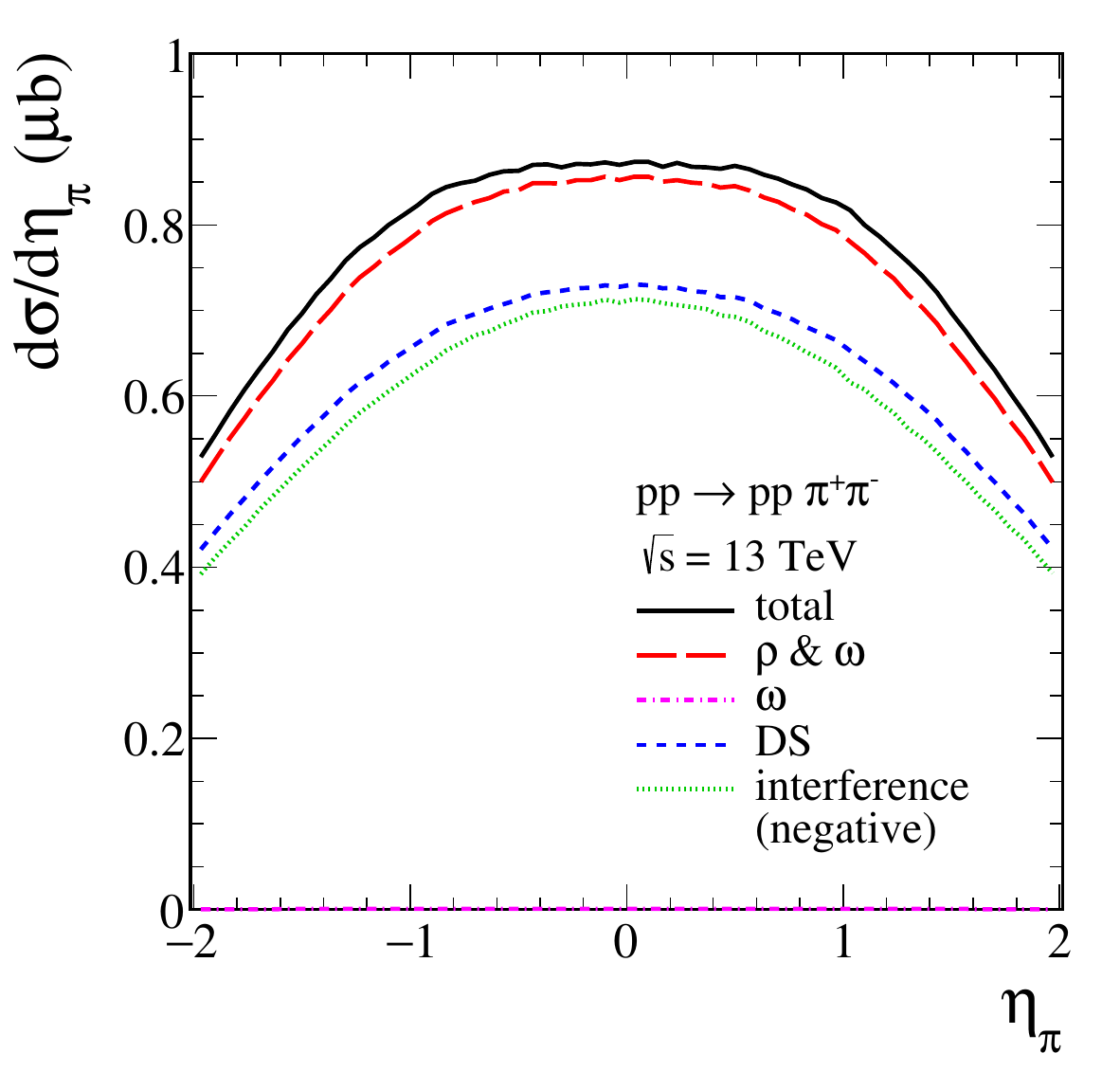}
\caption{\label{fig:2}
The differential distributions
for the $pp \to pp \pi^{+} \pi^{-}$ reaction
at $\sqrt{s}=13$~TeV and for $|\eta_{\pi}| < 2$.
The meaning of the lines is the same as in figure~\ref{fig:2a}~(b).
In the panels (b) and (c) the interference term is destructive.
No absorption effects are included here.
}
\end{figure}

In figure~\ref{fig:2} we show the distributions
in transverse momentum of the pion,
in transverse momentum of the $\pi^{+} \pi^{-}$ pair, 
and in pseudorapidity of the pion.
We predict the total cross section $\sigma_{\rm tot} = 3.09$~$\mu$b
for $|\eta_{\pi}| < 2$, neglecting absorptive corrections.
This result is obtained by using 
the pomeron/reggeon-pion-pion vertices
as given in (\ref{A17})--(\ref{A19})
with the form factor $F_{M}(t)$
from (\ref{A21b}) with 
$m_{0}^{2} = 0.50$~GeV$^{2}$.
Our discussion in Appendix~\ref{sec:appendixB}
suggests to use a slighty different form factor
in (\ref{A17})--(\ref{A19}), 
setting $m_{0}^{2} = 0.75$~GeV$^{2}$
in (\ref{A21b}).
Taking the larger value of $m_0^2 = 0.75$~GeV$^2$ in the DS term gives $\sigma_{\rm tot} = 3.22$~$\mu$b.
For completeness, we also provide the cross-section values
for the DS contribution: $\sigma_{\rm DS} = 2.51$~$\mu$b
for $m_{0}^{2} = 0.50$~GeV$^{2}$ and 
$\sigma_{\rm DS} = 2.75$~$\mu$b for $m_{0}^{2} = 0.75$~GeV$^{2}$.
%------------------------------------------------------------------------
%---------------------------------

\begin{figure}[tbp]
\centering
\includegraphics[width=0.46\textwidth]{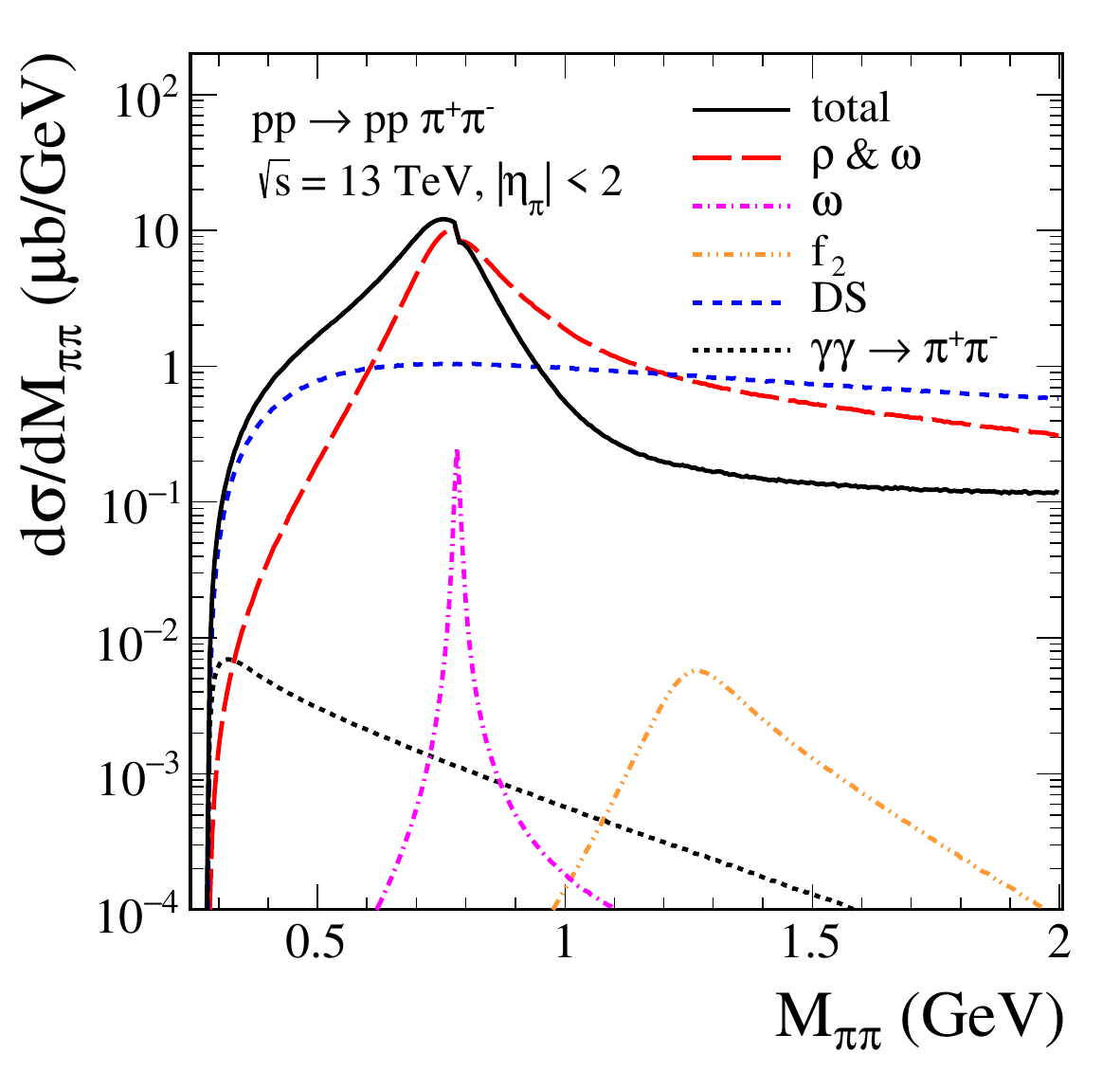}
\includegraphics[width=0.46\textwidth]{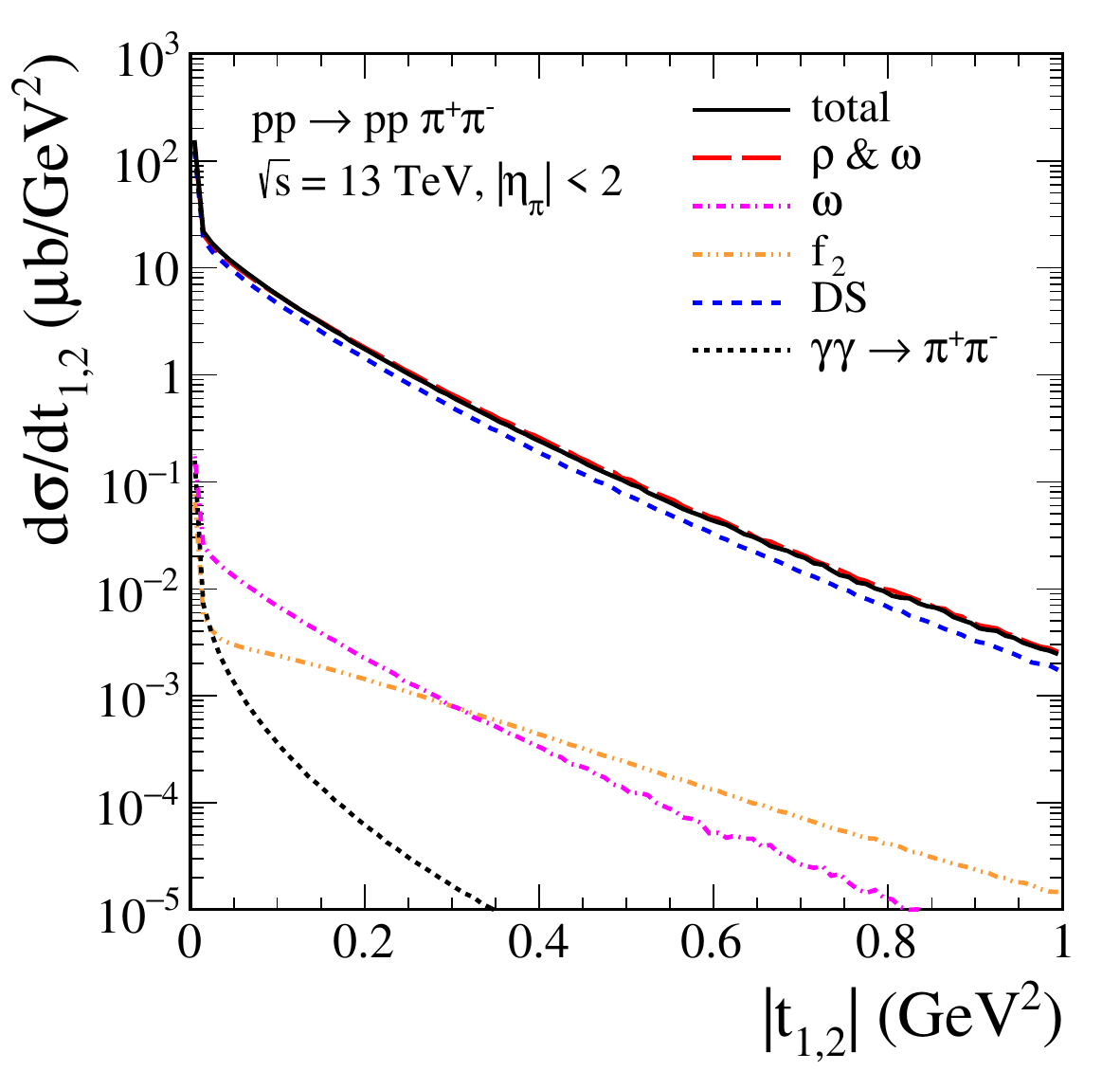}
  \caption{\label{fig:3}
The differential distributions for 
the $pp \to pp \pi^{+} \pi^{-}$ reaction
at $\sqrt{s}=13$~TeV and for $|\eta_{\pi}| < 2$.
The full model (total) and individual contributions
from vector-meson production, 
$f_{2}$ production,
and non-resonant processes are shown.
%The calculations were done 
%Results are shown for the Drell-S\"oding term
%for the revised/improved model 
%for the different conditions explained in the text
%(see the black lines)}
%and for the model from \cite{Lebiedowicz:2014bea} (see the red line).
No absorption effects are included here.}
\end{figure}

Figure~\ref{fig:3} shows the distributions in the two-pion invariant mass
and in the four-momentum transfer squared from one of the proton vertices
(we have $t_{1,2} = t_{1}$ or $t_{2}$).
Again we show the complete result (total), and
the resonant and the non-resonant (DS) contributions.
The left panel shows that the $f_{2}(1270)$ production 
is more than three orders
of magnitude smaller than the production of $\rho(770)$
and that the $\gamma \gamma \to \pi^{+} \pi^{-}$ term 
is negligibly small.
The right panel shows that the $|t_{1,2}|$ distributions
are strongly peaked at very small $|t_{1,2}|$.
This is caused by the factors $1/t_{1,2}$ from the photon propagators.
The low-$|t_{1,2}|$ region, up to $|t_{1,2}| < 0.05$~GeV$^{2}$,
is dominated by the photon exchange.
This peaking of the $\gamma \Pom$ fusion contributions
at very small $|t_{1,2}|$ is the reason that the highly interesting
experimental results of \cite{TOTEM:2024aso}
cannot be compared with our theory.
In \cite{TOTEM:2024aso} a cut on the transverse momenta
of the scattered protons of 
$0.2~{\rm GeV} \leqslant p_{t} \leqslant 0.8~{\rm GeV}$
was applied which corresponds to
$0.04~{\rm GeV}^{2} \leqslant |t| \leqslant 0.64~{\rm GeV}^{2}$.
For these $t$ values the $\gamma \Pom$ fusion contributions
are very small compared to the $\Pom \Pom$ ones; 
see \cite{Lebiedowicz:2016ioh}.
%------------------------------------------------------------------------
%--------------------------------------------------------------------------

\begin{figure}[tbp]
\centering
\includegraphics[width=0.46\textwidth]{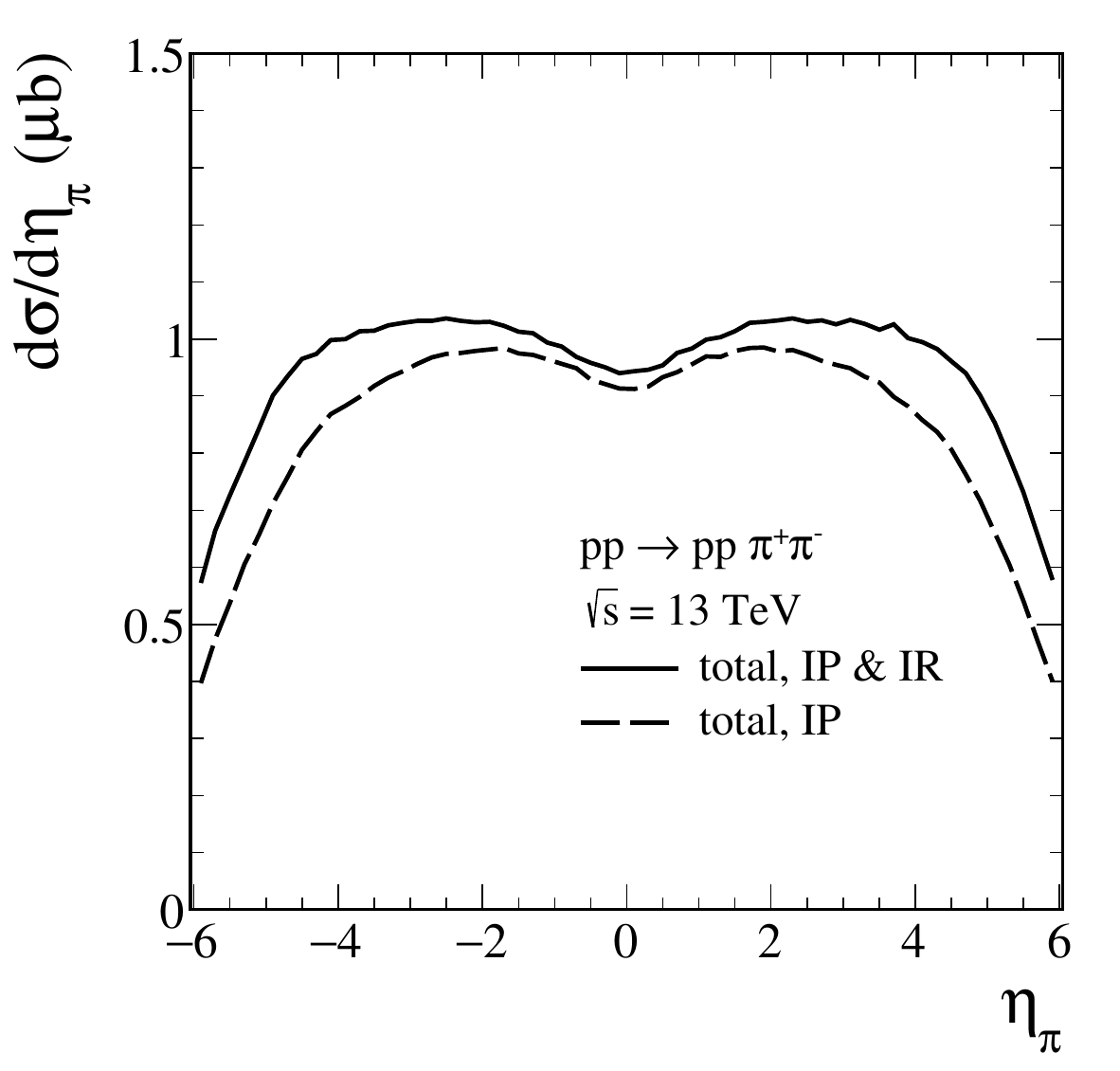}
\includegraphics[width=0.46\textwidth]{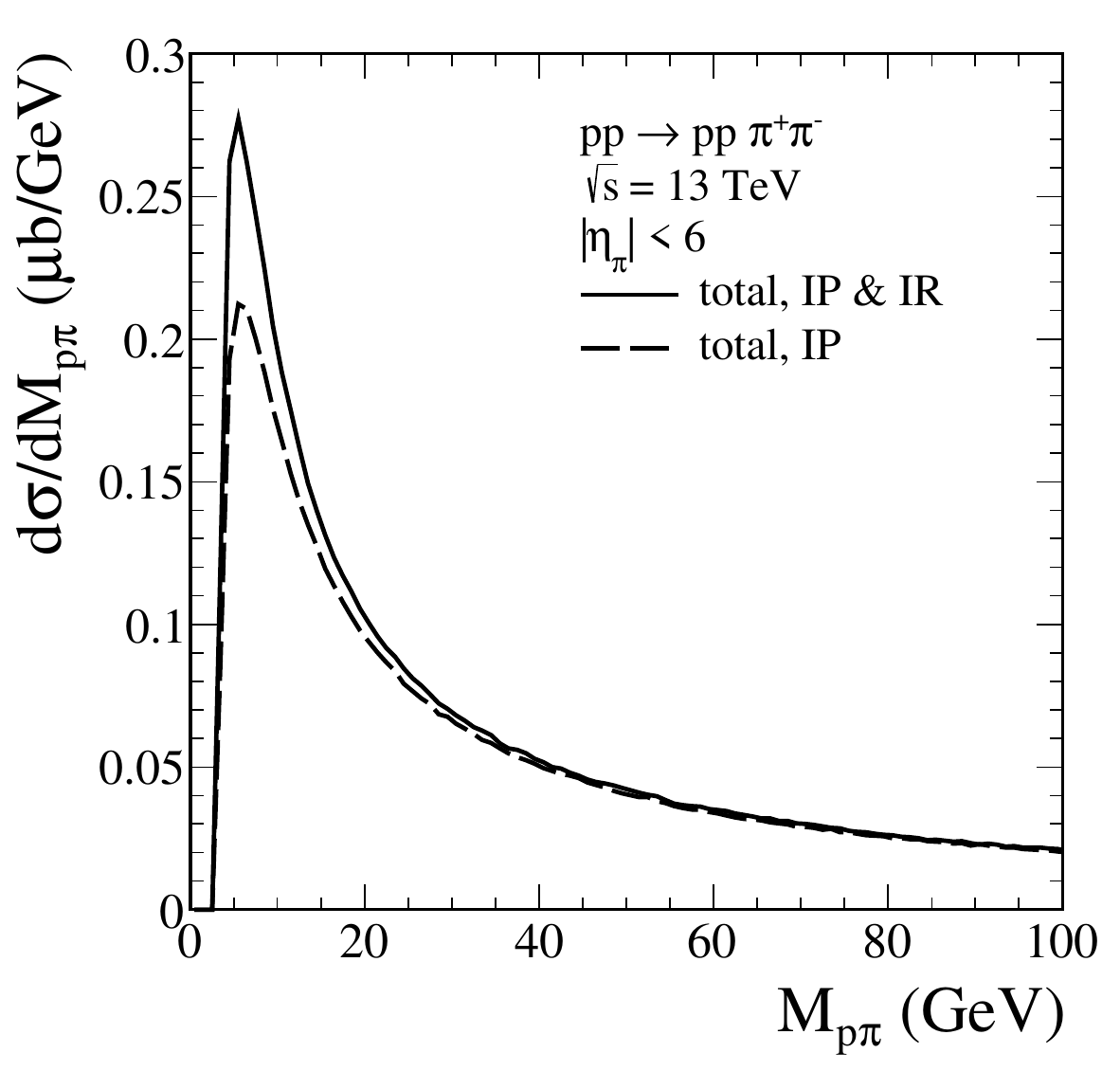}
  \caption{\label{fig:4}
The differential distributions 
for the $pp \to pp \pi^{+} \pi^{-}$ reaction
at $\sqrt{s}=13$~TeV and for $|\eta_{\pi}| < 6$.
No absorption effects are included here.}
\end{figure}

In figure~\ref{fig:4} we present the distributions
in pseudorapidity $\eta_{\pi}$ of the pion and
in the proton-pion invariant mass for $|\eta_{\pi}| < 6$.
The solid lines correspond to the tensor pomeron plus reggeon exchanges
in the amplitude (denoted by $\Pom \,\& \,\Reg$) 
while the dashed lines correspond to the pomeron exchange alone.
We can see that the secondary reggeon exchanges contribute mainly
at backward and forward pion pseudorapidity regions
that correspond to low $M_{p \pi}$ regions.
In figure~\ref{fig:1} we apply a cut $|\eta_{\pi}| < 2$ 
which eliminates small subenergies
$M_{p \pi}$ and we have $M_{p \pi} > 20$~GeV.
%($M_{\pi \pi}$),
%in transverse momentum of the pion ($p_{t, \pi}$),
%in transverse momentum of the proton ($p_{t, p}$),
%and in the azimuthal angle between the outgoing protons ($\phi_{pp}$).
%The calculation were done for the complete model.
%We show the complete result 
%and the results corresponding to the exchange of only $\Pom$.
%------------------------------------------------------------------------
%--------------------------------------------------------------------------

Now we come to the discussion of the uncertainties
of the parameters of our model.
In our opinion, the parameters for the $\rho^{0}$ production are relatively well-fixed by comparing the model results for the $\gamma p \to \rho^{0} p$ reaction with the HERA data;
see appendix~\ref{sec:appendixB}.
The uncertainties of the Drell-S\"oding contribution (the continuum model) due to the parameters of the $\pi^{\pm} p$ interactions (coupling constants and form factors) are estimated to be less than 15\%.
Therefore, they are irrelevant for the large factor of 3.5 
for the Drell-S\"oding term of our present calculation
compared to \cite{Bolz:2014mya,Lebiedowicz:2014bea}.
To summarize: 
the predicted cross section for the $pp \to pp \pi^{+} \pi^{-}$ reaction is given with a precision of about 10\%
(taking into account the impact of absorption corrections).
The cross section for the pure continuum, 
which is difficult to measure, 
is changed by a large factor of 3.5
with respect to the previous model 
from \cite{Bolz:2014mya,Lebiedowicz:2014bea}.
The interference effect between the resonance and the continuum is then modified less (by about 80\%, 
see figure~\ref{fig:2a}).
Let us recall that the interference effect in the models \cite{Bolz:2014mya,Lebiedowicz:2014bea}
was not big enough and could not explain the asymmetric spectral shape of the $M_{\pi\pi}$ distribution
in the region of the $\rho^{0}$ measured by the H1 Collaboration \cite{H1:2020lzc,Bolz_Meson2021};
see figure~\ref{fig:H1}.

%--------------------------------------------------
\section{Conclusions}
\label{sec:5}
%--------------------------------------------------

In the present paper we have
presented theoretical calculations for the reactions
$\gamma^{(*)}p \to \pi^{+} \pi^{-} p$ and
$pp \to pp \pi^{+} \pi^{-}$
using the tensor-pomeron approach \cite{Ewerz:2013kda}.
This model, which was developed to describe soft
high-energy diffractive reactions, 
has its roots in \cite{Nachtmann:1991ua}
where functional integral methods in QCD were used
to investigate this class of reactions.
In particular, it was found that the pomeron-exchange amplitude
for quark-quark, quark-antiquark, and antiquark-antiquark
high-energy scattering could be represented as a superposition
of elementary exchanges of spin $2+4+6+\ldots$;
see section 6.2 of \cite{Nachtmann:1991ua}.
This same structure for the pomeron-exchange amplitude
is realized in the tensor-pomeron approach which, therefore,
has quite a solid root in QCD,
the theory of quarks and gluons.
See appendix~B of \cite{Ewerz:2013kda}.
Thus, the parameters of the model, for instance the $\Pom pp$
coupling constant $\beta_{\Pom NN}$ (\ref{A21})
should be calculable in QCD.
This is certainly true.
But it is clear that the constant $(\beta_{\Pom NN})^{-1}$
of dimension GeV is a nonperturbative quantity like,
for instance, the proton mass.
Therefore, only with nonperturbative methods of QCD
do we have a chance to calculate $\beta_{\Pom NN}$.
Indeed, in \cite{Meggiolaro:1996hf,Meggiolaro:2006mx,Giordano:2008ua} 
the Pisa group has made a brave effort
to calculate properties of high-energy elastic scattering
from lattice QCD using the results of \cite{Nachtmann:1991ua}
and further developments, reviewed in \cite{Nachtmann:1997hei}.
So much for the QCD basis of the tensor-pomeron model
for soft reactions \cite{Ewerz:2013kda}.
In the present paper we only deal with soft diffractive reactions.
But it is legitimate to ask how this model relates to hard diffractive
processes where perturbative QCD methods are frequently applied.
This question really goes beyond the scope of the present article.
But let us just mention that in \cite{Britzger:2019lvc}
and \cite{Lebiedowicz:2022xgi} 
the original tensor-pomeron model of \cite{Ewerz:2013kda}
was supplemented with a hard tensor-pomeron exchange.
The resulting theory is very successful in describing the DIS 
structure functions at low-$x$ and $Q^{2} \lesssim 60$~GeV$^{2}$
\cite{Britzger:2019lvc}
and the available data for deeply virtual Compton scattering 
\cite{Lebiedowicz:2022xgi}.
Concerning the connection of this hard pomeron, for instance,
to the BFKL (Balitsky-Fadin-Kuraev-Lipatov) pomeron
the authors of \cite{Britzger:2019lvc} write that this needs
more studies.
But they note that the value found by them for the hard-pomeron
intercept of $1.3008(^{+73}_{-84})$ is very close
to typical values obtained from BFKL dynamics
in next-to-leading logarithmic approximation \cite{Fadin:1998py,Ciafaloni:1998gs}.
Finally we refer to \cite{Ewerz:2016onn} for some remarks
on the history of the various views of the pomeron exchange.

A main purpose of the present paper was to give,
in the tensor-pomeron approach,
a calculation of the Drell-S\"oding term which is
satisfactory from the QFT point of view.
To explain this we can look at figure~\ref{fig:100} 
for pomeron exchange.
In \cite{Bolz:2014mya} and \cite{Lebiedowicz:2014bea} 
the parameters $(s,t)$ of the effective
pomeron propagators occurring in all three diagrams of
figure~\ref{fig:100} were taken to be the same.
This clearly is an approximation since the energies squared
$s_{1}$ and $s_{2}$ are different in general.
But by simply putting in $s_{1}$ and $s_{2}$ for the diagrams
of figures~\ref{fig:100}(b) and \ref{fig:100}(a),
respectively, gauge invariance is violated.
The key for the solution of this problem was found
in our investigations of bremsstrahlung in \cite{Lebiedowicz:2022nnn}.
In the present work we have used techniques similar to the ones
applied in \cite{Lebiedowicz:2022nnn} to calculate
the non-resonant (Drell-S\"oding, DS) contribution
to the reactions
$\gamma^{(*)}p \to \pi^{+} \pi^{-} p$ and
$pp \to pp \pi^{+} \pi^{-}$.
We have in this way given an essential improvement of
the corresponding calculations presented in 
\cite{Bolz:2014mya} and \cite{Lebiedowicz:2014bea}.
%We have made estimates of the cross sections 
%and we have shown several differential distributions
%from the $\rho(770)$ resonance and the two-pion continuum.
The calculations have been done using the tensor-pomeron approach
including the secondary reggeon exchanges.
We have discussed in detail our assumptions which we use
in the evaluation of the Drell-S\"oding term
in the above two reactions; see (\ref{1.1}) and (\ref{1.2}).
Again, our improvement compared to \cite{Bolz:2014mya,Lebiedowicz:2014bea} is best explained looking at figure~\ref{fig:100}.
In figures~\ref{fig:100}(a) and \ref{fig:100}(b) we have Regge exchanges
with different subenergies squared $s_{2}$, and $s_{1}$, respectively.
This leads to different Regge factors, where, in fact,
we replace $s_{1,2}$ by the correct Regge variables $2 \nu_{1,2}$
which are defined in (\ref{A3}).
We determined the result for the diagram of figure~\ref{fig:100}(c)
from the gauge-invariance condition.
This turned out to be a highly nontrivial task.
Then we have given the complete set of amplitudes,
resonant and nonresonant, for the 
$\gamma^{(*)}p \to \pi^{+} \pi^{-} p$ reaction
in section~\ref{sec:2B} and for the
$pp \to pp \pi^{+} \pi^{-}$ reaction in section~\ref{sec:3}.
In section~\ref{sec:4} we have given numerical results
for the $pp \to pp \pi^{+} \pi^{-}$ reaction
at c.m. energy $\sqrt{s} = 13$~TeV.
The improved two-pion continuum model 
for the $pp \to pp \pi^{+} \pi^{-}$ reaction
gives a larger cross section 
for the DS contribution
by a factor of about 3.5 compared to the previous one
obtained in \cite{Lebiedowicz:2014bea}.
As a result, the interference effect, arising from
the resonant $\rho(770)$ and the two-pion continuum contributions,
is more pronounced
and leads to larger skewing of the observed 
spectral shape of $\rho^{0}$.
Our findings should be important for the ongoing and planned
measurements of the $pp \to pp \pi^{+} \pi^{-}$ reaction by
the ALICE, ATLAS, CMS, and LHCb Collaborations at the LHC, even
when the leading protons are not detected 
and only rapidity-gap conditions are checked.
Our findings are also important for the resonant ($\rho^{0} \to \pi^{+} \pi^{-}$) and non-resonant dipion production
in ultra-peripheral $p$A and AA collisions.

We emphasize that a detailed comparison of our improved model
for real photoproduction 
$\gamma p \to \pi^{+} \pi^{-} p$
with the corresponding HERA data \cite{H1:2020lzc}
would be most welcome.
This can only be done by the corresponding experimentalists.

In the present paper we have only discussed
the preliminary H1 results shown in \cite{Bolz_Meson2021}.
We found that our new Drell-S\"oding calculation should be
quite consistent with the two-pion mass distribution
from \cite{Bolz_Meson2021};
see the discussion of figure~\ref{fig:H1}.
A detailed comparison of theory and experiment for the total cross section,
the two-pion invariant mass distribution, and other
differential cross sections,
would allow to get precise numbers for our model parameters.
These are in essence the intercepts and slope parameters
of the pomeron and reggeons and their couplings
to the proton and the pions.
Most of these parameters, 
except for those connected with the odderon,
are already quite well known from other reactions;
see for instance \cite{Donnachie:2002,Ewerz:2013kda}.
With the experimental study of the reactions considered by us
here we shall, thus, get a detailed check 
of the basic structures of the tensor-pomeron model.
The production of $f_{2}$ decaying to $\pi^{+} \pi^{-}$
can occur by photon plus odderon fusion,
and thus offers the chance to detect and study odderon effects.
All these topics can also be studied in the future at
the electron-ion colliders, e.g.,
EIC \cite{Accardi:2012qut,Aschenauer:2017jsk,AbdulKhalek:2021gbh,Anderle:2021wcy} 
and LHeC \cite{LHeCStudyGroup:2012zhm,Bordry:2018gri}.

Finally we note that in many studies of the reactions
$\gamma p \to \pi^{+} \pi^{-} p$ and
$p p \to pp \pi^{+} \pi^{-}$ the Drell-S\"oding contribution
is represented by some simple function with parameters
to be determined by experiment.
We hope to have shown in the present paper that the DS term
can and should be calculated with reliable methods of QFT
using the tensor-pomeron approach. Thus, in our opinion
the DS term deserves to be studied experimentally for
its own sake.

%Comparison of the model with the H1 data  
%for the $\gamma p \to \pi^{+} \pi^{-} p$ reaction,
%in particular, description of measured $\sigma(W_{\gamma p})$
%and of the $d\sigma/dM_{\pi\pi}$ and $d\sigma/dt$ differential cross sections,
%would allow tuning of the model parameters 
%and precise determination of their uncertainty ranges.
%In \cite{Bolz_Meson2021} the first attempt of comparison was made.

%--------------------
\appendix
%--------------------

%\appendix % label section numbers alphabetically: "A", "B", etc
%\counterwithin*{equation}{section} 
\renewcommand\theequation{\thesection.\arabic{equation}} 
% reset 'equation' counter whenever '\section' is executed
% how to display the equation "number" - ohne "dot", also C1 statt C.1

%--------------------------------------------------
\section{Kinematic relations, propagators, and vertices}
\label{sec:appendixA}
%--------------------------------------------------

We start with kinematic relations for the amplitude (\ref{2.2a}).
We have energy-momentum conservation
\begin{eqnarray}
q + p = k_{1} + k_{2} + p'\,,
\label{A1}
\end{eqnarray}
and we define the following variables:
\begin{eqnarray}
s &=& (q + p)^{2} = (k_{1} + k_{2} + p')^{2}\,,
\qquad \sqrt{s} = W_{\gamma p}\,, \nonumber \\
t &=& (p - p')^{2} = (q - k_{1} - k_{2})^{2}\,, \nonumber \\
s_{1} &=& (p' + k_{1})^{2} = (p + q - k_{2})^{2}\,, \nonumber \\
u_{1} &=& (p - k_{1})^{2} = (p' - q + k_{2})^{2}\,, \nonumber \\
s_{2} &=& (p' + k_{2})^{2} = (p + q - k_{1})^{2}\,, \nonumber \\
u_{2} &=& (p - k_{2})^{2} = (p' - q + k_{1})^{2}\,, \nonumber \\
M_{\pi \pi}^{2} &=& (k_{1} + k_{2})^{2}
= (p - p' + q)^{2}
\,;
\label{A2}\\
\nu_{1} &=& \frac{1}{4}(s_{1} - u_{1}) \nonumber\\
        &=& \frac{1}{4}\left[ (p+p',k_{1}-k_{2}) + (p+p',q) \right]\,, \nonumber\\
\nu_{2} &=& \frac{1}{4}(s_{2} - u_{2}) \nonumber\\
        &=& \frac{1}{4}\left[ (p+p',k_{2}-k_{1}) + (p+p',q) \right]\,.
\label{A3}
\end{eqnarray}
We have
\begin{eqnarray}
\nu_{1}^{2} &=& \frac{1}{16}[(p+p',q)^{2} + (p+p',k_{1}-k_{2})^{2} + 2(p+p',k_{1}-k_{2})(p+p',q)]\,, \nonumber\\
\nu_{2}^{2} &=& \frac{1}{16}[(p+p',q)^{2} + (p+p',k_{1}-k_{2})^{2} - 2(p+p',k_{1}-k_{2})(p+p',q)]\,,
\label{A4}
\end{eqnarray}
and we define
\begin{eqnarray}
\bar{\nu}^{2} &=& \frac{1}{2}(\nu_{1}^{2} + \nu_{2}^{2})\,, \nonumber\\
\varkappa &=& \frac{2(q,p+p')(p+p',k_{1}-k_{2})}{16 \bar{\nu}^{2}}\,.
\label{A5}
\end{eqnarray}
Note that $\nu_{1}^{2}$ and $\nu_{2}^{2}$ differ 
only in the sign of the term which is linear in $q$. 
This is important for our calculations.

From (\ref{A2}) and (\ref{A3}) we find
\begin{eqnarray}
\nu_{1} &=& \frac{1}{2}(p+p',k_{1}) \geqslant m_{p} m_{\pi} > 0\,, \nonumber\\
\nu_{2} &=& \frac{1}{2}(p+p',k_{2}) \geqslant m_{p} m_{\pi} > 0\,.
\label{A5a}
\end{eqnarray}
We have
\begin{eqnarray}
| \varkappa | & \leqslant & 1\,, 
\label{A5b}\\
16 \nu_{1}^{2} &=& 16 \bar{\nu}^{2} (1 + \varkappa) \,, \nonumber \\
16 \nu_{2}^{2} &=& 16 \bar{\nu}^{2} (1 - \varkappa) \,.
\label{A5c}
\end{eqnarray}
For real $\lambda$ we have
\begin{eqnarray}
(16 \nu_{2}^{2})^{-\lambda} 
&=& (16 \bar{\nu}^{2})^{-\lambda}
  + (16 \nu_{2}^{2})^{-\lambda} 
  - (16 \bar{\nu}^{2})^{-\lambda} \nonumber \\
&=& (16 \bar{\nu}^{2})^{-\lambda} \left[ 1 + (1-\varkappa)^{\lambda}-1 \right] \nonumber \\
&=& (16 \bar{\nu}^{2})^{-\lambda} \left[ 1 + \lambda \varkappa \,g(\lambda,\varkappa) \right]\,,
\label{A5d}
\end{eqnarray}
where we define
\begin{eqnarray}
g(\lambda, \varkappa) = \frac{(1 - \varkappa)^{-\lambda} - 1}{\lambda \varkappa} \,.
\label{A5e}
\end{eqnarray}
For $|\varkappa| < 1$ we can expand $g(\lambda, \varkappa)$
in a power series in $\varkappa$ which shows that there
$g(\lambda, \varkappa)$ is an analytic function in $\varkappa$
for all complex and real values of $\lambda$:

\begin{eqnarray}
g(\lambda, \varkappa) = 
1 + \frac{\lambda + 1}{2!} \varkappa +
\frac{(\lambda + 1)(\lambda + 2)}{3!} \varkappa^{2} +  \ldots \,.
\label{A5f}
\end{eqnarray}
Inserting the explicit expression for $\varkappa$ from (\ref{A5})
in (\ref{A5d}) we get
\begin{eqnarray}
(16 \nu_{2}^{2})^{-\lambda} 
&=&
(16 \bar{\nu}^{2})^{-\lambda} 
\left[ 1 + \lambda \varkappa \,g(\lambda, \varkappa) \right]\nn \\
&=&
(16 \bar{\nu}^{2})^{-\lambda} 
\left[ 1 + \lambda
\frac{2(q,p+p')(p+p',k_{1}-k_{2})}{16 \bar{\nu}^{2}} g(\lambda,\varkappa) \right].
\label{A5g}
\end{eqnarray}
In a completely analogous way we get
\begin{eqnarray}
(16 \nu_{1}^{2})^{-\lambda} 
=
(16 \bar{\nu}^{2})^{-\lambda} 
\left[ 1 - \lambda
\frac{2(q,p+p')(p+p',k_{1}-k_{2})}{16 \bar{\nu}^{2}} g(\lambda,-\varkappa) \right].
\label{A5h}
\end{eqnarray}

Next we list and discuss propagators and vertices which we need
for our work.
We use the framework of QCD plus electromagnetism to the lowest relevant order.
Thus, all our propagators and vertices respect
the symmetries: 
parity ($P$),
charge conjugation ($C$), 
time reversal ($T$), 
and $\Theta = CPT$.

We start with the full pion propagator and 
the full pion-photon vertex function:
\begin{align}
\includegraphics[width=120pt]{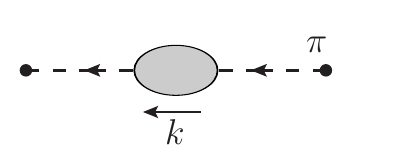}
i\Delta_{F}(k^{2})\,,
\label{A6}
\end{align}
\begin{align}
&\includegraphics[width=120pt]{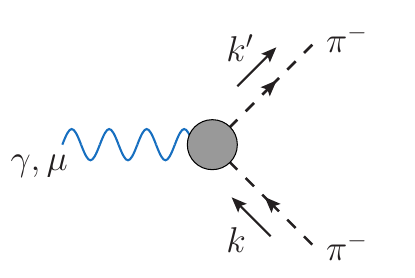}
\includegraphics[width=120pt]{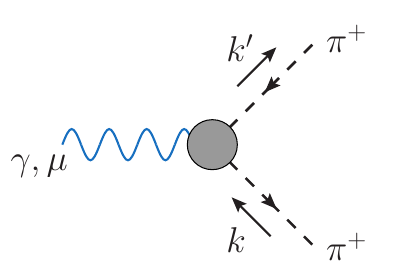}
\nn \\
&
i \Gamma_{\mu}^{(\gamma \pi^{-}\pi^{-})}(k',k)
= -i \Gamma_{\mu}^{(\gamma \pi^{+}\pi^{+})}(k',k)
= i e \widehat{\Gamma}_{\mu}^{(\gamma \pi\pi)}(k',k) \,.
%& = ie (k' + k)_{\mu}
%C: here we discuss the full vertex
\label{A7}
\end{align}
We take $e = \sqrt{4 \pi \alpha_{\rm em}} > 0$ and we have
\begin{eqnarray}
\Delta_{F}^{-1}(m_{\pi}^{2}) &=& 0\,, \quad
\frac{\partial}{\partial k^{2}}
\Delta_{F}^{-1}(k^{2})|_{k^{2} = m_{\pi}^{2}} = 1\,,
\label{A7a} \\
\widehat{\Gamma}_{\mu}^{(\gamma \pi \pi)}(k',k)&=& 
\widehat{\Gamma}_{\mu}^{(\gamma \pi \pi)}(k,k')
= -\widehat{\Gamma}_{\mu}^{(\gamma \pi \pi)}(-k',-k)\,;
\label{A8}
\end{eqnarray}
see (B12) and (B28)--(B33) of \cite{Lebiedowicz:2023mlz}.

The generalized Ward identity
\cite{Ward:1950xp,Takahashi:1957xn}
for the pion field reads
\begin{eqnarray}
(k'-k)^{\mu} \, \widehat{\Gamma}_{\mu}^{(\gamma \pi\pi)}(k',k) =
\Delta_{F}^{-1}(k'^{2}) - \Delta_{F}^{-1}(k^{2})\,;
\label{A9}
\end{eqnarray}
see (B39) of \cite{Lebiedowicz:2023mlz}.
Using (\ref{A7a})--(\ref{A9}) 
we find from (B13) and (B50) of \cite{Lebiedowicz:2023mlz} 
the following relations
\begin{eqnarray}
\Delta_{F}^{-1}(k^{2}) &=&
(k^{2} - m_{\pi}^{2}) + (k^{2} - m_{\pi}^{2})^{2} \, C(k^{2} - m_{\pi}^{2})\,,
\label{A18a}\\
\Delta_{F}(k'^{2})\,
\widehat{\Gamma}_{\lambda}^{(\gamma \pi \pi)}(k',k)|_{k^{2} = m_{\pi}^{2}}
&=&
\frac{(k'+k)_{\lambda}}{k'^{2} - m_{\pi}^{2} + i\varepsilon}
-
\left[(k'+k)_{\lambda} q'^{2} - q'_{\lambda}(k'^{2}-m_{\pi}^{2})\right] \nn\\
&&
\times 
\left(k'^{2} - m_{\pi}^{2}  + i\varepsilon \right)^{-1}
\left[ 1+(k'^{2}-m_{\pi}^{2})\,C(k'^{2} - m_{\pi}^{2}) \right]^{-1}
\widetilde{B}(k'^{2}-m_{\pi}^{2}, 0, q'^{2})\,, 
\nn \\
q' &=& k' - k \,.
\label{A18b}
\end{eqnarray}
Here $C(k'^{2} - m_{\pi}^{2})$ and $\widetilde{B}(k'^{2}-m_{\pi}^{2}, k^{2} - m_{\pi}^{2}, q'^{2})$, 
as defined in appendix~B of \cite{Lebiedowicz:2023mlz},
are analytic functions, free of singularities for
\begin{eqnarray}
|k'^{2} - m_{\pi}^{2}| < 8 m_{\pi}^{2}\,, \quad
|k^{2} - m_{\pi}^{2}| < 8 m_{\pi}^{2}\,, \quad
|q'^{2}| < 4 m_{\pi}^{2}\,.
\label{A18c}
\end{eqnarray}
The function 
$\widetilde{B}(k'^{2}-m_{\pi}^{2}, k^{2} - m_{\pi}^{2}, (k' - k)^{2})$ is symmetric under the exchange $k \leftrightarrow k'$.

From (\ref{A18a}) and (\ref{A18b}) we obtain
\begin{eqnarray}
\widehat{\Gamma}_{\lambda}^{(\gamma \pi \pi)}(k',k)|_{k^{2} = m_{\pi}^{2}}
&=&
(k'+k)_{\lambda}
\left[ 
1 - q'^{2} \,\widetilde{B}(k'^{2}-m_{\pi}^{2}, 0, q'^{2}) +
(k'^{2}-m_{\pi}^{2})\,C(k'^{2} - m_{\pi}^{2}) \right] \nn \\
&& + \,q'_{\lambda} (k'^{2}-m_{\pi}^{2})\,
\widetilde{B}(k'^{2}-m_{\pi}^{2}, 0, q'^{2})\,.
\label{A18e}
\end{eqnarray}
Specializing in (\ref{A18e}) to $k'^{2} = m_{\pi}^{2}$
we get the relation to the electromagnetic form factor of the pion
\begin{eqnarray}
\widehat{\Gamma}_{\lambda}^{(\gamma \pi \pi)}(k',k)|_{k^{2} = k'^{2} = m_{\pi}^{2}}
&=&
(k'+k)_{\lambda}
\left[ 
1 - q'^{2} \,\widetilde{B}(0, 0, q'^{2}) \right] \nn \\
&=&  (k'+k)_{\lambda} \, F_{\rm em}^{(\pi)}(q'^{2})\,,
\label{A18f} \\
F_{\rm em}^{(\pi)}(q'^{2}) &=& 
1 - q'^{2} \,\widetilde{B}(0, 0, q'^{2})\,,
\label{A18g} \\
\widetilde{B}(0, 0, q'^{2}) &=& 
\frac{1 - F_{\rm em}^{(\pi)}(q'^{2})}{q'^{2}}\,.
\label{A18h}
\end{eqnarray}

For our calculations of the Drell-S{\"o}ding term
we set in (\ref{A18b})
$k = k_{1}$, $k_{1}^{2} = m_{\pi}^{2}$,
\mbox{$q'=-q$},
$k' = k_{1} - q$,
and use (\ref{A8}) and 
the symmetry relation of $\widetilde{B}$. We get then
\begin{eqnarray}
&&\widehat{\Gamma}_{\mu}^{(\gamma \pi \pi)}(k_{1}, k_{1}-q)\,
\Delta_{F}\big{[}(k_{1}-q)^{2}\big{]}|_{k_{1}^{2} 
= m_{\pi}^{2}}
\nn\\
&&\quad =
\frac{(2k_{1} -q)_{\mu}}{-2k_{1} \cdot q + q^{2} + i\varepsilon}
\bigg{\lbrace}
1 - q^{2} \,\widetilde{B}(0, -2k_{1} \cdot q + q^{2} ,q^{2})
\left[
1 + (-2k_{1} \cdot q + q^{2})\,
C(-2k_{1} \cdot q + q^{2}) \right]^{-1}
\bigg{\rbrace}
\nn\\
&&\qquad - \, q_{\mu}\,
\widetilde{B}(0, -2k_{1} \cdot q + q^{2} ,q^{2})
\left[
1 + (-2k_{1} \cdot q + q^{2})\,
C(-2k_{1} \cdot q + q^{2}) \right]^{-1}
\,.
\label{A18i}
\end{eqnarray}
%

%In our case the terms with explicit $q_{\mu}$
%do not contribute.
%For real photons in (\ref{1.1}) this is clear.
%For virtual photons in electroproduction and in (\ref{1.2})
%this follows from current conservation at the electron-photon and proton-photon vertices.

We see that for real photons where $q^{2} = 0$
%only the first term on the r.h.s. of (\ref{A10}) survives.
%For photons of small virtuality (\ref{2.2b}) we shall neglect
%the second term on the r.h.s. of (\ref{A10}) which,
%in essence, is a correction proportional 
%to $q^{2}$ to the first term.
%Thus, in our present paper we set
we get as exact results
\begin{eqnarray}
\widehat{\Gamma}_{\mu}^{(\gamma \pi \pi)}(k_{1}, k_{1}-q)\,
\Delta_{F}\big{[}(k_{1}-q)^{2}\big{]}|_{k_{1}^{2} = m_{\pi}^{2}\,, \; q^{2} = 0}
=
\frac{(2k_{1} -q)_{\mu}}{-2k_{1} \cdot q + q^{2} + i\varepsilon} + \; {\rm a \; gauge \; term} \propto q_{\mu}\,, \nn \\
\label{A20}
\end{eqnarray}
and similarly
\begin{eqnarray}
\widehat{\Gamma}_{\mu}^{(\gamma \pi \pi)}(k_{2}, k_{2}-q)\,
\Delta_{F}\big{[}(k_{2}-q)^{2}\big{]}|_{k_{2}^{2} = m_{\pi}^{2}\,, \; q^{2} = 0}
=
\frac{(2k_{2} -q)_{\mu}}{-2k_{2} \cdot q + q^{2} + i\varepsilon} + \; {\rm a \; gauge \; term} \propto q_{\mu}\,. \nn \\
\label{A21_aux}
\end{eqnarray}

For photons of small virtuality (\ref{2.2b}),
\begin{eqnarray}
-0.5~{\rm GeV}^{2} < q^{2} < 0\,,
\label{A19a}
\end{eqnarray}
we shall approximate the term in the curly brackets of (\ref{A18i})
as follows [see eq.~(\ref{A18g})]
\begin{eqnarray}
&&1 - q^{2} \,\widetilde{B}(0, -2k_{1} \cdot q + q^{2} ,q^{2})
\left[
1 + (-2k_{1} \cdot q + q^{2})\,
C(-2k_{1} \cdot q + q^{2}) \right]^{-1} \nn \\
&&\qquad \approx 1 - q^{2} \,
\widetilde{B}(0, 0, q^{2}) = F_{\rm em}^{(\pi)}(q^{2})
\approx F_{M}(q^{2}) \,.
\label{A21a}
\end{eqnarray}
Here 
\begin{eqnarray}
F_{M}(q^{2}) = \frac{m_{0}^{2}}{m_{0}^{2}-q^{2}}\,, \quad 
m_{0}^{2} = 0.50~{\rm GeV}^{2}\,,
\label{A21b}
\end{eqnarray}
is a simple, but rather satisfactory, representation
of $F_{\rm em}^{(\pi)}(q^{2})$ in the region (\ref{A19a});
see e.g. eq.~(3.22) of \cite{Donnachie:2002}.
For a detailed description of $F_{\rm em}^{(\pi)}(q^{2})$
over a wider $q^{2}$ range see \cite{Melikhov:2003hs}.
Thus, in the present paper we set
\begin{eqnarray}
&&
%\left.
\widehat{\Gamma}_{\mu}^{(\gamma \pi \pi)}(k_{i}, k_{i}-q)\,
\Delta_{F}\big{[}(k_{i}-q)^{2}\big{]}
|_
{k_{i}^{2} = m_{\pi}^{2},\;
-0.5~{\rm GeV}^{2} < q^{2} \leqslant 0} \nn \\
%\right|_
%{\mathop{^{k_{i}^{2} = m_{\pi}^{2},}_{-0.5~{\rm GeV}^{2} < q^{2} \leqslant 0}}}
&&\qquad =
\frac{(2k_{i} -q)_{\mu}}{-2k_{i} \cdot q + q^{2} + i\varepsilon}
F_{M}(q^{2}) - q_{\mu} \frac{1 - F_{M}(q^{2})}{q^{2}}
\,, \nn \\
&& (i = 1, 2)\,.
\label{A21c}
\end{eqnarray}
To repeat: for $q^{2} = 0$ this is an exact result
up to an irrelevant gauge term.
For small photon virtualities this represents our model assumption.

For the photon-proton vertex we take the standard expression 
as in (B.47)--(B.53) of \cite{Bolz:2014mya}:
\begin{align}
&\includegraphics[width=140pt]{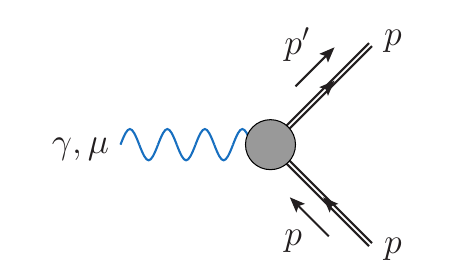}\nn \\
i\Gamma_{\mu}^{(\gamma pp)}(p',p)
&= -i e 
\left[\gamma_{\mu} F_{1}(t) + \frac{i}{2 m_{p}} \sigma_{\mu \nu}(p'-p)^{\nu}F_{2}(t) \right]\,,\nn \\
e &> 0, \qquad t = (p' - p)^{2}, \qquad 
\sigma_{\mu \nu} = \frac{i}{2}( \gamma_{\mu}\gamma_{\nu} - \gamma_{\nu}\gamma_{\mu})\,, \nn \\
F_{1}(t) &= \left( 1 - \frac{t}{4 m_{p}^{2}} \frac{\mu_{p}}{\mu_{N}} \right)
\left(1 - \frac{t}{4 m_{p}^{2}} \right)^{-1} G_{D}(t)\,, \nn\\
F_{2}(t) &= \left( \frac{\mu_{p}}{\mu_{N}} - 1 \right)
\left(1 - \frac{t}{4 m_{p}^{2}} \right)^{-1} G_{D}(t)\,, \nn\\
\mu_{N}  &= \frac{e}{2 m_{p}}, \qquad 
\frac{\mu_{p}}{\mu_{N}} = 2.7928\,,  \nn\\
G_{D}(t) &= \left(1 - \frac{t}{m_{D}^{2}} \right)^{-2} , \qquad 
m_{D}^{2} = 0.71~{\rm GeV}^{2}\,.
\label{A12a}
\end{align}
The proton's Dirac and Pauli form factors and the dipole form factor
are denoted by $F_{1}$, $F_{2}$, and $G_{D}$, respectively.
%see for instance Chapter~2 in \cite{Close:2007}.

\begin{figure}[tbp]
\centering
\includegraphics[width=300pt]{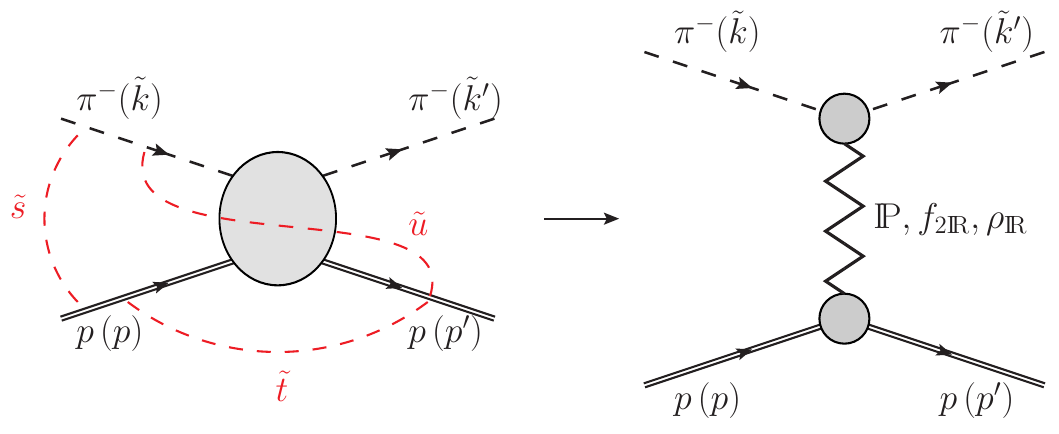}
\caption{\label{fig:200}
The reaction $\pi^{-} p \to \pi^{-} p$ and its representation
as a sum of Regge exchanges
$\Pom$, $f_{2 \Reg}$, and $\rho_{\Reg}$.
For $\pi^{+} p \to \pi^{+} p$ the diagrams are analogous.}
\end{figure}

The next topic to discuss is the hadronic reaction of an
on- or off-shell pion scattering on an on-shell proton
(see figure~\ref{fig:200})
\begin{eqnarray}
\pi^{\pm}(\tilde{k}) + p(p,\mathfrak{s}) \to \pi^{\pm}(\tilde{k}') + p(p',\mathfrak{s'})\,.
\label{A13}
\end{eqnarray}
Here $\tilde{k}$, $\tilde{k}'$  are the momenta of the on- and off--shell pions
and $p$, $p'$, $\mathfrak{s}$, $\mathfrak{s'}$ are the proton momenta and spin indices, respectively.
We have
\begin{eqnarray}
\tilde{k} + p = \tilde{k}' + p'\,.
\label{A14}
\end{eqnarray}
%
%------------------------------------------------------------------
%-------------------------------------------------------------------

Here we set
\begin{eqnarray}
\tilde{s} &=& (\tilde{k} + p)^{2} = (\tilde{k}' + p')^{2} \,, \nonumber \\
\tilde{t} &=& (\tilde{k} - \tilde{k}')^{2} = (p - p')^{2}\,, \nonumber \\
\tilde{u} &=& (\tilde{k} -p')^{2} = (p - \tilde{k}')^{2}\,, \nonumber \\
\tilde{\nu} &=& \frac{1}{4}(\tilde{s} - \tilde{u})
= \frac{1}{4} (\tilde{k} + \tilde{k}', p + p')
\,.
\label{A15}
\end{eqnarray}

The hadronic scattering amplitudes for (\ref{A13}) are denoted by
\begin{eqnarray}
{\cal M}_{\mathfrak{s'},\mathfrak{s}}^{(\pi^{\pm})}(\tilde{k}',p',\tilde{k},p)|_{h}\,.
\label{A16}
\end{eqnarray}
We shall consider these amplitudes in the Regge regime
where $\tilde{s}$ is large and $|\tilde{t}|$ stays limited.
As is well known, however, the correct Regge variables
to use are $2 \tilde{\nu}$ and $\tilde{t}$,
due to their crossing properties;
see for instance Chapter~6.4 of \cite{Collins:1977}
and section~6 of \cite{Ewerz:2013kda}.

We take the effective vertices involving 
the $\Pom$, $f_{2 \Reg}$, and $\rho_{\Reg}$ exchanges
in figure~\ref{fig:200}
as given in \cite{Ewerz:2013kda,Bolz:2014mya}:

%-----------------------------------------------------
\includegraphics[width=120pt]{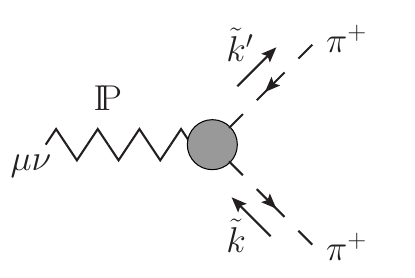} 
\includegraphics[width=120pt]{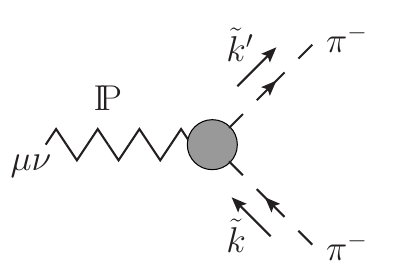} 
%\vspace*{-0.45cm}
\begin{align}
i\Gamma_{\mu \nu}^{(\Pom \pi \pi)}(\tilde{k}',\tilde{k})
&= -i 2 \beta_{\Pom \pi \pi} 
F_{M}[(\tilde{k}'-\tilde{k})^{2}]
\left[ (\tilde{k}' + \tilde{k})_{\mu} 
       (\tilde{k}' + \tilde{k})_{\nu}
- \frac{1}{4} g_{\mu \nu} (\tilde{k}' + \tilde{k})^{2} \right]\,,
\nn \\
\beta_{\Pom \pi \pi} &= 1.76 \; \mathrm{GeV}^{-1}\,,
\label{A17}
\end{align}
%-----------------------------------------------------

%-----------------------------------------------------
\includegraphics[width=120pt]{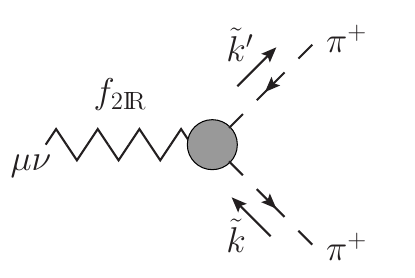} 
\includegraphics[width=120pt]{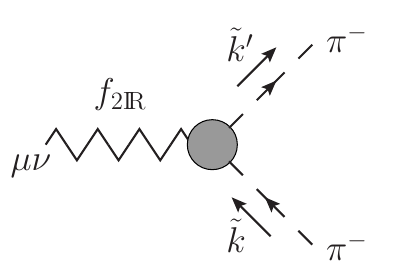} 
%\vspace*{-0.45cm}
\begin{align}
i\Gamma_{\mu \nu}^{(f_{2 \Reg} \pi \pi)}(\tilde{k}',\tilde{k})
&= -i \frac{g_{f_{2 \Reg}\pi \pi}}{2 M_{0}} 
F_{M}[(\tilde{k}'-\tilde{k})^{2}]
\left[ (\tilde{k}' + \tilde{k})_{\mu} 
       (\tilde{k}' + \tilde{k})_{\nu}
- \frac{1}{4} g_{\mu \nu} (\tilde{k}' + \tilde{k})^{2} \right]\,,
\nn \\
g_{f_{2 \Reg}\pi \pi} &= 9.30\,, \qquad M_{0} = 1\; {\rm GeV}\,,
\label{A18}
\end{align}
%-----------------------------------------------------

%-----------------------------------------------------
\includegraphics[width=120pt]{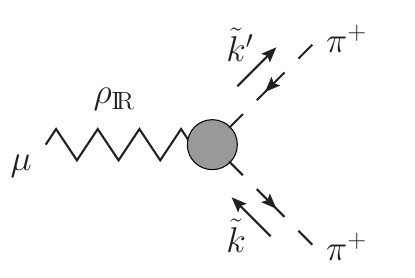} 
\includegraphics[width=120pt]{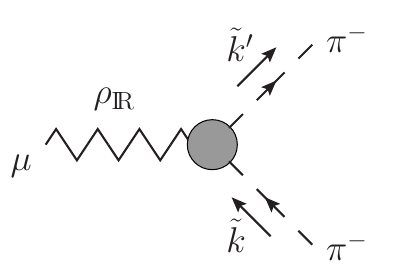} 
%\vspace*{-0.45cm}
\begin{align}
i\Gamma_{\mu}^{(\rho_{\Reg} \pi^{+} \pi^{+})}(\tilde{k}',\tilde{k})
&= 
-i\Gamma_{\mu}^{(\rho_{\Reg} \pi^{-} \pi^{-})}(\tilde{k}',\tilde{k})\nn\\
&=
-\frac{i}{2} g_{\rho_{\Reg}\pi \pi}
F_{M}[(\tilde{k}'-\tilde{k})^{2}]
(\tilde{k}' + \tilde{k})_{\mu} \,,
\nn \\
g_{\rho_{\Reg} \pi \pi} &= 15.63\,.
\label{A19}
\end{align}
%-----------------------------------------------------
Here $F_{M}(.)$ is a form factor which we choose,
as in (3.34) of \cite{Ewerz:2013kda},
for simplicity as the electromagnetic pion form factor (\ref{A21b}).

The vertices on the proton side of the diagram in figure~\ref{fig:200}
are again taken from \cite{Ewerz:2013kda}:

%----------------------------------------------------
\includegraphics[width=120pt]{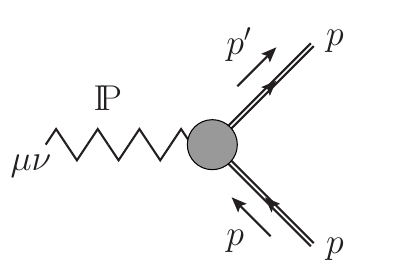} 
%\vspace*{-0.45cm}
\begin{align}
i\Gamma_{\mu \nu}^{(\Pom pp)}(p',p)
&=-i 3 \beta_{\Pom NN} F_{1}[(p'-p)^{2}]
\left[ \frac{1}{2} \gamma_{\mu}(p'+p)_{\nu} 
     + \frac{1}{2} \gamma_{\nu}(p'+p)_{\mu} 
- \frac{1}{4} g_{\mu \nu} (\slash{p}' + \slash{p}) \right], \nn \\
\beta_{\Pom NN} &= 1.87 \; \mathrm{GeV}^{-1}\,,
\label{A21}
\end{align}
%----------------------------------------------------

%%%%%%%%%%%%%%%%%%%%%%%%%%%%%%%%%%%%%%%
%\item $\Reg_{+} p p$ vertex, 
%where $\Reg_{+} = f_{2\Reg}, a_{2\Reg}$
%[see (3.49)--(3.52) 
%and section~6.3 of \cite{Ewerz:2013kda}]\\
%%%%%%%%%%%%%%%%%%%%%%%%%%%%%%%%%%%%%%%
\includegraphics[width=120pt]{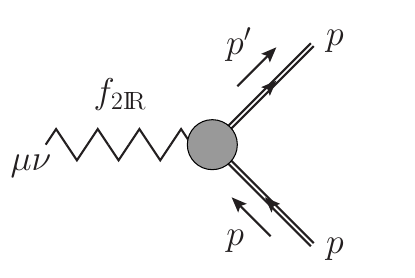} 
%\vspace*{-0.45cm}
\begin{align}
i\Gamma_{\mu \nu}^{(f_{2 \Reg} pp)}(p',p)
&=-i  \frac{g_{f_{2 \Reg} pp}}{M_{0}}F_{1}[(p'-p)^{2}]
\left[ \frac{1}{2} \gamma_{\mu}(p'+p)_{\nu} 
     + \frac{1}{2}  \gamma_{\nu}(p'+p)_{\mu} 
- \frac{1}{4} g_{\mu \nu} (\slash{p}' + \slash{p})\right],\nn\\
g_{f_{2 \Reg} pp} &= 11.04\,, \qquad M_{0} = 1 \; \mathrm{GeV}\,,
\label{A22}
\end{align}
%----------------------------------------------------

%%%%%%%%%%%%%%%%%%%%%%%%%%%%%%%%%%%%%%%
%\item $\Reg_{-} p p$ vertex,
%where $\Reg_{-} = \omega_{\Reg}, \rho_{\Reg}$
%[see (3.59)--(3.62) 
%and section~6.3 of \cite{Ewerz:2013kda}]\\
%%%%%%%%%%%%%%%%%%%%%%%%%%%%%%%%%%%%%%%
\includegraphics[width=120pt]{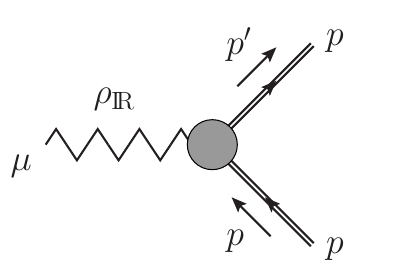} 
%\vspace*{-0.5cm}
\begin{align}
i\Gamma_{\mu}^{(\rho_{\Reg} pp)}(p',p)
&=-i g_{\rho_{\Reg} pp} F_{1}[(p'-p)^{2}] \gamma_{\mu}\,, \nn\\
g_{\rho_{\Reg} pp} 
&= 2.02\,.
\label{A23}
\end{align}
Here $F_{1}(.)$ is a form factor conventionally taken
as the Dirac electromagnetic form factor
of the proton; see (\ref{A12a}).

Finally we discuss the effective propagators for the Regge exchanges
$\Pom$, $f_{2 \Reg}$, and $\rho_{\Reg}$.
Here we follow again \cite{Ewerz:2013kda}
but we replace $\tilde{s}$ by $2 \tilde{\nu}$
which is the correct Regge variable to use
(see Chapter~6.4 of \cite{Collins:1977}).
The effective propagators are as follows:

\includegraphics[width=100pt]{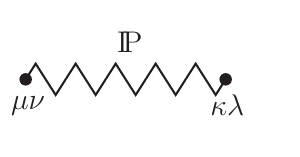}
\begin{eqnarray}
%\vspace*{-2cm}
&&i\Delta^{(\Pom)}_{\mu \nu, \kappa \lambda}(2 \tilde{\nu}, \tilde{t}) = 
\frac{1}{8 \tilde{\nu}} \left( g_{\mu \kappa} g_{\nu \lambda} 
                             + g_{\mu \lambda} g_{\nu \kappa}
- \frac{1}{2} g_{\mu \nu} g_{\kappa \lambda} \right)
(-i \,2 \tilde{\nu} \,\alpha'_{\Pom})^{\alpha_{\Pom}(\tilde{t})-1}\,, 
\label{A24}\\
&&\alpha_{\Pom}(\tilde{t}) = 1 + \epsilon_{\Pom}
+ \alpha'_{\Pom}\tilde{t}\,, \nn \\
%\qquad \alpha_{\Pom}(0) = 1 + \epsilon_{\Pom}\,, \nonumber\\
&&\epsilon_{\Pom} = 0.0808\,, \qquad
\alpha'_{\Pom} = 0.25 \; \mathrm{GeV}^{-2}\,;
\label{A25}
\end{eqnarray}

\includegraphics[width=100pt]{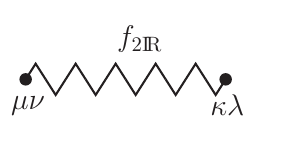} 
\begin{eqnarray}
%\vspace*{-1.8cm}
&&i\Delta^{(f_{2 \Reg})}_{\mu \nu, \kappa \lambda}(2 \tilde{\nu}, \tilde{t}) = 
\frac{1}{8 \tilde{\nu}} \left( g_{\mu \kappa} g_{\nu \lambda} 
                             + g_{\mu \lambda} g_{\nu \kappa}
- \frac{1}{2} g_{\mu \nu} g_{\kappa \lambda} \right)
(-i \,2 \tilde{\nu} \,\alpha'_{f_{2 \Reg}})^{\alpha_{f_{2 \Reg}}(\tilde{t})-1}\,,
\label{A26} \\
&&\alpha_{f_{2 \Reg}}(\tilde{t}) = \alpha_{f_{2 \Reg}}(0)
+ \alpha'_{f_{2 \Reg}}\tilde{t}\,, \nn \\
%\qquad \alpha_{\Pom}(0) = 1 + \epsilon_{\Pom}\,, \nonumber\\
&&\alpha_{f_{2 \Reg}}(0) = 0.5475\,, \qquad
\alpha'_{f_{2 \Reg}} = 0.9 \; \mathrm{GeV}^{-2}\,;
\label{A27} 
\end{eqnarray}

\includegraphics[width=100pt]{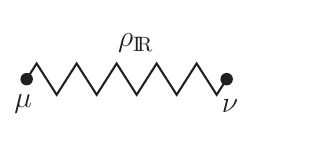} 
\begin{eqnarray}
%\vspace*{-1.8cm}
&&i\Delta^{(\rho_{\Reg})}_{\mu \nu}(2 \tilde{\nu}, \tilde{t}) = 
i g_{\mu \nu} \frac{1}{M_{-}^{2}}
(-i \,2 \tilde{\nu} \,\alpha'_{\rho_{\Reg}})^{\alpha_{\rho_{\Reg}}(\tilde{t})-1}\,,
\label{A28} \\
&&\alpha_{\rho_{\Reg}}(\tilde{t}) = \alpha_{\rho_{\Reg}}(0)
+ \alpha'_{\rho_{\Reg}}\tilde{t}\,, \nn \\
%\qquad \alpha_{\Pom}(0) = 1 + \epsilon_{\Pom}\,, \nonumber\\
&&\alpha_{\rho_{\Reg}}(0) = 0.5475\,, \qquad
\alpha'_{\rho_{\Reg}} = 0.9 \; \mathrm{GeV}^{-2}\,, \qquad
M_{-} = 1.41 \; \mathrm{GeV}\,.
\label{A29}
\end{eqnarray}
In (\ref{A25}), (\ref{A27}), and (\ref{A29})
we list the default values of the parameters 
for the $\Pom$, $f_{2 \Reg}$, and $\rho_{\Reg}$ exchanges 
from \cite{Ewerz:2013kda}.

Now we are ready to calculate the amplitudes (\ref{A16})
in the Regge limit.
The pomeron exchange term of figure~\ref{fig:200} gives
\begin{eqnarray}
i{\cal M}_{\mathfrak{s'},\mathfrak{s}}^{(\pi^{\pm})}(\tilde{k}',p',\tilde{k},p)|_{\Pom}
&=&
i\Gamma_{\mu \nu}^{(\Pom \pi \pi)}(\tilde{k}',\tilde{k})\,
i\Delta^{(\Pom) \mu \nu,\kappa \lambda}(2\tilde{\nu},\tilde{t})\;
\bar{u}_{\mathfrak{s'}}(p')
i\Gamma_{\kappa \lambda}^{(\Pom pp)}(p',p)
u_{\mathfrak{s}}(p)\,, \qquad
\label{A30}\\
{\cal M}_{\mathfrak{s'},\mathfrak{s}}^{(\pi^{\pm})}(\tilde{k}',p',\tilde{k},p)|_{\Pom}
&=&
i2\beta_{\Pom \pi \pi}3\beta_{\Pom NN} 
F_{M}(\tilde{t}) F_{1}(\tilde{t})
\frac{1}{8 \tilde{\nu}}(-i \,2 \tilde{\nu} \,\alpha'_{\Pom})^{\alpha_{\Pom}(\tilde{t})-1} \nn \\
&& \times \bar{u}_{\mathfrak{s'}}(p')
\left[
2(\tilde{k}' + \tilde{k})^{\nu} \gamma_{\nu} (\tilde{k}' + \tilde{k}, p'+p) - (\tilde{k}' + \tilde{k})^{2} m_{p}
\right]
u_{\mathfrak{s}}(p)\,.  \nn \\
\label{A31}
\end{eqnarray}
Similar expressions are easily derived for the $f_{2 \Reg}$
and $\rho_{\Reg}$ exchange contributions
to the amplitudes (\ref{A16}).
To write them in a convenient way we introduce 
the following functions
(cf. (2.9)--(2.14) of \cite{Lebiedowicz:2022nnn} for analogous functions)
\begin{eqnarray}
{\cal F}_{\Pom \pi p}(2 \tilde{\nu}, \tilde{t}) 
&=&
2 \beta_{\Pom \pi \pi}\,
3 \beta_{\Pom NN}\,
F_{M}(\tilde{t}) \,
F_{1}(\tilde{t}) 
\frac{1}{8 \tilde{\nu}}(-i \, 2 \tilde{\nu}\, \alpha'_{\Pom})^{\alpha_{\Pom}(\tilde{t})-1} \,, \nonumber\\
{\cal F}_{f_{2 \Reg} \pi p}(2 \tilde{\nu}, \tilde{t}) &=&
\frac{g_{f_{2 \Reg} \pi \pi} \,
g_{f_{2 \Reg} pp}}{2 M_{0}^{2}}\,
F_{M}(\tilde{t}) \,
F_{1}(\tilde{t}) 
\frac{1}{8 \tilde{\nu}}(-i \, 2 \tilde{\nu} \,
\alpha'_{f_{2 \Reg}})^{\alpha_{f_{2 \Reg}}(\tilde{t})-1} \,, \nonumber\\
{\cal F}_{\rho_{\Reg} \pi p}(2 \tilde{\nu}, \tilde{t}) &=&
\frac{g_{\rho_{\Reg} \pi \pi} \,
g_{\rho_{\Reg} pp}}{2 M_{-}^{2}}\,
F_{M}(\tilde{t}) \,
F_{1}(\tilde{t}) 
(-i \, 2 \tilde{\nu} \, \alpha'_{\rho_{\Reg}})^{\alpha_{\rho_{\Reg}}(\tilde{t})-1} \,.
\label{A32}
\end{eqnarray}
We get then
\begin{eqnarray}
{\cal M}_{\mathfrak{s'},\mathfrak{s}}^{(\pi^{\pm})}(\tilde{k}',p',\tilde{k},p)|_{\Pom}
&=& i {\cal F}_{\Pom \pi p}(2 \tilde{\nu}, \tilde{t}) \nn \\
&&\times
\bar{u}_{\mathfrak{s'}}(p')
\left[
2(\tilde{k}' + \tilde{k})^{\nu} \gamma_{\nu} (\tilde{k}' + \tilde{k}, p'+p) - (\tilde{k}' + \tilde{k})^{2} m_{p}
\right]
u_{\mathfrak{s}}(p),\qquad \quad
\label{A33}\\
{\cal M}_{\mathfrak{s'},\mathfrak{s}}^{(\pi^{\pm})}(\tilde{k}',p',\tilde{k},p)|_{f_{2 \Reg}}
&=& i {\cal F}_{f_{2 \Reg} \pi p}(2 \tilde{\nu}, \tilde{t})\nn \\
&&\times
\bar{u}_{\mathfrak{s'}}(p')
\left[
2(\tilde{k}' + \tilde{k})^{\nu} \gamma_{\nu} (\tilde{k}' + \tilde{k}, p'+p) - (\tilde{k}' + \tilde{k})^{2} m_{p}
\right]
u_{\mathfrak{s}}(p),\qquad \quad
\label{A34}\\
{\cal M}_{\mathfrak{s'},\mathfrak{s}}^{(\pi^{\pm})}(\tilde{k}',p',\tilde{k},p)|_{\rho_{\Reg}}
&=& \mp {\cal F}_{\rho_{\Reg} \pi p}(2 \tilde{\nu}, \tilde{t})\,
\bar{u}_{\mathfrak{s'}}(p')
(\tilde{k}' + \tilde{k})^{\nu} \gamma_{\nu} 
u_{\mathfrak{s}}(p)\,.
\label{A35}
\end{eqnarray}

Now we list our expressions for the odderon propagators and vertices.
For a single-pole odderon we have the ansatz
as in (B.37), (B.38) of \cite{Bolz:2014mya}
and (A7), (A8) of \cite{Lebiedowicz:2022nnn}:
\begin{eqnarray}
\includegraphics[width=130pt]{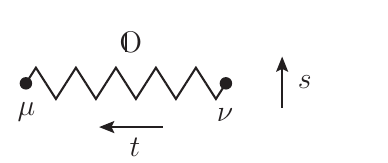} 
%\vspace*{-1.8cm}
&&i\Delta^{(\Ode)}_{\mu \nu}(s, t) = 
-i g_{\mu \nu} \frac{\eta_{\Ode}}{M_{0}^{2}}
(-i \,s \,\alpha'_{\Ode})^{\alpha_{\Ode}(t)-1}\,,
\label{A46} \\
&&
\alpha_{\Ode}(t) = \alpha_{\Ode}(0)
+ \alpha'_{\Ode}t\,, \nn \\
&&\alpha_{\Ode}(0) = 1 + \epsilon_{\Ode}\,, \qquad
\alpha'_{\Ode} = 0.25 \; \mathrm{GeV}^{-2}\,, \nn \\
&&
\eta_{\Ode} = \pm 1\,, \qquad
M_{0} = 1\; \mathrm{GeV}\,.
\label{A47}
\end{eqnarray}
For a double-pole odderon our ansatz is as in (A9) and (A10) of \cite{Lebiedowicz:2022nnn}:
\begin{equation}
i\widetilde{\Delta}^{(\Ode)}_{\mu \nu}(s,t) = i\Delta^{(\Ode)}_{\mu \nu}(s,t)
\left[
C_{1} + C_{2} \ln \left( -i \,s \,\alpha_{\Ode}' \right)
\right] \,,
\label{A47a}
\end{equation}
where $C_{1}$ and $C_{2}$ are real constants.
In \cite{Lebiedowicz:2022nnn} we found
%From the comparison with the $\rho$ values of $pp$ and $p \bar{p}$
%scattering we find that the following values give 
a reasonable description of the TOTEM data 
\cite{TOTEM:2016lxj,TOTEM:2017sdy}
for the following values of our parameters
[see (A10) and figure~6 of \cite{Lebiedowicz:2022nnn}]
\begin{eqnarray}
&&\eta_{\Ode} = -1\,, \quad 
%\alpha'_{\Ode} = 0.25 \; \mathrm{GeV}^{-2} \,, \quad 
\epsilon_{\Ode} = 0.0800 \,, \nonumber \\
&&(C_{1},C_{2}) = (-1.0,0.1), (-1.5,0.2), (-2.0,0.3)\,.
\label{A48}
\end{eqnarray}
In the present paper, we use $(C_{1},C_{2}) = (-1.5,0.2)$.

For the odderon vertices we make the same ansatz
as in \cite{Ewerz:2013kda,Bolz:2014mya}.
For the $\Ode pp$ vertex we have from
(3.68), (3.69) of \cite{Ewerz:2013kda}
and (B.103) of \cite{Bolz:2014mya}:
\begin{align}
\includegraphics[width=120pt]{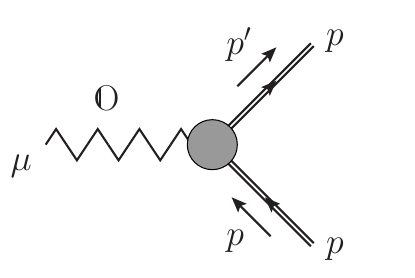} 
%\vspace*{-0.5cm}
i\Gamma_{\mu}^{(\Ode pp)}(p',p)
&=-i 3 \beta_{\Ode pp}  M_{0} F_{1}[(p'-p)^{2}] \gamma_{\mu}\,.
\label{A49}
\end{align}
For our study here we shall assume
$\beta_{\Ode pp} = 0.2$~GeV$^{-1}$;
see (A23) and figure~6 of \cite{Lebiedowicz:2022nnn}.
The form factor $F_{1}(.)$
is as in (\ref{A12a}).

The $\Ode \gamma f_{2}$ vertex is taken from
(B.104) and (B.105) of \cite{Bolz:2014mya}:
\begin{align}
\includegraphics[width=120pt]{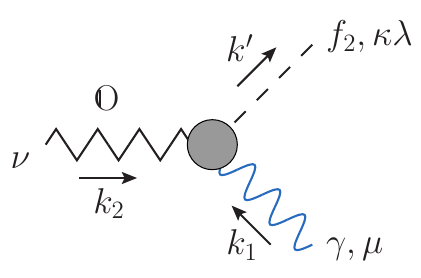}
i\Gamma_{\mu \nu \kappa \lambda}^{(\Ode \gamma f_{2})}(k',k_{1})
&=i e 
\tilde{F}^{(f_{2} \gamma \Ode)}(k'^{2},k_{1}^{2},k_{2}^{2})
%F_{M}(k_{1}^{2}) 
%F_{M}(k_{2}^{2})
%F^{(f_{2} \pi \pi)}(k'^{2}) 
\nn \\
& \quad \times
\left[
2 \hat{a}_{\Ode \gamma f_{2}} \Gamma^{(0)}_{\mu \nu \kappa \lambda}(k_{1},k_{2})
- \hat{b}_{\Ode \gamma f_{2}} \Gamma^{(2)}_{\mu \nu \kappa \lambda}(k_{1},k_{2})
\right],\nn \\
k' &= k_{1} + k_{2}\,.
\label{A50}
\end{align}
Here $\hat{a}_{\Ode \gamma f_{2}}$ and $\hat{b}_{\Ode \gamma f_{2}}$ 
are unknown coupling parameters.
%with dimensions GeV$^{-3}$  and GeV$^{-1}$, respectively.
The default values for them,
quoted in Table~1 of \cite{Bolz:2014mya}, are
$\hat{a}_{\Ode \gamma f_{2}} = 0.5$~GeV$^{-3}$,
$\hat{b}_{\Ode \gamma f_{2}} = 0.5$~GeV$^{-1}$.
The functions $\Gamma^{(0,2)}_{\mu \nu \kappa \lambda}$ are defined in 
(3.18), (3.19) of \cite{Ewerz:2013kda}
and (B.39), (B.40) of \cite{Bolz:2014mya}.
In the present paper, 
we assume the form factor 
$\tilde{F}^{(f_{2} \gamma \Ode)}(.)$
in the form
\begin{align}
\tilde{F}^{(f_{2} \gamma \Ode)}(k'^{2},k_{1}^{2},k_{2}^{2})
=
F^{(f_{2} \gamma \gamma)}(k'^{2})\,
F_{M}(k_{1}^{2}) \,
F_{M}(k_{2}^{2})
\label{A51}
\end{align}
with $F_{M}(k_{1}^{2})$,
$F_{M}(k_{2}^{2})$ as in (\ref{A21b})
and $F^{(f_{2} \gamma \gamma)}(k'^{2})$ as in (\ref{formfactor_f2gg}).

%--------------------------------------------------
\section{Comparison of the tensor-pomeron model with data for the $\gamma p \to \rho^{0} p$ and $\pi^{-} p \to \pi^{-} p$ reactions}
\label{sec:appendixB}
%--------------------------------------------------

In this section, we first compare our model results
with a compilation of data for the $\gamma p \to \rho^{0} p$ reaction 
from fixed-target and HERA measurements.
In the second part, we focus on the $\pi^{-} p \to \pi^{-} p$ reaction.
%in particular from the H1 experiment \cite{H1:2020lzc}.

For a detailed description of the amplitude for the 
$\gamma p \to \rho^{0} p$ reaction
within the tensor-pomeron model 
see section~II of \cite{Lebiedowicz:2014bea}.
In the following, 
we shall discuss the uncertainty of the model parameters,
such as the $\Pom \rho \rho$ and $f_{2 \Reg} \rho \rho$ coupling constants and the $\epsilon_{\Pom}$ parameter,
focusing primarily on the new H1 data \cite{H1:2020lzc}.

\begin{figure}[tbp]
\centering
\includegraphics[width=0.46\textwidth]{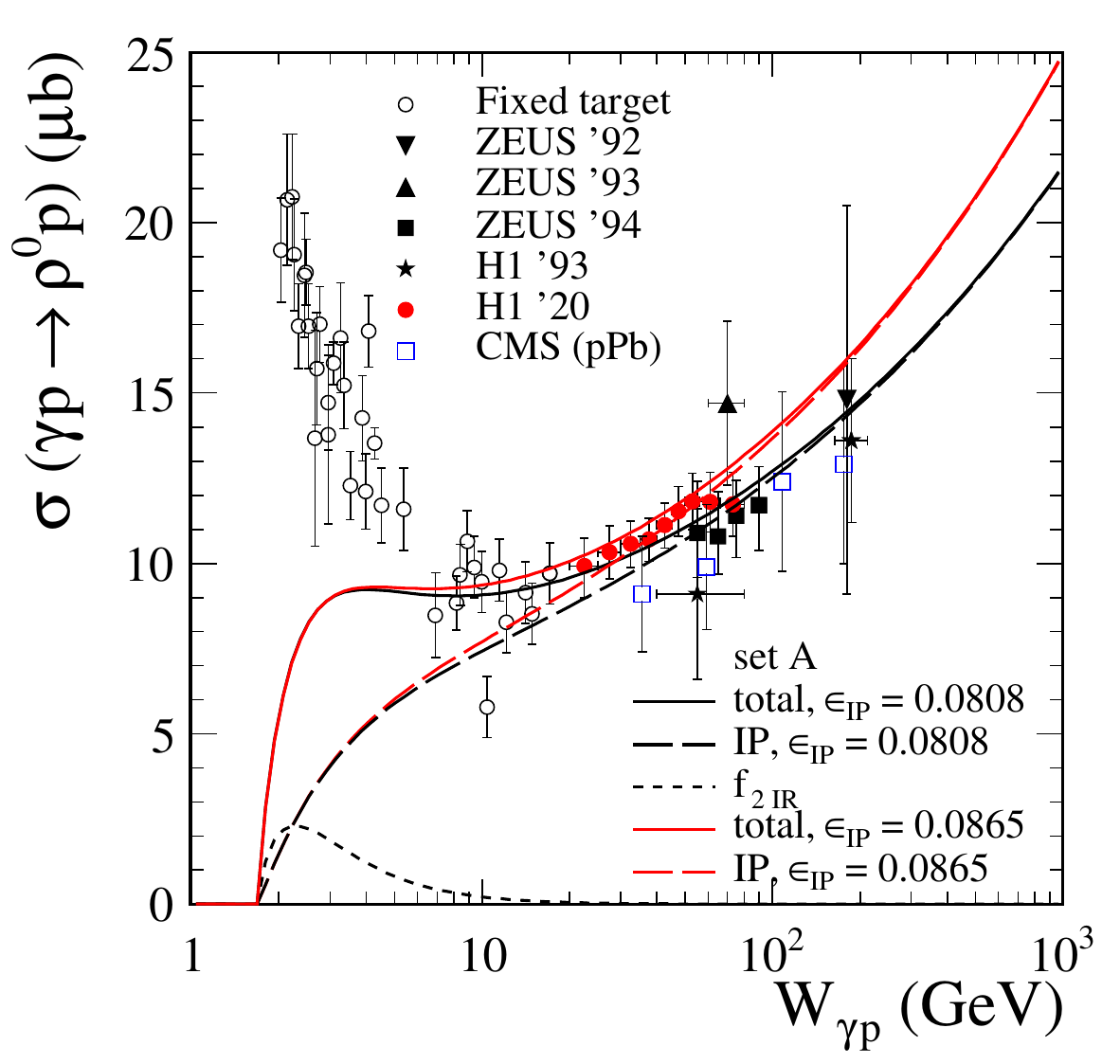}
\includegraphics[width=0.46\textwidth]{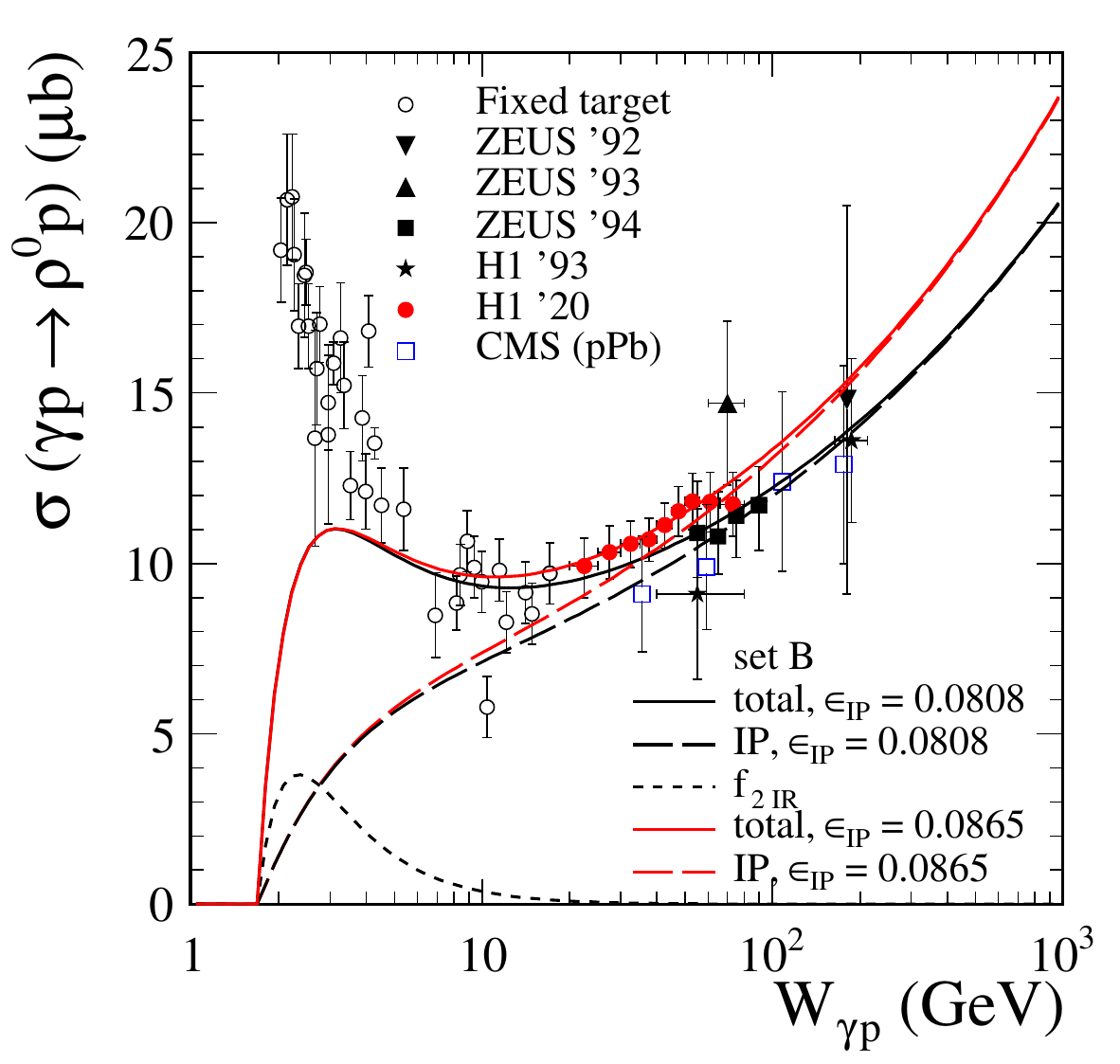}
\includegraphics[width=0.46\textwidth]{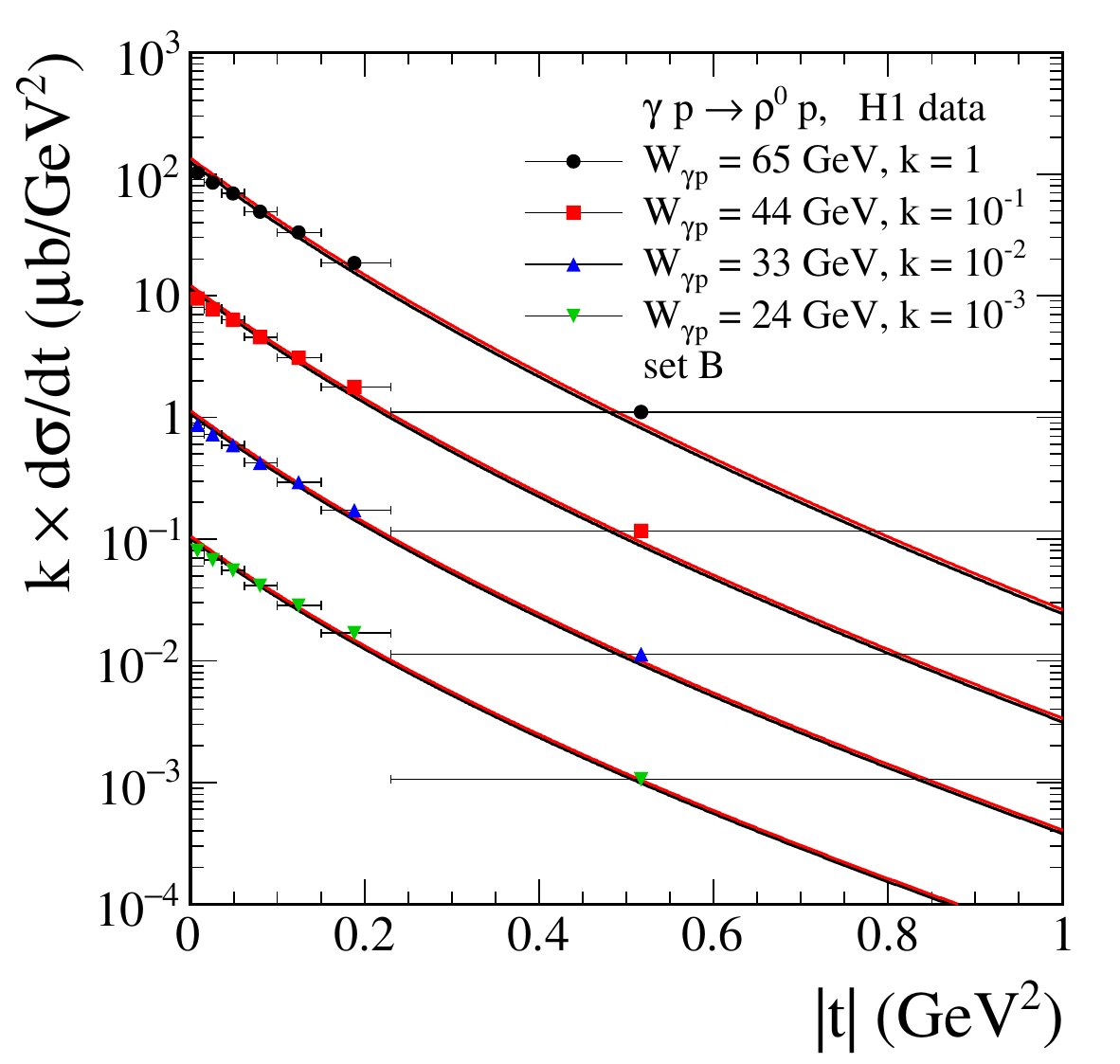}
  \caption{\label{fig:appendixB}
Top panels: Our results for the elastic cross section 
for the $\gamma p \to \rho^{0} p$ reaction 
as a function of the photon-proton c.m. energy $W_{\gamma p}$
compared to a compilation of fixed-target 
\cite{Ballam:1971wq,Park:1971ts,Struczinski:1975ik,Egloff:1979mg,Aston:1982hr}
and HERA data
\cite{ZEUS:1995bfs,H1:1996prv,ZEUS:1997rof,H1:2020lzc}.
The data points marked by the red circles come 
from the H1 measurement \cite{H1:2020lzc}.
Data obtained in ultraperipheral $p$-Pb collisions 
by the CMS Collaboaration \cite{CMS:2019awk} 
are shown by the blue open squares.
The left panel shows theoretical results 
obtained with the parameter set~A
of coupling constants given by (\ref{B1}),
while the right panel shows results 
obtained with the parameter set~B (\ref{B2}).
The solid line corresponds to the complete model result 
including both the pomeron and $f_{2 \Reg}$ exchanges. 
The individual pomeron and reggeon
exchange contributions are denoted by the long-dashed line 
and short-dashed line, respectively.
We show calculations for two values 
of the $\epsilon_{\Pom}$ parameter given in (\ref{B0}).
Bottom panel:
The differential cross section $d\sigma/dt$
together with the H1 data from \cite{H1:2020lzc}.
Here we show the results scaled by a factor $k$ 
(specified in the figure legend) 
for displaying purposes.
In the calculations we used parameter set B of coupling constants 
given by (\ref{B2}).
The meaning of the solid lines is the same as in the top right panel.}
\end{figure}

Figure~\ref{fig:appendixB} shows 
the complete theoretical results and individual components
to the cross sections $\sigma(W_{\gamma p})$
and $d\sigma/dt$
for the $\gamma p \to \rho^{0} p$ reaction
together with experimental data.
The complete cross section is a coherent sum
of pomeron and $f_{2 \Reg}$ exchange contributions 
in the amplitude.
These terms have different energy dependence.
At high energies, the dominant contribution
comes from the pomeron exchange. 
We see from the top panels of figure~\ref{fig:appendixB}
that at $W_{\gamma p} \gtrsim 100$~GeV
the $\gamma p \to \rho^{0} p$ cross section is completely
dominated by the pomeron exchange.
We show calculations for two values 
of the $\epsilon_{\Pom}$ parameter
in the pomeron trajectory:
\begin{eqnarray}
\epsilon_{\Pom} = 0.0808 \quad {\rm and} \quad \epsilon_{\Pom} = 0.0865\,.
\label{B0}
\end{eqnarray}
The larger value of $\epsilon_{\Pom}$ given in (\ref{B0})
is motivated by our analysis of the elastic $pp$ scattering
and by comparing the corresponding model to the LHC data
(see figure~5 of \cite{Lebiedowicz:2022nnn}).
For comparison, we show the results for two sets of parameters
for the pomeron and $f_{2 \Reg}$ coupling constants, 
set~A and set~B.
The parameter set~A corresponds to the default values 
of the $\Pom \rho \rho$ and $f_{2 \Reg} \rho \rho$ 
coupling constants as given in 
table~1 of \cite{Bolz:2014mya}.
This set of parameters was also used in \cite{Bolz_Meson2021} to compare the model results
with the H1 data.
For set B, we used the values from eq.~(2.14) of \cite{Lebiedowicz:2014bea}.
We have respectively:
\begin{align}
&{\rm set~A}:
a_{\Pom \rho \rho} = 0.45~{\rm GeV}^{-3}\,,\;
b_{\Pom \rho \rho} = 6.5~{\rm GeV}^{-1}\,, \;
a_{f_{2 \Reg} \rho \rho} = 2.92~{\rm GeV}^{-3}\,,\;
b_{f_{2 \Reg} \rho \rho} = 5.02~{\rm GeV}^{-1}\,;
\label{B1} \\
&{\rm set~B}: 
a_{\Pom \rho \rho} = 0.7~{\rm GeV}^{-3}\,,\;
b_{\Pom \rho \rho} = 6.2~{\rm GeV}^{-1}\,,\;
a_{f_{2 \Reg} \rho \rho} = 0\,,\;
b_{f_{2 \Reg} \rho \rho} = 9.3~{\rm GeV}^{-1}\,.
\label{B2}
\end{align}
Figure~\ref{fig:appendixB} shows that the model uncertainties
related to photoproduction of $\rho^{0}$ meson
are of order of 10\%.

%-------------------------------------------------------------
%--------------------------------------------------------------

\begin{figure}[tbp]
\centering
\includegraphics[width=0.7\textwidth]{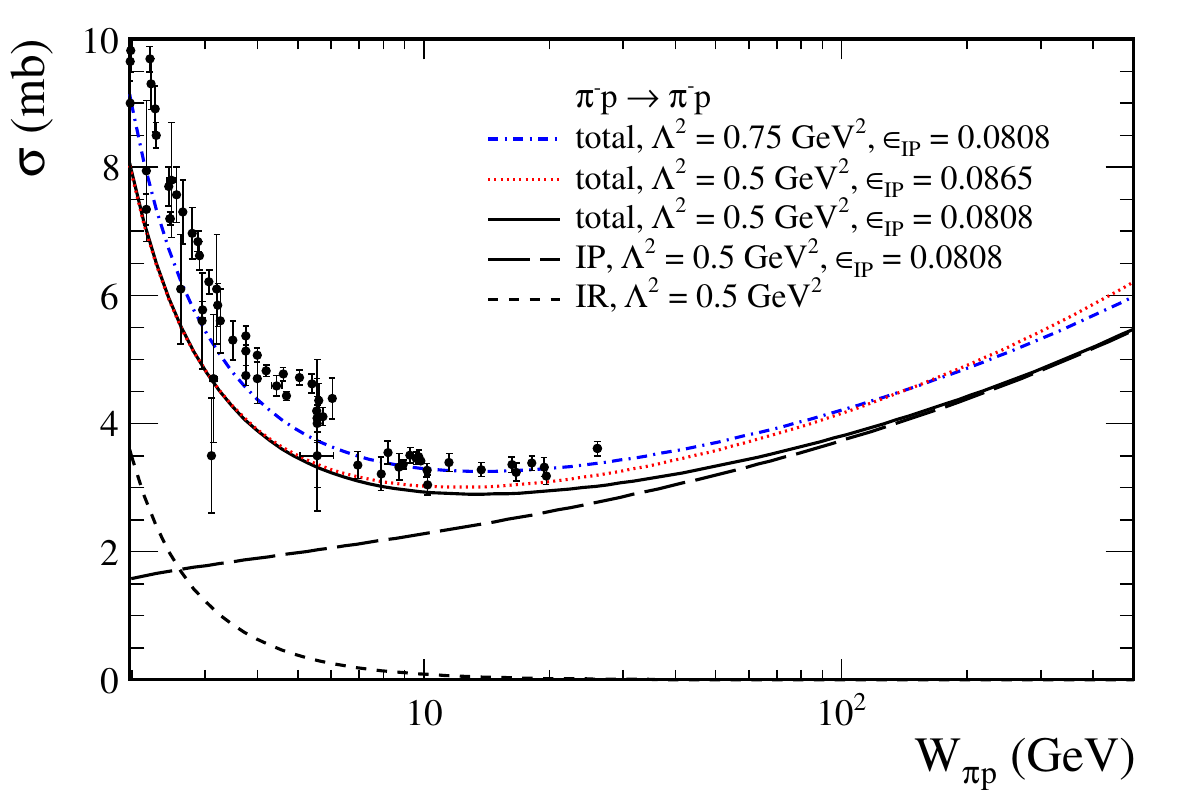}
\includegraphics[width=0.46\textwidth]{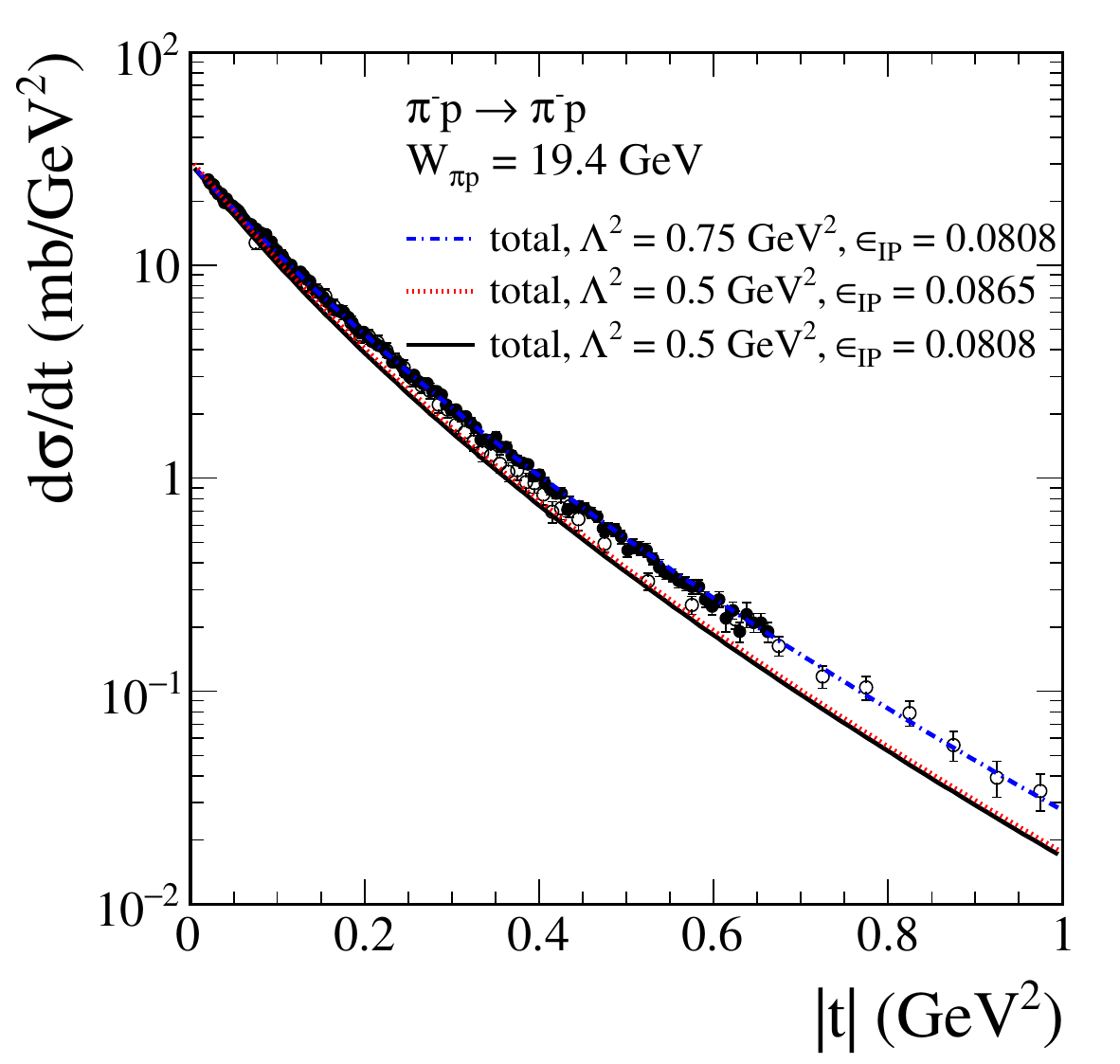}
  \caption{\label{fig:pimp}
Top panel: Our results for the elastic cross section
for the $\pi^{-} p \to \pi^{-} p$ reaction 
as a function of the pion-proton c.m. energy
$W_{\pi p}$ compared to
the experimental data from the Particle Data Group \cite{PDG}.
Shown are the complete theoretical results (total)
and individual components, $\Pom$ and $\Reg$,
obtained with 
$\Lambda^{2} = 0.50$~GeV$^{2}$
and $\epsilon_{\Pom} = 0.0808$ (see the black lines).
Here $\Lambda^{2}$ means $m_{0}^{2}$ in (\ref{A21b}).
Shown are also results calculated for 
$\Lambda^{2} = 0.75$~GeV$^{2}$ and $\epsilon_{\Pom} = 0.0808$
(see the blue dash-dotted line),
and for $\Lambda^{2} = 0.50$~GeV$^{2}$ and $\epsilon_{\Pom} = 0.0865$
(see the red dotted line).
Bottom panel:
The differential cross section $d\sigma/dt$
at $W_{\pi p} = 19.4$~GeV
together with the experimental data 
from \cite{Akerlof:1976gk,Schiz:1979rh}.
The meaning of the lines is the same as in the top panel.}
\end{figure}

Finally, figure~\ref{fig:pimp} shows the comparison of our model results with the experimental data
for the $\pi^{-} p \to \pi^{-} p$ reaction;
see (\ref{A13}) for on-shell pions.
In the top panel, the integrated cross section for
the $\pi^{-} p$ elastic scattering
versus the pion-proton c.m. energy $W_{\pi p}$
is presented. 
In the bottom panel, the differential distribution,
$d\sigma/dt$, at $W_{\pi p} = 19.4$~GeV is shown.
The amplitude for the process is given by
(\ref{A30})--(\ref{A35})
with the effective vertices and propagators
for the pomeron, $f_{2 \Reg}$, and $\rho_{\Reg}$
exchanges; see (\ref{A17})--(\ref{A29}).
Theoretical results are calculated 
for two parameters
$\Lambda^{2} \equiv m_{0}^{2}$ in (\ref{A21b}).
We choose $\Lambda^{2} = 0.50$~GeV$^{2}$,
the default value, and $\Lambda^{2} = 0.75$~GeV$^{2}$.
We see from figure \ref{fig:pimp} that
a better description of the data can be obtained by assuming 
$\Lambda^{2} = 0.75$~GeV$^{2}$ instead of 
$\Lambda^{2} = 0.50$~GeV$^{2}$;
compare the blue dash-dotted line to the black solid line,
respectively.
For comparison, we show the complete result 
corresponding to $\Lambda^{2} = 0.50$~GeV$^{2}$
and $\epsilon_{\Pom} = 0.0865$,
see the red dotted line.
In summary, the uncertainties of the model resulting from
comparison to the $\pi^{-} p$ scattering are less than 15\%.
%-------------------------------------------------------------
%--------------------------------------------------------------

\newpage
\textbf{Note added.}
In this article we have emphasized the importance of using the correct Regge variable $2 \nu$ instead of the squared energy variable $s$ for the calculations
of the Drell-S{\"o}ding diagrams (see figure~\ref{fig:100}).
In this way we could take into account that
$s_{2}$ and $s_{1}$ in figures \ref{fig:100}(a) and \ref{fig:100}(b),
respectively, are different in general.
In addition we got from the gauge invariance relation
a solution for the diagram~(c) of figure~\ref{fig:100} which is satisfactory from the QFT point of view. On the other hand, for the diagrams of figures~\ref{fig:400} and \ref{fig:401}, we approximated $2 \nu$ by the corresponding~$s$.
Here we want to show that this is legitimate for the high-energy regime which we are discussing. Let us, for instance, consider figure~\ref{fig:400} for the reaction $\gamma p \to \rho p$.
This is a two-body reaction where we have for $s$, $t$, $u$, and $2 \nu$ the following relations:
\begin{align}
s+t+u &= 2 m_{p}^{2} + m_{\rho}^{2}\,,\nonumber\\
2 \nu &= \frac{1}{2}(s-u) = s - \frac{1}{2}(-t+2m_{p}^{2}+m_{\rho}^{2})\,.\nonumber
\end{align}
We are considering diffractive reactions with $|t| \leqslant 1$~GeV$^{2}$.
Taking the energy range of HERA as an example,
roughly $\sqrt{s} = 10$~GeV to 300~GeV, we find
that $2 \nu$ and $s$ differ at most by 1.7\% to 0.02\textperthousand~for $|t| \leqslant 1$~GeV$^{2}$.

To conclude: for the high-energy diffractive reactions of figures~\ref{fig:400} and \ref{fig:401} the difference between the corresponding variables $2 \nu$ and $s$ can be neglected. If one wants to use our formulas for lower energies, for instance
$\sqrt{s} \approx 5$~GeV, it is no problem to replace $s$ in the appropriate places by $2 \nu$.

%--------------------
\acknowledgments
%--------------------
The authors would like to thank C. Ewerz and R. McNulty for useful discussions.
Furthermore, we would like to thank S. Schmitt and 
A. Bolz for correspondence and for agreeing that we use a figure
from \cite{Bolz_Meson2021} as our figure~\ref{fig:H1}.
The work of A.S. was partially supported by the Centre for Innovation and
Transfer of Natural Sciences and Engineering Knowledge in Rzesz\'ow (Poland).

%\paragraph{Note added.} ...

% Create the reference section using BibTeX:\clearpage

\end{document}